\pdfoutput=1
\documentclass[prd,showpacs,letterpaper,10pt,aps,notitlepage,nofootinbib,superscriptaddress]{revtex4-1}

\usepackage[utf8]{inputenc}
\usepackage{amsfonts}
\usepackage{amsmath}
\usepackage{graphicx}
\usepackage{hyperref}
\usepackage{placeins}
\usepackage{mathrsfs}
\usepackage{bm}
\usepackage[utf8]{inputenc}
\usepackage{multirow}
\usepackage{slashed}
\usepackage{dsfont}
\usepackage{cleveref}
\usepackage[export]{adjustbox}
\hypersetup{colorlinks, linkcolor = [rgb]{0,0.0,0.75}, citecolor = [rgb]{0,0.0,0.75}, urlcolor = [rgb]{0,0.0,0.75}}

\renewcommand{\hat}{\widehat}
\renewcommand{\tilde}{\widetilde}

\newcommand{\nb}{\phantom{0}}
\newcommand{\wm}{\phantom{-}}
\newcommand{\bs}[1]{\ensuremath{{\boldsymbol{#1}}}}

\begin{document}

\title{\texorpdfstring{$\bm{\Lambda_b \to \Lambda^*(1520)\ell^+\ell^-}$}{Lambdab to Lambda*(1520)l+l-} form factors from lattice QCD}

\author{Stefan Meinel}
\affiliation{Department of Physics, University of Arizona, Tucson, AZ 85721, USA}

\author{Gumaro Rendon}
\affiliation{Physics Department, Brookhaven National Laboratory, Upton, NY 11973, USA}

\date{April 15, 2021}

\begin{abstract}
We present the first lattice QCD determination of the $\Lambda_b \to \Lambda^*(1520)$ vector, axial vector, and tensor form factors that are relevant for the rare decays $\Lambda_b \to \Lambda^*(1520)\ell^+\ell^-$. The lattice calculation is performed in the $\Lambda^*(1520)$ rest frame with nonzero $\Lambda_b$ momenta, and is limited to the high-$q^2$ region. An interpolating field with covariant derivatives is used to obtain good overlap with the $\Lambda^*(1520)$. The analysis treats the $\Lambda^*(1520)$ as a stable particle, which is expected to be a reasonable approximation for this narrow resonance. A domain-wall action is used for the light and strange quarks, while the $b$ quark is implemented with an anisotropic clover action with coefficients tuned to produce the correct $B_s$ kinetic mass, rest mass, and hyperfine splitting. We use three different ensembles of lattice gauge-field configurations generated by the RBC and UKQCD collaborations, and perform extrapolations of the form factors to the continuum limit and physical pion mass. We give Standard-Model predictions for the $\Lambda_b \to \Lambda^*(1520)\ell^+\ell^-$ differential branching fraction and angular observables in the high-$q^2$ region.
\end{abstract}

\maketitle

\FloatBarrier
\section{Introduction}
\FloatBarrier

Decays of $b$-hadrons that proceed through the flavor-changing neutral current transition $b\to s\ell^+\ell^-$ play an important role in searching for physics beyond the Standard Model \cite{Blake:2016olu}. Global analyses of the increasingly precise experimental data point to lepton-flavor-nonuniversal shifts in one or more of the Wilson coefficients with respect to their Standard-Model values \cite{Alguero:2019ptt, Aebischer:2019mlg}. These deviations, along with further hints for violation of lepton-flavor universality in $b\to c\tau\bar{\nu}$ decays, have led to significant activity in constructing models of new fundamental physics, as reviewed for example in Ref.~\cite{Buttazzo:2017ixm}.

When searching for new physics in weak decays, it is important to consider multiple decay modes involving different species of hadrons. Different decay modes may be sensitive to different combinations of operators in the effective Hamiltonian, and will also differ in their experimental and theoretical systematic uncertainties. The benefits of $\Lambda_b$ baryon decays in constraining $\Delta B = \Delta S=1$ Wilson coefficients have been discussed by several authors \cite{Gremm:1995nx,Mannel:1997xy,Huang:1998ek,Hiller:2001zj,Chen:2002rg,Legger:2006cq,Hiller:2007ur,Boer:2014kda,Meinel:2016grj,Blake:2017une,Das:2018sms,Yan:2019tgn,Descotes-Genon:2019dbw,Blake:2019guk,Das:2020cpv}. Experimental data are available for the differential branching fraction and angular observables of $\Lambda_b\to\Lambda(\to p \pi^-)\mu^+\mu^-$ \cite{Aaltonen:2011qs,Aaij:2013mna,Aaij:2015xza,Aaij:2018gwm}, as well as the branching fraction of $\Lambda_b\to\Lambda\gamma$ \cite{Aaij:2019hhx}. In Ref.~\cite{Blake:2019guk}, an analysis of $b\to s\mu^+\mu^-$ Wilson coefficients using all 33 independent angular observables of $\Lambda_b\to\Lambda(\to p \pi^-)\mu^+\mu^-$ decays \cite{Aaij:2018gwm} and using $\Lambda_b\to \Lambda$ form factors from lattice QCD \cite{Detmold:2016pkz} was reported. Within the present uncertainties, the results are consistent both with the anomalies seen in $B$ meson decays and with the Standard Model \cite{Blake:2019guk}.

Going beyond the lightest $\Lambda$ baryon in the final state, the LHCb Collaboration has also reported first measurements of $\Lambda_b \to p K^- \ell^+ \ell^-$ decays, including $CP$ asymmetries \cite{Aaij:2017mib} and the muon-versus-electron ratio $R_{p K^-}$ \cite{Aaij:2019bzx}. The $\Lambda_b \to p K^- \mu^+ \mu^-$ $CP$ asymmetries were measured in the kinematic region with $m_{p K^-}<2350\:{\rm MeV}$ and $q^2=m^2_{\ell^+\ell^-}\notin [0.98,1.1] \cup [8.0,11] \cup [12.5,15]\:{\rm GeV^2}$ \cite{Aaij:2017mib} to avoid large contributions from the $\phi$, $J/\psi$, and $\psi^\prime$ resonances; the ratio $R_{p K^-}$ was measured for $m_{p K^-}<2600\:{\rm MeV}$ and $q^2\in [0.1,6.0]\:{\rm GeV^2}$ \cite{Aaij:2019bzx}.

The $p K^-$-invariant-mass distribution of $\Lambda_b \to p K^- \ell^+ \ell^-$ for $q^2$ away from the $\phi$, $J/\psi$, and $\psi^\prime$ resonances is expected to be similar to the distribution with $q^2$ on-resonance. This $p K^-$-invariant-mass distribution has been observed in $\Lambda_b \to p K^- J/\psi(\to \ell^+\ell^-)$ \cite{Aaij:2015tga}. As can be seen in Fig.~3 of Ref.~\cite{Aaij:2015tga}, a large number of $\Lambda^*$ baryon resonances contribute to this decay in overlapping mass regions. However, one
resonance produces a narrow peak that clearly stands out above the other contributions: the $\Lambda^*(1520)$, which has a width of $15.6\pm 1.0$ MeV \cite{Tanabashi:2018oca} and is the lightest resonance with $J^P=\frac32^-$. Thus, it may be feasible for LHCb to measure the $\Lambda_b \to \Lambda^*(1520)(\to p K^-)\ell^+\ell^-$ decay rate and angular observables for $q^2$ in the nonresonant (rare-decay) region.

The phenomenology of $\Lambda_b \to \Lambda^*(1520)(\to p K^-)\ell^+\ell^-$ was discussed in Refs.~\cite{Descotes-Genon:2019dbw,Das:2020cpv}, where the expressions for the complete angular distribution were given (for unpolarized $\Lambda_b$), approximate relations among the $\Lambda_b\to\Lambda^*(1520)$ form factors based on effective field theories were obtained, and numerical studies of the differential decay rate and angular observables were performed using form factors from a quark model \cite{Mott:2011cx}. The prospects for measurements of $\Lambda_b \to \Lambda^*(1520)(\to p K^-)\ell^+\ell^-$ angular observables at LHCb were recently studied in Ref.~\cite{Amhis:2020phx}. Earlier work had also considered the decay mode $\Lambda_b \to \Lambda^*(1520)(\to p K^-)\gamma$, primarily as a probe of the photon polarization in $b\to s \gamma$ \cite{Legger:2006cq,Hiller:2007ur}; the formalism for an amplitude analysis of $\Lambda_b \to p K^-\gamma$ was recently discussed also in Ref.~\cite{Albrecht:2020azd}. The authors of Ref.~\cite{Legger:2006cq} pointed out that this mode may be easier to reconstruct in hadron-collider experiments than $\Lambda_b \to \Lambda(\to p \pi^-)\gamma$, since the $\Lambda$ has a long lifetime of $c\tau \approx 7.9$ cm \cite{Tanabashi:2018oca} and, like the photon, often escapes the innermost vertex locator without leaving any trace.

To make predictions for the $\Lambda_b \to \Lambda^*(1520)(\to p K^-)\ell^+\ell^-$ decay observables in the Standard Model and beyond, the $\Lambda_b \to \Lambda^*(1520)$ form factors corresponding to the matrix elements of the $b\to s$ vector, axial vector, and tensor currents are required. These form factors have previously been studied in a quark model \cite{Pervin:2005ve,Mott:2011cx}. In the following, we present the first, exploratory lattice-QCD determination of the $\Lambda_b \to \Lambda^*(1520)$ form factors (we reported preliminary results in Ref.~\cite{Meinel:2016cxo}). The lattice calculation of $\frac12^+\to \frac32^-$ form factors is substantially more challenging than the calculation of $\frac12^+\to \frac12^+$ form factors, even when neglecting the strong decay of the $\frac32^-$ baryon in the analysis, as we do here. Correlation functions for negative-parity baryons have more statistical noise than correlation functions for the lightest positive-parity baryons. Furthermore, at nonzero momenta, the irreducible representations of the lattice symmetry groups mix positive and negative parities and also mix $J=\frac12$ and $J=\frac32$. To avoid having to deal with this mixing, we perform our calculation in the $\Lambda^*(1520)$ rest frame and give the $\Lambda_b$ nonzero momentum (since the $\Lambda_b$ is the ground state, the mixing with other $J^P$ values does not cause difficulties in isolating it). This has the effect that our calculation is limited to a relatively small kinematic region near $q^2_{\rm max}$.

This paper is organized as follows. Our definition of the $\Lambda_b \to \Lambda^*(1520)$ form factors is presented in Sec.~\ref{sec:FFdefs}. The lattice actions and parameters are given in Sec.~\ref{sec:latticeparams}. Section \ref{sec:twopoint} explains our choices of the baryon interpolating fields and contains numerical results for the hadron masses. The three-point functions and our method for extracting the individual form factors are described in Sec.~\ref{sec:threept}. We perform simple chiral, continuum, and kinematic extrapolations of the form factors as discussed in Sec.~\ref{sec:FFextrap}. We then use the extrapolated form factors to calculate the $\Lambda_b \to \Lambda^*(1520)\mu^+\mu^-$ differential decay rate and angular observables in the Standard Model, presented in Sec.~\ref{sec:observables}. Conclusions are given in Sec.~\ref{sec:conclusions}. Appendix \ref{sec:FFrelations} contains relations between our form factor definition and other definitions that have been used in the literature.

\FloatBarrier
\section{Definitions of the form factors}
\label{sec:FFdefs}
\FloatBarrier

The $\Lambda^*(1520)$ is the lightest of the strange baryon resonances with $I=0$ and $J^P=\frac32^-$. It has a mass of $1519.5\pm1.0$ MeV, a width of $15.6\pm 1.0$ MeV, and decays mainly into $N\bar{K}$, $\Sigma \pi$, or $\Lambda\pi\pi$ \cite{Tanabashi:2018oca}. In this work, we treat the $\Lambda^*(1520)$ as if it is a stable single-particle state. We expect this to be a reasonable approximation, given the relatively small width and given the other sources of uncertainty in our calculation. In the following, we denote the $\Lambda^*(1520)$ as simply $\Lambda^*$.

We are interested in the matrix elements $\langle \Lambda^*(\mathbf{p^\prime},s^\prime) |\, \bar{s}\Gamma b \, |  \Lambda_b(\mathbf{p},s) \rangle$ for $\Gamma\in\{ \gamma^\mu, \gamma^\mu\gamma_5, i\sigma^{\mu\nu}q_\nu, i\sigma^{\mu\nu}q_\nu\gamma_5\}$ with $q=p-p^\prime$. These matrix elements are described by fourteen independent form factors that are functions of $q^2$ only. Possible definitions of these form factors were given, for example, in Refs.~\cite{Leibovich:1997az,Pervin:2005ve,Meinel:2016cxo,Mott:2011cx,Boer:2018vpx,Descotes-Genon:2019dbw}. Here we use a helicity-based definition. We first presented such a definition in Ref.~\cite{Meinel:2016cxo}; the choice used here differs from that in Ref.~\cite{Meinel:2016cxo} only by a $q^2$-dependent rescaling to avoid divergences in the form factors at the endpoint $q^2_{\rm max}= (m_{\Lambda_b}- m_{\Lambda^*})^2$. We use the standard relativistic normalization of states,
\begin{eqnarray}
 \langle \Lambda_b(\mathbf{k},r) |  \Lambda_b(\mathbf{p},s)\rangle &=&  \delta_{rs} 2E_{\Lambda_b} (2\pi)^3 \delta^3(\mathbf{k}-\mathbf{p}), \\
 \langle \Lambda^*(\mathbf{k}^\prime,r^\prime) |  \Lambda^*(\mathbf{p}^\prime,s^\prime)\rangle &=&  \delta_{r^\prime s^\prime} 2E_{\Lambda^*} (2\pi)^3 \delta^3(\mathbf{k}^\prime-\mathbf{p}^\prime),
\end{eqnarray}
and introduce Dirac and Rarita-Schwinger spinors satisfying
\begin{eqnarray}
 \sum_s u(m_{\Lambda_b},\mathbf{p}, s) \bar{u}(m_{\Lambda_b},\mathbf{p}, s) &=& m_{\Lambda_b} + \slashed{p}, \\
 \sum_{s^\prime} u_\mu(m_{\Lambda^*},\mathbf{p}^\prime, s^\prime) \bar{u}_\nu(m_{\Lambda^*},\mathbf{p}^\prime, s^\prime) &=& -(m_{\Lambda^*}+\slashed{p}^\prime)\left(g_{\mu\nu}-\frac13\gamma_\mu\gamma_\nu - \frac{2}{3m_{\Lambda^*}^2}p^\prime_\mu p^\prime_\nu - \frac{1}{3m_{\Lambda^*}}(\gamma_\mu p^\prime_\nu - \gamma_\nu p^\prime_\mu)\right).
\end{eqnarray}
We introduce the notation
\begin{eqnarray}
 \langle \Lambda^*(\mathbf{p^\prime},s^\prime) |\, \bar{s}\Gamma b \, |  \Lambda_b(\mathbf{p},s) \rangle &=& \bar{u}_\lambda(m_{\Lambda^*},\mathbf{p^\prime}, s^\prime)  \:\mathscr{G}^{\lambda}[\Gamma]\:  u(m_{\Lambda_b},\mathbf{p}, s), \label{eq:Glambda}
\end{eqnarray}
and
\begin{equation}
 s_\pm = (m_{\Lambda_b}\pm m_{\Lambda^*})^2 - q^2.
\end{equation}
The form factors $f_0$, $f_+$, $f_\perp$, $f_{\perp^\prime}$, $g_0$, $g_+$, $g_\perp$, $g_{\perp^\prime}$, $h_+$, $h_\perp$, $h_{\perp^\prime}$, $\tilde{h}_+$, $\tilde{h}_\perp$, and $\tilde{h}_{\perp^\prime}$ are defined via
\begin{eqnarray}
\nonumber \mathscr{G}^{\lambda}[\gamma^\mu] &=&
 f_0 \frac{ m_{\Lambda^*}}{s_+}\,\frac{(m_{\Lambda_b}-m_{\Lambda^*})\,p^\lambda q^\mu}{q^2}   \\
\nonumber &&  + f_+ \frac{m_{\Lambda^*}}{s_-} \,\frac{(m_{\Lambda_b}+m_{\Lambda^*})\, p^\lambda ( q^2(p^\mu+p^{\prime \mu}) - (m_{\Lambda_b}^2-m_{\Lambda^*}^2) q^\mu   )}{q^2\, s_+}   \\
 \nonumber && + f_\perp \frac{m_{\Lambda^*}}{s_-} \left(p^\lambda \gamma^\mu - \frac{2\, p^\lambda(m_{\Lambda_b}p^{\prime \mu} + m_{\Lambda^*} p^\mu)}{s_+}    \right)   \\
 && + f_{\perp^\prime} \frac{m_{\Lambda^*}}{s_-} \left( p^\lambda \gamma^\mu - \frac{2\, p^\lambda p^{\prime \mu}}{m_{\Lambda^*}}
 + \frac{2\, p^\lambda(m_{\Lambda_b}p^{\prime \mu} + m_{\Lambda^*} p^\mu  )}{s_+} + \frac{s_-\,  g^{\lambda\mu}}{m_{\Lambda^*}}\right), \label{eq:Ggmu}
\end{eqnarray}
\begin{eqnarray}
\nonumber \mathscr{G}^{\lambda}[\gamma^\mu\gamma_5] &=&
 - g_0\gamma_5\,\frac{m_{\Lambda^*}}{s_-}\frac{(m_{\Lambda_b}+m_{\Lambda^*})\,p^\lambda q^\mu}{q^2}   \\
\nonumber &&  - g_+\gamma_5\,\frac{m_{\Lambda^*}}{s_+}\frac{(m_{\Lambda_b}-m_{\Lambda^*})\, p^\lambda ( q^2(p^\mu+p^{\prime \mu}) - (m_{\Lambda_b}^2-m_{\Lambda^*}^2) q^\mu   )}{q^2\, s_-}   \\
 \nonumber && - g_\perp\gamma_5 \frac{m_{\Lambda^*}}{s_+}\left(p^\lambda \gamma^\mu - \frac{2\, p^\lambda(m_{\Lambda_b}p^{\prime \mu} - m_{\Lambda^*} p^\mu  )}{s_-}    \right)   \\
 && - g_{\perp^\prime}\gamma_5 \frac{m_{\Lambda^*}}{s_+}\left(p^\lambda \gamma^\mu + \frac{2\, p^\lambda p^{\prime \mu}}{m_{\Lambda^*}}
 + \frac{2\, p^\lambda(m_{\Lambda_b}p^{\prime \mu}  - m_{\Lambda^*} p^\mu )}{s_-} - \frac{s_+\,  g^{\lambda\mu}}{m_{\Lambda^*}}\right),
\end{eqnarray}
\begin{eqnarray}
\nonumber \mathscr{G}^{\lambda}[i\sigma^{\mu\nu}q_\nu] &=&
 - h_+\frac{m_{\Lambda^*}}{s_-}\,\frac{ p^\lambda ( q^2(p^\mu+p^{\prime \mu}) - (m_{\Lambda_b}^2-m_{\Lambda^*}^2) q^\mu   )}{s_+}   \\
 \nonumber && - h_\perp\frac{m_{\Lambda^*}}{s_-} (m_{\Lambda_b}+m_{\Lambda^*}) \left(p^\lambda \gamma^\mu - \frac{2\, p^\lambda(m_{\Lambda_b}p^{\prime \mu} + m_{\Lambda^*} p^\mu)}{s_+}    \right)   \\
 && - h_{\perp^\prime}\frac{m_{\Lambda^*}}{s_-} (m_{\Lambda_b}+m_{\Lambda^*}) \left( p^\lambda \gamma^\mu - \frac{2\, p^\lambda p^{\prime \mu}}{m_{\Lambda^*}}
 + \frac{2\, p^\lambda(m_{\Lambda_b}p^{\prime \mu} + m_{\Lambda^*} p^\mu  )}{s_+} + \frac{s_-\,  g^{\lambda\mu}}{m_{\Lambda^*}}\right),
\end{eqnarray}
\begin{eqnarray}
\nonumber \mathscr{G}^{\lambda}[i\sigma^{\mu\nu}q_\nu\gamma_5] &=&
 - \tilde{h}_+\gamma_5\frac{m_{\Lambda^*}}{s_+}\,\frac{ p^\lambda ( q^2(p^\mu+p^{\prime \mu}) - (m_{\Lambda_b}^2-m_{\Lambda^*}^2) q^\mu   )}{s_-}   \\
 \nonumber && - \tilde{h}_\perp\gamma_5\frac{m_{\Lambda^*}}{s_+} (m_{\Lambda_b}-m_{\Lambda^*}) \left(p^\lambda \gamma^\mu - \frac{2\, p^\lambda(m_{\Lambda_b}p^{\prime \mu} - m_{\Lambda^*} p^\mu)}{s_-}    \right)   \\
 && - \tilde{h}_{\perp^\prime}\gamma_5\frac{m_{\Lambda^*}}{s_+} (m_{\Lambda_b}-m_{\Lambda^*}) \left( p^\lambda \gamma^\mu + \frac{2\, p^\lambda p^{\prime \mu}}{m_{\Lambda^*}}
 + \frac{2\, p^\lambda(m_{\Lambda_b}p^{\prime \mu} - m_{\Lambda^*} p^\mu  )}{s_-} - \frac{s_+\,  g^{\lambda\mu}}{m_{\Lambda^*}}\right), \label{eq:Gsmunug5}
\end{eqnarray}
where $\sigma^{\mu\nu}=\frac{i}{2}(\gamma^\mu\gamma^\nu-\gamma^\nu\gamma^\mu)$. The requirement that physical matrix elements are non-singular for $q^2 \to q^2_{\rm max}= (m_{\Lambda_b}- m_{\Lambda^*})^2$ imposes certain requirements on the behavior of the form factors in this limit \cite{Descotes-Genon:2019dbw}. More information on this behavior can be obtained from heavy-quark effective theory \cite{Boer:2018vpx} if the strange quark is treated as a heavy quark. For our definition, we expect all form factors to be finite and nonzero at $q^2=q^2_{\rm max}$. Relations between our form factors and other definitions used in the literature are given in Appendix \ref{sec:FFrelations}.

\FloatBarrier
\section{Lattice actions and parameters}
\label{sec:latticeparams}
\FloatBarrier

Our calculation utilizes three different ensembles of gauge-field configurations generated by the RBC and UKQCD collaborations \cite{Aoki:2010dy, Blum:2014tka}. These ensembles include the effects of 2+1 flavors of sea quarks, implemented with a domain-wall action \cite{Kaplan:1992bt, Furman:1994ky, Shamir:1993zy}; the gauge action used is the Iwasaki action \cite{Iwasaki:1984cj}. The main parameters of the ensembles and valence-quark actions are listed in Table \ref{tab:latticeparams}; see Table \ref{tab:hadronmasses} for the resulting hadron masses. To compute the $u$, $d$, and $s$-quark propagators, we use the same domain-wall action as for the sea-quarks, with valence light-quark masses equal to the sea light-quark masses, and valence strange-quark masses tuned to the physical values, which are slightly lower than the sea strange-quark masses. For the $b$-quark propagators, we use the anisotropic clover action discussed in Ref.~\cite{Aoki:2012xaa}, but with parameters newly tuned by us to obtain the correct $B_s$ kinetic mass, rest mass, and hyperfine splitting.

\begin{table}
 \begin{tabular}{lccccccccccc}
\hline\hline
Label & $N_s^3\times N_t$ & $\beta$  & $a$ [fm] & $am_{u,d}$ & $am_{s}^{(\mathrm{sea})}$ 
& $am_{s}^{(\mathrm{val})}$ & $a m_Q^{(b)}$ & $\nu^{(b)}$ & $c_{E,B}^{(b)}$   & $N_{\rm ex}$ & $N_{\rm sl}$ \\
\hline
C01  & $24^3\times64$ & $2.13$   & $0.1106(3)$  & $0.01\nb$  & $0.04$      & $0.0323$  & $7.3258$ & $3.1918$ & $4.9625$  & 283 & 9056  \\
C005 & $24^3\times64$ & $2.13$   & $0.1106(3)$  & $0.005$    & $0.04$      & $0.0323$  & $7.3258$ & $3.1918$ & $4.9625$  & 311 & 9952  \\
F004 & $32^3\times64$ & $2.25$   & $0.0828(3)$  & $0.004$    & $0.03$      & $0.0248$  & $3.2823$ & $2.0600$ & $2.7960$  & 251 & 8032  \\
\hline\hline
\end{tabular}
\caption{\label{tab:latticeparams} Lattice parameters for the three different ensembles of gauge-field configurations. The values of the lattice spacing, $a$, were determined in Ref.~\cite{Blum:2014tka}. The bottom quark is implemented with the action described in Ref.~\cite{Aoki:2012xaa}, where the parameters are denoted as $m_0=m_Q$, $\zeta=\nu$, $c_P=c_E=c_B$. Here we newly tuned the parameters to obtain the correct $B_s$ kinetic mass, rest mass, and hyperfine splitting. The last two columns give the numbers of exact (ex) and sloppy (sl) samples used for the calculation of the correlation functions with all-mode averaging \cite{Blum:2012uh,Shintani:2014vja}.}
\end{table}

Our calculation employs all-mode averaging \cite{Blum:2012uh,Shintani:2014vja} to reduce the cost for the light and strange quark propagators. On each gauge-configuration, we computed one exact sample for the relevant correlation functions (discussed in the following sections), as well as 32 ``sloppy'' samples with reduced conjugate-gradient iteration count in the computation of the light and strange quark propagators. For the light quarks, we also used deflation based on the lowest 400 eigenvectors to reduce the cost and improve the accuracy of the propagators. On a given gauge-field configuration, the different samples correspond to different source locations on a four-dimensional grid, with a randomly chosen overall offset.

\FloatBarrier
\section{Two-point functions and hadron masses}
\label{sec:twopoint}
\FloatBarrier

We now proceed to the discussion of the baryon interpolating fields. Our lattice calculation uses $m_u=m_d$ and neglects QED, which means that we have exact isospin symmetry, and the $\Lambda_b$ and $\Lambda^*(1520)$ both have $I=0$. The continuous space-time symmetries on the other hand are reduced to discrete symmetries by the cubic lattice. At zero momentum, the relevant symmetry group is $^2 O$, the double cover of the cubic group \cite{Johnson:1982yq}, and we still have the full parity symmetry. At zero momentum, the continuum $J^P=\frac12^{\pm}$ and $J^P=\frac32^{\pm}$ irreps subduce identically to the $G_{1}^{g/u}$ and $H^{g/u}$ irreps; the next-higher values of $J$ that appear in these irreps are $J=\frac72$ and $J=\frac52$, respectively. In this case we can therefore safely construct the interpolating fields for both the $\Lambda_b$ and the $\Lambda^*(1520)$ using continuum symmetries. At nonzero momenta, we no longer have parity symmetry, and the relevant symmetry groups are Little Groups of $^2 O$ \cite{Gockeler:2012yj, Morningstar:2013bda, Paul:2018yev}. An interpolating field that would have $J^P=\frac32^-$ in the continuum then also couples to $J^P=\frac32^+$, and in some cases even $J^P=\frac12^+$ (for example, for momentum direction $(0,1,1)$, the only irrep containing $J=\frac32$ also contains $J=\frac12$), which would make isolating the $\Lambda^*(1520)$ extremely difficult. For this reason, we perform the lattice calculation in the $\Lambda^*(1520)$ rest frame, giving nonzero momentum to the $\Lambda_b$ instead. Since the $\Lambda_b$ is the lightest baryon with quark content $udb$, any contributions from mixing with opposite parity and higher $J$ only appear as excited-state contamination, which will be suppressed exponentially for large Euclidean time separations.

We take the interpolating field for the $\Lambda_b$ in position space to be
\begin{eqnarray}
 \nonumber (O_{\Lambda_b})_\gamma &=& \frac12\epsilon^{abc}\:(C\gamma_5)_{\alpha\beta}\left(\tilde{d}^a_\alpha\:\tilde{u}^b_\beta\: \tilde{b}^c_\gamma - \tilde{u}^a_\alpha\:\tilde{d}^b_\beta\: \tilde{b}^c_\gamma \right) \\
 &=& \epsilon^{abc}\:(C\gamma_5)_{\alpha\beta}\:\tilde{d}^a_\alpha\:\tilde{u}^b_\beta\: \tilde{b}^c_\gamma,
\end{eqnarray}
where $\tilde{q}$ denotes a smeared quark field. We use gauge-covariant Gaussian smearing of the form 
\begin{equation}
\tilde{q} = \left( 1 +\frac{\sigma_\textrm{Gauss}^2}{4 N_\textrm{Gauss} }\tilde{\Delta}\right)^{N_\textrm{Gauss}} q, \label{eq:GaussSmearing}
\end{equation}
where
\begin{equation}
 \tilde{\Delta} q(x) = \frac{1}{a^2}\sum_{j=1}^3 \left[ \tilde{U}_j(x) q(x+a\hat{j}) - 2q(x) + \tilde{U}_j^\dag(x-a\hat{j}) q(x-a\hat{j})  \right],
\end{equation}
and the gauge links $\tilde{U}$ are APE-smeared (in the case of the up, down, and strange quarks) or Stout-smeared (in the case of the bottom quark). The values used for the smearing parameters are given in Table \ref{tab:smearingparams}. We average over ``forward'' and ``backward'' two-point functions given by
\begin{eqnarray}
 C^{(2,\Lambda_b,\mathrm{fw})}_{\alpha\beta}(\mathbf{p}, t) &=&  \sum_{\mathbf{y}} e^{-i\mathbf{p}\cdot (\mathbf{y}-\mathbf{x})} \left\langle (O_{\Lambda_b})_{\alpha}(x_0+t,\mathbf{y})\: \overline{(O_{\Lambda_b})}_{\beta}(x_0,\mathbf{x})  \right\rangle, \\
C^{(2,\Lambda_b,\mathrm{bw})}_{\alpha\beta}(\mathbf{p}, t) &=& \sum_{\mathbf{y}} e^{-i\mathbf{p}\cdot (\mathbf{x}-\mathbf{y})} \left\langle (O_{\Lambda_b})_{\alpha}(x_0,\mathbf{x})\: \overline{(O_{\Lambda_b})}_{\beta}(x_0-t,\mathbf{y}) \right\rangle.
\end{eqnarray}
The $\Lambda_b$ masses obtained from single-exponential fits in the time region of ground-state dominance are given in the last column of Table \ref{tab:hadronmasses}.

\begin{table}
	\begin{tabular}{lccccccccc} \hline \hline 
    & \multicolumn{4}{c}{Up, down, and strange quarks} & \hspace{2ex} & \multicolumn{4}{c}{Bottom quarks} \\
               & $N_\textrm{Gauss}$ & $\sigma_\textrm{Gauss}/a$ & $N_\textrm{APE}$ & $\alpha_\textrm{APE}$ && $N_\textrm{Gauss}$ & $\sigma_\textrm{Gauss}/a$ & $N_\textrm{Stout}$ & $\rho_\textrm{Stout}$  \\ \hline
    Coarse     & $\phantom{0}30$ & $4.350$ & $25$ & $2.5$           && $10$ & $2.000$ & $10$ & $0.08$ \\
    Fine       & $\phantom{0}60$ & $5.728$ & $25$ & $2.5$           && $16$ & $2.667$ & $10$ & $0.08$ \\ \hline \hline
	\end{tabular}
	\caption{\label{tab:smearingparams}Parameters for the smearing of the quark fields in the baryon interpolating fields. A single sweep of APE smearing \cite{Albanese:1987ds} with parameter $\alpha_\textrm{APE}$ is defined as in Eq.~(8) of Ref.~\cite{Bonnet:2000dc}, and we apply $N_\textrm{APE}$ such sweeps. The Stout smearing is defined in Ref.~\cite{Morningstar:2003gk}.}
\end{table}

\begin{figure}
 \includegraphics[width=0.45\linewidth]{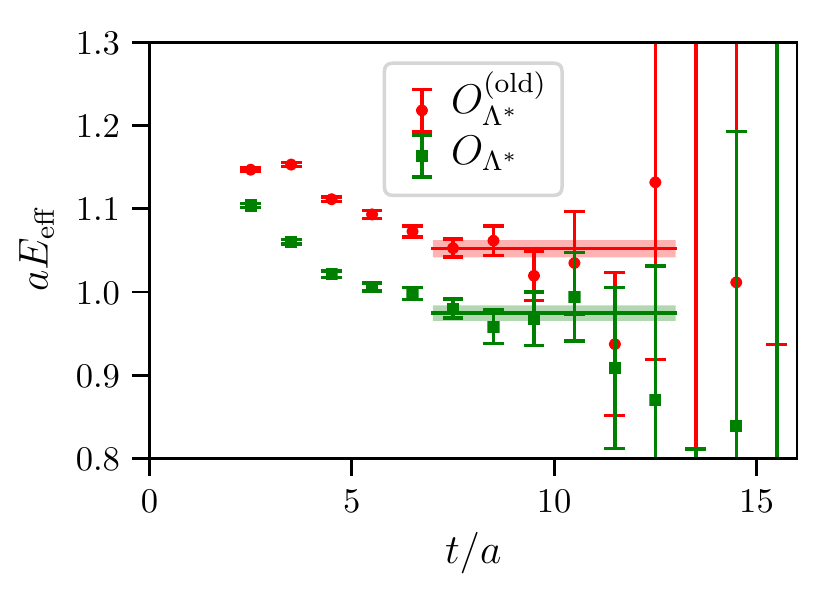}
 \caption{\label{fig:Lambdastar2pt}The effective masses computed for the two-point functions with the old and new $\Lambda^*$ interpolating fields, on the C005 ensemble. The horizontal lines indicate the time ranges used and energies obtained from single-exponential fits.}
\end{figure}

Even at zero momentum, constructing an interpolating field with a good overlap to the $\Lambda^*(1520)$ proved to be nontrivial. In a first, unsuccessful attempt, we tried the form
\begin{equation}
 (O_{\Lambda^*})_{j\gamma}^{({\rm old})} =  \epsilon^{abc}\:(C\gamma_j)_{\alpha\beta}\Big(\frac{1-\gamma_0}{2}\Big)_{\gamma\delta}\left( \tilde{u}^a_\alpha\:\tilde{s}^b_\beta\: \tilde{d}^c_\delta - \tilde{d}^a_\alpha\:\tilde{s}^b_\beta\: \tilde{u}^c_\delta \right), \label{eq:Lambda1}
\end{equation}
which can be projected to the $H^u$ irrep by contracting the index $j$ (which runs over the spatial directions) with\footnote{We use the Minkowski-space metric tensor $(g_{\mu\nu})=\mathrm{diag}(1,-1,-1,-1)$ and Minkowski-space gamma matrices throughout this paper, except where indicated with a subscript ``E.''}
\begin{equation}
 P^{kj}_{(3/2)} = g^{kj}-\frac13\gamma^k\gamma^j.
\end{equation}
Even though the resulting interpolating field has the correct values for all exactly conserved quantum numbers, it is found to have poor overlap with the $\Lambda^*(1520)$ and much greater overlap with higher-mass $J^P=\frac32^-$ states. The effective mass for the two-point function computed with $O_{\Lambda^*}^{({\rm old})}$ on the C005 ensemble is shown with the red circles in Fig.~\ref{fig:Lambdastar2pt}, and shows a ``false plateau'' at higher mass before the signal is swamped by noise. A previous lattice QCD study of $\Lambda^*$-baryon spectroscopy using interpolating fields similar to Eq.~(\ref{eq:Lambda1}) also did not find a $\Lambda^*(1520)$-like state \cite{Engel:2012qp}. The problem is that $O_{\Lambda^*}^{({\rm old})}$ [after projection with $P^{kj}_{(3/2)}$] has an internal structure corresponding to total quark spin $S=3/2$, total quark orbital angular momentum $L=0$, and flavor-$SU(3)$ octet, while quark models suggest that the $\Lambda^*(1520)$ dominantly has an $L=1$, $S=1/2$, and flavor-$SU(3)$-singlet structure \cite{Gromes:1982ze}. To obtain $L=1$, a suitable spatial structure of the interpolating field is needed, which can be achieved using covariant derivatives \cite{Edwards:2012fx}. For the main calculations in this work we use the form
\begin{equation}
 (O_{\Lambda^*})_{j\gamma} = \epsilon^{abc}\:(C\gamma_5)_{\alpha\beta}\Big(\frac{1+\gamma_0}{2}\Big)_{\gamma\delta}\left[ \tilde{s}^a_\alpha\:\tilde{d}^b_\beta\: (\tilde{\nabla}_j \tilde{u})^c_\delta  -  \tilde{s}^a_\alpha\:\tilde{u}^b_\beta\: (\tilde{\nabla}_j \tilde{d})^c_\delta  + \tilde{u}^a_\alpha\:(\tilde{\nabla}_j \tilde{d})^b_\beta\: \tilde{s}^c_\delta - \tilde{d}^a_\alpha\:(\tilde{\nabla}_j \tilde{u})^b_\beta\: \tilde{s}^c_\delta  \right], \label{eq:Lambda2}
\end{equation}
which has $L=1$, $S=1/2$, and is a flavor-$SU(3)$ singlet. The covariant derivatives, which are defined as
\begin{equation}
 \tilde{\nabla}_j \tilde{q}\,(x) = \frac{1}{2a} \left[ \tilde{U}_j(x) \tilde{q}(x+a\hat{j}) - \tilde{U}_j^\dag(x-a\hat{j}) \tilde{q}(x-a\hat{j})  \right],
\end{equation}
change the parity, so the projector $(1+\gamma_0)/2$ is used to obtain negative overall parity. As we did previously for $O_{\Lambda^*}^{({\rm old})}$, we project the two-point functions
\begin{eqnarray}
 C^{(2,\Lambda^*,\mathrm{fw})}_{jk\alpha\beta}(t) &=&  \sum_{\mathbf{y}} \left\langle (O_{\Lambda^*})_{j\alpha}(x_0+t,\mathbf{y})\: \overline{(O_{\Lambda^*})}_{k\beta}(x_0,\mathbf{x})  \right\rangle, \\
C^{(2,\Lambda^*,\mathrm{bw})}_{jk\alpha\beta}(t) &=& \sum_{\mathbf{y}}  \left\langle (O_{\Lambda^*})_{j\alpha}(x_0,\mathbf{x})\: \overline{(O_{\Lambda^*})}_{k\beta}(x_0-t,\mathbf{y}) \right\rangle
\end{eqnarray}
to the $H^u$ irrep with $P^{kj}_{(3/2)}$. In Eq.~(\ref{eq:Lambda2}), we eliminated covariant derivatives acting on the strange-quark fields using ``integration by parts,'' which is possible only at zero momentum. In this way, the calculation requires propagators with derivative sources only for the light quarks. The effective mass for $C^{(2,\Lambda^*)}$ computed on the C005 ensemble is shown with the green squares in Fig.~\ref{fig:Lambdastar2pt}, and shows a plateau at a significantly lower mass, which we identify (in the single-hadron/narrow-width approximation) with the $\Lambda^*(1520)$ resonance. The $\Lambda^*(1520)$ masses obtained from single-exponential fits in the plateau regions for all ensembles are given in the second-to-last column of Table \ref{tab:hadronmasses}.

\begin{table}
 \begin{tabular}{lccccccc}
\hline\hline
Label & $m_\pi$ [GeV]  & $m_K$ [GeV]   & $m_N$ [GeV]   & $m_\Lambda$ [GeV]  & $m_\Sigma$ [GeV] & $m_{\Lambda^*}$ [GeV] & $m_{\Lambda_b}$ [GeV] \\
\hline
C01  & $0.4312(13)$    & $0.5795(19)$  & $1.2647(51)$  & $1.3494(61)$       & $1.3877(61)$     & $1.825(16)$           & $5.793(17)$ \\
C005 & $0.3400(11)$    & $0.5501(19)$  & $1.1649(58)$  & $1.2659(66)$       & $1.3173(60)$     & $1.740(17)$           & $5.726(17)$ \\
F004 & $0.3030(12)$    & $0.5361(24)$  & $1.1197(59)$  & $1.2382(54)$       & $1.303(12)\nb$   & $1.757(15)$           & $5.722(23)$ \\
\hline\hline
\end{tabular}
\caption{\label{tab:hadronmasses} Hadron masses obtained from single-exponential fits to the respective two-point functions computed on the three different ensembles.}
\end{table}

We also computed the pion, kaon, nucleon, Lambda, and Sigma two-point functions and obtained the masses given in the same table. For the three ensembles we have, the mass differences $m_{\Lambda^*}-m_\Sigma-m_\pi$ are found to be in the range from approximately 80 to 150 MeV (physical value: 192 MeV), while $m_{\Lambda^*}-m_N-m_K$ ranges from approximately $-20$ to $+100$ MeV (physical value: 89 MeV). These results support our identification of the extracted energy level with the $\Lambda^*(1520)$ in the narrow-width approximation. A proper finite-volume scattering analysis with L\"uscher's method \cite{Briceno:2017max} is beyond the scope of this work. Here we just note that the lowest noninteracting $N$-$K$ and $\Sigma$-$\pi$ scattering states in the $H^u$ irrep must have nonzero back-to-back momenta and their energies are well above $m_{\Lambda^*}$ for our lattice volumes (this is another benefit of working in the $\Lambda^*$ rest frame).

For later reference, we also define overlap factors of the interpolating fields with the baryon states of interest as
\begin{eqnarray}
 \langle 0 | O_{\Lambda_b}  | \Lambda_b(\mathbf{p},s) \rangle &=& (Z_{\Lambda_b}^{(1)}+Z_{\Lambda_b}^{(2)}\gamma^0) \: u(m_{\Lambda_b},\mathbf{p}, s),
\end{eqnarray}
and
\begin{eqnarray}
 \langle 0 | (O_{\Lambda^*})_j  | \Lambda^*(\mathbf{0},s^\prime) \rangle &=& Z_{\Lambda^*}\:\frac{1+\gamma_0}{2} u_j(m_{\Lambda^*}, \mathbf{0}, s^\prime).
\end{eqnarray}
As everywhere in this paper, $|\Lambda^*(\mathbf{0},s^\prime) \rangle$ denotes the lowest-energy $3/2^-$ state. For the $\Lambda_b$ at nonzero momentum, it is necessary to have the two separate coefficients $Z_{\Lambda_b}^{(1)}$ and $Z_{\Lambda_b}^{(2)}$ that may also depend on $\mathbf{p}$, because the spatial-only smearing of the quark fields breaks hypercubic symmetry (and because the lattice itself also breaks the Lorentz symmetry). The spectral decomposition of $C^{(2,\:\Lambda_b)}(\mathbf{p},t)$ then reads
\begin{eqnarray}
\nonumber C^{(2,\:\Lambda_b)}(\mathbf{p},t) &=& \frac{1}{2v^0} (Z_{\Lambda_b}^{(1)}+Z_{\Lambda_b}^{(2)}\gamma^0) (1+\slashed{v}) (Z_{\Lambda_b}^{(1)}+Z_{\Lambda_b}^{(2)}\gamma^0)\, e^{-E_{\Lambda_b}t} \\
&& + \:\:(\text{excited-state contributions})
\end{eqnarray}
with $v^\mu=p^\mu/m_{\Lambda_b}$, while the spectral decomposition of $C^{(2,\Lambda^*)}(t)$ after projection with $P_{(3/2)}$ becomes
\begin{eqnarray}
\nonumber P^{jl}_{(3/2)} C^{(2,\Lambda^*)}_{lk}(t)  &=&  -\frac12 Z_{\Lambda^*}^2 (1+\gamma_0) \left(g^j_{\:\:k}-\frac13\gamma^j\gamma_k \right) \, e^{-m_{\Lambda^*} t} \\
&& + \:\:(\text{excited-state contributions}).
\end{eqnarray}
The excited-state contributions decay exponentially faster with $t$ than the ground-state contributions shown here.

\FloatBarrier
\section{Three-point functions and form factors}
\label{sec:threept}
\FloatBarrier

To determine the form factors, we compute forward and backward three-point functions
\begin{eqnarray}
C^{(3,{\rm fw})}_{j\,\gamma\,\delta}(\mathbf{p},\Gamma, t, t^\prime) &=& \sum_{\mathbf{y},\mathbf{z}} e^{-i\mathbf{p}\cdot(\mathbf{y}-\mathbf{z})} \left\langle (O_{\Lambda^*})_{j\gamma}(x_0,\mathbf{x})\:\: J_\Gamma(x_0-t+t^\prime,\mathbf{y})\:\: (\overline{O_{\Lambda_b}})_{\delta} (x_0-t,\mathbf{z}) \right\rangle, \label{eq:threeptfw} \\
C^{(3,\mathrm{bw})}_{j\,\delta\,\gamma}(\mathbf{p},\Gamma, t, t-t^\prime) &=& \sum_{\mathbf{y},\mathbf{z}}
e^{-i\mathbf{p}\cdot(\mathbf{z}-\mathbf{y})} \Big\langle (O_{\Lambda_b})_{\delta}(x_0+t,\mathbf{z})\:\: J_\Gamma^\dag(x_0+t^\prime,\mathbf{y})
\:\: (\overline{O_{\Lambda^*}})_{j\gamma} (x_0,\mathbf{x}) \Big\rangle, \label{eq:threeptbw}
\end{eqnarray}
where $\mathbf{p}$ is the momentum of the $\Lambda_b$, $\Gamma$ is the Dirac matrix in the $b\to s$ current $J_\Gamma$, $t$ is the source-sink separation, and $t^\prime$ is the current-insertion time. To match the currents to the continuum $\overline{\rm MS}$ scheme, we employ the mostly nonperturbative method described in Refs.~\cite{Hashimoto:1999yp, ElKhadra:2001rv}. Specifically, we use
\begin{equation}
 J_\Gamma=\rho_\Gamma\sqrt{Z_V^{(ss)} Z_V^{(bb)}} \left[ \bar{s}\: \Gamma\: b + a\, d_1\,\bar{s}\: \Gamma\: \bs{\gamma}_{\rm E}\cdot\bs{\nabla}  b \right], \label{eq:improvedcurrent}
\end{equation}
where $Z_V^{(ss)}$ and $Z_V^{(bb)}$ are the matching factors of the temporal components of the $s\to s$ and $b\to b$ vector currents, determined nonperturbatively using charge conservation, $\rho_\Gamma$ are residual matching factors that are numerically close to 1 and are computed using one-loop lattice perturbation theory \cite{Lehner:2012bt}, and the term with coefficient $d_1$ removes $\mathcal{O}(a)$ discretization errors at tree level. In Eq.~(\ref{eq:improvedcurrent}), $\bs{\gamma}_{\rm E}$ denotes the three Euclidean spatial gamma matrices, $\gamma_{\rm E}^j=-i\gamma^j$. The values of $Z_V^{(ss)}$, $Z_V^{(bb)}$, and $d_1$ are given in Table \ref{tab:matching}. For the residual matching factors $\rho_\Gamma$ of the vector and axial-vector currents, we use the one-loop values given in Table III of Ref.~\cite{Detmold:2015aaa}. These matching factors were computed for slightly different values of the parameters in the $b$-quark action \cite{Aoki:2012xaa}, but are not expected to depend strongly on these parameters. For the residual matching factors of the tensor currents, one-loop results were not available and we set them to the tree-level values equal to unity. Following Ref.~\cite{Detmold:2016pkz}, we estimate the resulting systematic uncertainty in the tensor form factors at scale $\mu=m_b$ to be equal to 2 times the maximum value of $|\rho_{\gamma^{\mu}}-1|$, $|\rho_{\gamma^{\mu}\gamma_5}-1|$, which is $0.05316$. Note that the contributions from the operator $O_7$ in the weak Hamiltonian to the $\Lambda_b \to \Lambda^*(1520)\ell^+\ell^-$ differential decay rate at high $q^2$ are relatively small, so the larger systematic uncertainty in the tensor form factors is unproblematic.

\begin{table}[b]
 \begin{tabular}{llccc}
\hline\hline
       & & $Z_V^{(bb)}$  & $Z_V^{(ss)}$    & $d_1^{(b)}$  \\
\hline
Coarse & & $9.0631(84)$  & $0.71273(26)$   &  $0.0728$    \\
Fine   & & $4.7449(21)$  & $0.7440(18)\nb$ &  $0.0696$    \\
\hline\hline
\end{tabular}
\caption{\label{tab:matching} Matching parameters. We determined the values of $Z_V^{(bb)}$ using the charge-conservation condition from ratios of $B_s$ two-point and three-point functions. The values of $Z_V^{(ss)}$ are taken from Ref.~\cite{Blum:2014tka}. The $\mathcal{O}(a)$-improvement coefficients $d_1^{(b)}$ were computed at tree level in mean-field-improved perturbation theory.}
\end{table}

Both the forward and backward three-point functions are computed using light and strange quark propagators with sources (Gaussian-smeared, with and without derivatives) located at $(x_0,\mathbf{x})$. Given the more complicated interpolating field for the $\Lambda^*$ (compared to that for the $\Lambda$ in Ref.~\cite{Detmold:2016pkz}), here we apply the sequential-source method for the $b$-quark propagators through the weak current, and not through the $\Lambda_b$ interpolating field as was done in Ref.~\cite{Detmold:2016pkz}. This method fixes $t^\prime$ rather than $t$, but we only computed the three-point functions for $t=2t^\prime$, $t=2t^\prime+a$, and $t=2t^\prime-a$.  We generated data for nine different separations on the coarse lattices and ten different separations on the fine lattices, as shown in Table \ref{tab:seps}.

\begin{figure}
 \includegraphics[width=0.4\linewidth]{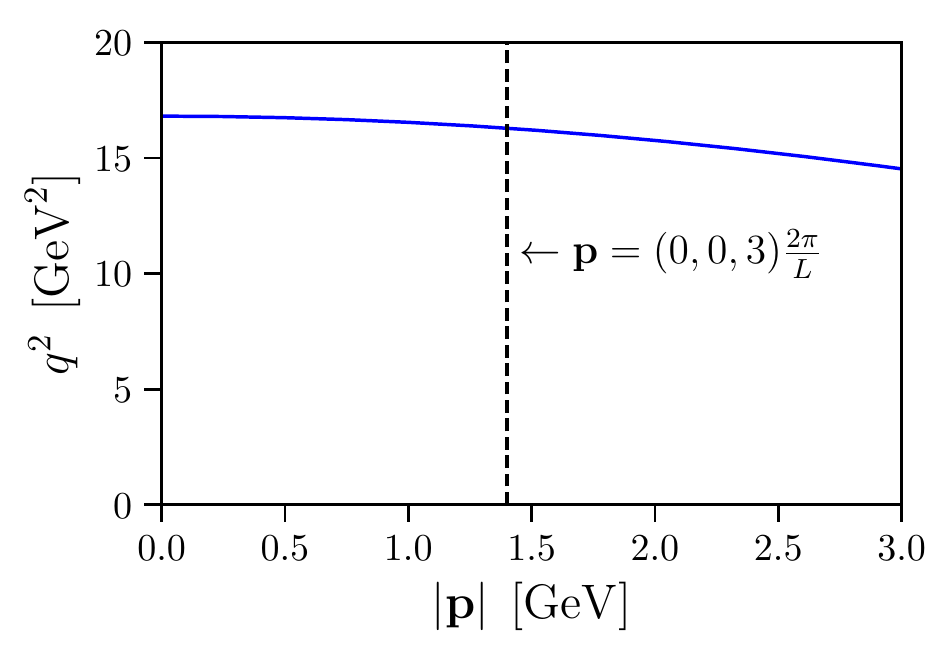}
 \caption{\label{fig:qsqrvsp}The value of the four-momentum transfer squared as a function of the $\Lambda_b$ momentum in the $\Lambda^*$ rest frame. The vertical dashed line indicates the largest momentum we use in this calculation.}
\end{figure}

Due to the large mass of the $\Lambda_b$, large values of $\mathbf{p}$ are needed to appreciably move $q^2$ away from $q^2_{\rm max}$, as shown in Fig.~\ref{fig:qsqrvsp}. At the same time, discretization errors are expected to grow with $\mathbf{p}$, and the number of $b$-quark sequential propagators that need to be computed is proportional to the number of choices for $\mathbf{p}$. In this first lattice study of the $\Lambda_b\to \Lambda^*$ form factors, we therefore used only two different choices: $\mathbf{p}=(0,0,2)\frac{2\pi}{L}$ and $\mathbf{p}=(0,0,3)\frac{2\pi}{L}$. Here, $L=N_s a$ are the spatial lattice extents, which are approximately 2.7 fm for all three ensembles.

\begin{table}[h]
\begin{tabular}{ccccc}
\hline\hline
       & & $t/a$ \\
\hline
Coarse & & $4,5,...,12$ \\
Fine   & & $5,6,...,14$ \\
\hline\hline
\end{tabular}
\caption{\label{tab:seps}The source-sink separations for which we computed the three-point functions on the coarse (C01, C005) and fine (F004) ensembles.}
\end{table}

After projection with $P_{(3/2)}$, the spectral decomposition of the forward three-point function reads
\begin{eqnarray}
\nonumber P_{(3/2)}^{jl} \:  C^{(3,{\rm fw})}_l(\mathbf{p},\Gamma, t, t^\prime) &=&  -\frac{1}{v^0}  Z_{\Lambda^*} \frac{1+\gamma_0}{2} \left(g^j_{\:\:\lambda}-\frac13\gamma^j\gamma_\lambda - \frac{1}{3}\gamma^j g_{0\lambda} \right) \:\mathscr{G}^{\lambda}[\Gamma]\: \frac{1+\slashed{v}}{2} (Z_{\Lambda_b}^{(1)}+Z_{\Lambda_b}^{(2)}\gamma^0) \:e^{-m_{\Lambda^*} (t-t^\prime)} e^{-E_{\Lambda_b}t^\prime} \\
&& + \:\:(\text{excited-state contributions}),
\end{eqnarray}
while the decomposition of the backward three-point function is given by the Dirac adjoint. Here, $\mathscr{G}^{\lambda}[\Gamma]$ are, up to small lattice-discretization and finite-volume effects, the linear combinations of form factors defined in Eqs.~(\ref{eq:Ggmu})-(\ref{eq:Gsmunug5}).

To extract the form factors, we utilize two different types of combinations of correlation functions. The first type (Sec.~\ref{sec:FFsquares}) allows us to extract the absolute magnitudes of individual form factors, but not their relative signs. The second type (Sec.~\ref{sec:FFratios}) allows us to extract ratios of different form factors in which the sign information is preserved.

\subsection{Extracting the squares of individual form factors}
\label{sec:FFsquares}

To remove the unwanted overlap factors and cancel the exponential time-dependence for the ground-state contribution, we form the ratios
\begin{eqnarray}
 {\mathscr{R}^{jk\mu\nu}(\mathbf{p},t,t^\prime)^X} &=& \frac{ \mathrm{Tr}\Big[ P_{(3/2)}^{jl} \:  C^{(3,{\rm fw})}_l(\mathbf{p},\Gamma_X^\mu, t, t^\prime) \:\: (1+\slashed{v}) \:\: C^{(3,{\rm bw})}_{m}(\mathbf{p},\Gamma_X^\nu, t, t-t^\prime) \: P_{(3/2)}^{mk} \Big] }{\mathrm{Tr}\Big[P_{(3/2)}^{lm} \: C^{(2,\Lambda^*)}_{lm}(t) \Big]\mathrm{Tr}\Big[(1+\slashed{v}) \:\: C^{(2,\Lambda_b)}(\mathbf{p},t)  \Big]},
\end{eqnarray}
where $X\in\{V,A,TV,TA\}$ and $\Gamma_V^\mu=\gamma^\mu$, $\Gamma_A^\mu=\gamma^\mu\gamma_5$, $\Gamma_{TV}^\mu=i\sigma^{\mu\nu}q_\nu$, $\Gamma_{TA}^\mu=i\sigma^{\mu\nu}\gamma_5q_\nu$, and the traces are over the Dirac indices. To isolate the individual helicity form factors, we then contract with the timelike, longitudinal, and transverse polarization vectors
\begin{equation}
 \epsilon^{(0)} = (\,q^0,\: \mathbf{q}\,), \hspace{3ex}
 \epsilon^{(+)} = (\,|\mathbf{q}|,\: (q^0/|\mathbf{q}|)\mathbf{q}\,), \hspace{3ex}
 \epsilon^{(\perp,\,j)} = (\,0,\: \mathbf{e}_j \times \mathbf{q}\,), \label{eq:polarizationvectors}
\end{equation}
and define
\begin{eqnarray}
{\mathscr{R}_{0}^X(\mathbf{p},t,t^\prime)} &=& g_{jk}\, {\epsilon^{(0)}_\mu  \epsilon^{(0)}_\nu} \, {\mathscr{R}^{jk\mu\nu}(\mathbf{p},t,t^\prime)^X}, \\
{\mathscr{R}_{+}^X(\mathbf{p},t,t^\prime)} &=&  g_{jk}\,{\epsilon^{(+)}_\mu  \epsilon^{(+)}_\nu} \, {\mathscr{R}^{jk\mu\nu}(\mathbf{p},t,t^\prime)^X}, \\
{\mathscr{R}_{\perp}^X(\mathbf{p},t,t^\prime)} &=&  p_j\,p_k\,{\epsilon^{(\perp,l)}_\mu  \epsilon^{(\perp,l)}_\nu} \, {\mathscr{R}^{jk\mu\nu}(\mathbf{p},t,t^\prime)^X}, \\
{\mathscr{R}_{\perp^\prime}^X(\mathbf{p},t,t^\prime)} &=& \left[{\epsilon^{(\perp,m)}_j\epsilon^{(\perp,m)}_k} - \frac12\, p_j\, p_k \right]{\epsilon^{(\perp,l)}_\mu  \epsilon^{(\perp,l)}_\nu} \, {\mathscr{R}^{jk\mu\nu}(\mathbf{p},t,t^\prime)^X}.
\end{eqnarray}
Repeated Latin indices are summed only over the spatial directions, while repeated Greek indices are summed over all four spacetime directions. The above quantities are equal to the squares of the individual form factors times certain combinations of the hadron masses and energies. For a given value of $t$, the excited-state contamination will be minimal for $t^\prime=t/2$. Using this choice and removing the kinematic factors, we evaluate \allowdisplaybreaks
\begin{eqnarray}
 {R_0^V(\mathbf{p}, t)} &=& \frac{48\, E_{\Lambda_b}}{(E_{\Lambda_b}-m_{\Lambda_b})(m_{\Lambda_b}-m_{\Lambda^*})^2}\:\mathscr{R}_{0}^V(\mathbf{p},t,t/2)  \:\: \nonumber\\
 &=&\:\: f_0^2 \:+\: (\text{excited-state contributions}), \hspace{2ex} \label{eq:R0V} \\\nonumber\\
 {R_+^V(\mathbf{p}, t)} &=& \frac{48\, E_{\Lambda_b}}{(E_{\Lambda_b}+m_{\Lambda_b})(m_{\Lambda_b}+m_{\Lambda^*})^2}\:\mathscr{R}_{+}^V(\mathbf{p},t,t/2) \:\:\nonumber\\
 &=&\:\: f_+^2 \:+\: (\text{excited-state contributions}), \hspace{2ex} \label{eq:RplusV} \\\nonumber\\
 {R_\perp^V(\mathbf{p}, t)} &=& -\frac{36\, E_{\Lambda_b}}{(E_{\Lambda_b}-m_{\Lambda_b})^2(E_{\Lambda_b}+m_{\Lambda_b})^3}\:\mathscr{R}_{\perp}^V(\mathbf{p},t,t/2)  \:\:\nonumber\\
 &=&\:\: f_\perp^2 \:+\: (\text{excited-state contributions}),  \\\nonumber\\
 {R_{\perp^\prime}^V(\mathbf{p}, t)} &=& -\frac{8\, E_{\Lambda_b}}{(E_{\Lambda_b}-m_{\Lambda_b})^2(E_{\Lambda_b}+m_{\Lambda_b})^3}\:\mathscr{R}_{\perp^\prime}^V(\mathbf{p},t,t/2)  \:\:\nonumber\\
 &=&\:\: f_{\perp^\prime}^2 \:+\: (\text{excited-state contributions}), \label{eq:RperpprimeV}\\\nonumber\\
 {R_0^A(\mathbf{p}, t)} &=& \frac{48\, E_{\Lambda_b}}{(E_{\Lambda_b}+m_{\Lambda_b})(m_{\Lambda_b}+m_{\Lambda^*})^2 }\:\mathscr{R}_{0}^A(\mathbf{p},t,t/2) \:\: \nonumber\\
 &=&\:\: g_0^2 \:+\: (\text{excited-state contributions}), \hspace{2ex} \label{eq:R0A} \\\nonumber\\
 {R_+^A(\mathbf{p}, t)} &=& \frac{48\, E_{\Lambda_b}}{(E_{\Lambda_b}-m_{\Lambda_b})(m_{\Lambda_b}-m_{\Lambda^*})^2}\:\mathscr{R}_{+}^A(\mathbf{p},t,t/2) \:\:\nonumber\\
 &=&\:\: g_+^2 \:+\: (\text{excited-state contributions}), \hspace{2ex} \\\nonumber\\
 {R_\perp^A(\mathbf{p}, t)} &=& -\frac{36\, E_{\Lambda_b} }{(E_{\Lambda_b}+m_{\Lambda_b})^2(E_{\Lambda_b}-m_{\Lambda_b})^3}\:\mathscr{R}_{\perp}^A(\mathbf{p},t,t/2)  \:\:\nonumber\\
 &=&\:\: g_\perp^2 \:+\: (\text{excited-state contributions}),  \\\nonumber\\
 {R_{\perp^\prime}^A(\mathbf{p}, t)} &=& -\frac{8\, E_{\Lambda_b} }{(E_{\Lambda_b}+m_{\Lambda_b})^2(E_{\Lambda_b}-m_{\Lambda_b})^3}\:\mathscr{R}_{\perp^\prime}^A(\mathbf{p},t,t/2)  \:\:\nonumber\\
 &=&\:\: g_{\perp^\prime}^2 \:+\: (\text{excited-state contributions}), \label{eq:RperpprimeA} \\\nonumber\\
 {R_+^{TV}(\mathbf{p}, t)} &=& \frac{48\, E_{\Lambda_b}}{(E_{\Lambda_b}+m_{\Lambda_b})\,q^4}\:\mathscr{R}_{+}^{TV}(\mathbf{p},t,t/2) \:\:\nonumber\\
 &=&\:\: h_+^2 \:+\: (\text{excited-state contributions}), \hspace{2ex} \\\nonumber\\
 {R_\perp^{TV}(\mathbf{p}, t)} &=&-\frac{36\, E_{\Lambda_b}}{(E_{\Lambda_b}+m_{\Lambda_b})^3(E_{\Lambda_b}-m_{\Lambda_b})^2(m_{\Lambda_b}+m_{\Lambda^*})^2}\:\mathscr{R}_{\perp}^{TV}(\mathbf{p},t,t/2)  \:\:\nonumber\\
 &=&\:\: h_\perp^2 \:+\: (\text{excited-state contributions}),  \\\nonumber\\
 {R_{\perp^\prime}^{TV}(\mathbf{p}, t)} &=&-\frac{8\, E_{\Lambda_b} }{(E_{\Lambda_b}+m_{\Lambda_b})^3(E_{\Lambda_b}-m_{\Lambda_b})^2(m_{\Lambda_b}+m_{\Lambda^*})^2}\:\mathscr{R}_{\perp^\prime}^{TV}(\mathbf{p},t,t/2)  \:\:\nonumber\\
 &=&\:\: h_{\perp^\prime}^2 \:+\: (\text{excited-state contributions}), \label{eq:RperpprimeTV} \\\nonumber\\
 {R_+^{TA}(\mathbf{p}, t)} &=& \frac{48\, E_{\Lambda_b}}{(E_{\Lambda_b}-m_{\Lambda_b})\,q^4}\:\mathscr{R}_{+}^{TA}(\mathbf{p},t,t/2) \:\:\nonumber\\
 &=&\:\: \tilde{h}_+^2 \:+\: (\text{excited-state contributions}), \hspace{2ex} \\\nonumber \\
 {R_\perp^{TA}(\mathbf{p}, t)} &=&-\frac{36\, E_{\Lambda_b}}{(E_{\Lambda_b}-m_{\Lambda_b})^3(E_{\Lambda_b}+m_{\Lambda_b})^2(m_{\Lambda_b}-m_{\Lambda^*})^2}\:\mathscr{R}_{\perp}^{TA}(\mathbf{p},t,t/2)  \:\:\nonumber\\
 &=&\:\: \tilde{h}_\perp^2 \:+\: (\text{excited-state contributions}), \label{eq:RperprimeTA}  \\\nonumber\\
 {R_{\perp^\prime}^{TA}(\mathbf{p}, t)} &=&-\frac{8\, E_{\Lambda_b} }{(E_{\Lambda_b}-m_{\Lambda_b})^3(E_{\Lambda_b}+m_{\Lambda_b})^2(m_{\Lambda_b}-m_{\Lambda^*})^2}\:\mathscr{R}_{\perp^\prime}^{TA}(\mathbf{p},t,t/2)  \:\:\nonumber\\
 &=&\:\: \tilde{h}_{\perp^\prime}^2 \:+\: (\text{excited-state contributions}). \label{eq:RperpprimeTA}
\end{eqnarray} \interdisplaylinepenalty=10000
Since $t^\prime$ and $t$ must both be integer multiples of the lattice spacing, here we imply an average over the two values of $t^\prime$ closest to $t/2$ for odd $t/a$. The excited-state contributions in the above quantities will decay exponentially as a function of the source-sink separation $t$.

\subsection{Extracting ratios of form factors}

\label{sec:FFratios}

To preserve the sign information, we define the following linear projections of three-point functions:

\begin{align}
 {\mathscr{S}_\lambda^{V,TV}(\mathbf{p},t,t^\prime)} &= \mathrm{Tr}\Big[ M^{(\lambda)}_{\mu j} P_{(3/2)}^{jl} \:  C^{(3,{\rm fw})}_l(\mathbf{p},\Gamma_{V,TV}^\mu, t, t^\prime) \:\: \frac{(1+\slashed{v})}{2} \Big] , \\
 {\mathscr{S}_\lambda^{A,TA}(\mathbf{p},t,t^\prime)} &= \mathrm{Tr}\Big[\gamma_5 M^{(\lambda)}_{\mu j} P_{(3/2)}^{jl} \:  C^{(3,{\rm fw})}_l(\mathbf{p},\Gamma_{A,TA}^\mu, t, t^\prime) \:\: \frac{(1+\slashed{v})}{2} \Big] ,
\end{align}
where $\lambda\in\{0,+,\perp,\perp^\prime\}$ and
\begin{align}
M^{(0)}_{\mu j}&=\epsilon_{\mu}^{(0)} \epsilon_{j}^{(0)}, \\
M^{(+)}_{\mu j}&=\epsilon_{\mu}^{(+)} \epsilon_{j}^{(0)}, \\
M^{(\perp^1)}_{\mu j}&=\sum^3_{i=1}\epsilon_{\mu}^{(\perp,i)} \epsilon_{j}^{(\perp,i)}, \\
M^{(\perp^2)}_{ij}&= i \epsilon_{l}^{(0)} \gamma^{l} \gamma_5 \epsilon^{(0)m}\epsilon_{mij}, \quad M^{(\perp^2)}_{0j}=0, \\
M^{(\perp)}_{\mu j}&=-M^{(\perp^1)}_{\mu j}+M^{(\perp^2)}_{\mu j}, \\
M^{(\perp^\prime)}_{\mu j}&=M^{(\perp^1)}_{\mu j}+M^{(\perp^2)}_{\mu j},
\end{align}
with the polarization vectors as defined in Eq.~(\ref{eq:polarizationvectors}). As before, repeated Latin indices are summed only over the spatial directions. To improve the signals, we use the average of the forward three-point function and the Dirac adjoint of the backward three-point function instead of just $C^{(3,{\rm fw})}$. We can isolate the form factors, up to common overlap factors and exponentials, in the following way: 
\begin{align}
  {S_0^V(\mathbf{p},t,t^\prime)}&=\frac{3 E_{\Lambda_b} m_{\Lambda_b}}{ (E_{\Lambda_b}-m_{\Lambda_b}) (E_{\Lambda_b}+m_{\Lambda_b}) (m_{\Lambda_b}-m_{\Lambda^*})}\:{\mathscr{S}_0^V(\mathbf{p},t,t^\prime)}\cr
 &=f_0  \: Z_{\Lambda^*} ( Z^{(1)}_{\Lambda_b} m_{\Lambda_b} + Z^{(2)}_{\Lambda_b} E_{\Lambda_b} ) e^{-m_{\Lambda^*}(t-t')} e^{-E_{\Lambda_b} t'} \cr
 &\hspace{3ex}+ \text{(excited-state contributions)},
\end{align}

\begin{align}
{S_+^V(\mathbf{p},t,t^\prime)}&=\frac{3 E_{\Lambda_b} m_{\Lambda_b}}{ (E_{\Lambda_b}-m_{\Lambda_b})^{1/2} (E_{\Lambda_b}+m_{\Lambda_b})^{3/2} (m_{\Lambda_b}+m_{\Lambda^*})}\:{\mathscr{S}_+^V(\mathbf{p},t,t^\prime)}\cr
 &=f_+  \: Z_{\Lambda^*} ( Z^{(1)}_{\Lambda_b} m_{\Lambda_b} + Z^{(2)}_{\Lambda_b} E_{\Lambda_b} ) e^{-m_{\Lambda^*}(t-t')} e^{-E_{\Lambda_b} t'} \cr
 &\hspace{3ex}+ \text{(excited-state contributions)}, \\ \nonumber \\
  {S_\perp^V(\mathbf{p},t,t^\prime)}&=\frac{3 E_{\Lambda_b} m_{\Lambda_b}}{ 2(E_{\Lambda_b}-m_{\Lambda_b}) (E_{\Lambda_b}+m_{\Lambda_b})^2}\:{\mathscr{S}_\perp^V(\mathbf{p},t,t^\prime)}\cr
 &=f_{\perp}  \: Z_{\Lambda^*} ( Z^{(1)}_{\Lambda_b} m_{\Lambda_b} + Z^{(2)}_{\Lambda_b} E_{\Lambda_b} ) e^{-m_{\Lambda^*}(t-t')} e^{-E_{\Lambda_b} t'} \cr
 &\hspace{3ex}+ \text{(excited-state contributions)}, \\ \nonumber \\ 
  {S_{\perp^\prime}^V(\mathbf{p},t,t^\prime)}&=\frac{ E_{\Lambda_b} m_{\Lambda_b}}{ 2(E_{\Lambda_b}-m_{\Lambda_b}) (E_{\Lambda_b}+m_{\Lambda_b})^2}\:{\mathscr{S}_{\perp^\prime}^V(\mathbf{p},t,t^\prime)}\cr
 &=f_{\perp^\prime}  \: Z_{\Lambda^*} ( Z^{(1)}_{\Lambda_b} m_{\Lambda_b} + Z^{(2)}_{\Lambda_b} E_{\Lambda_b} ) e^{-m_{\Lambda^*}(t-t')} e^{-E_{\Lambda_b} t'} \cr
 &\hspace{3ex}+ \text{(excited-state contributions)},  \\ \nonumber \\ 
  {S_0^A(\mathbf{p},t,t^\prime)}&=\frac{3 E_{\Lambda_b} m_{\Lambda_b}}{ (E_{\Lambda_b}+m_{\Lambda_b}) (E_{\Lambda_b}-m_{\Lambda_b}) (m_{\Lambda_b}+m_{\Lambda^*})}\:{\mathscr{S}_0^A(\mathbf{p},t,t^\prime)}\cr
 &=g_0  \: Z_{\Lambda^*} ( Z^{(1)}_{\Lambda_b} m_{\Lambda_b} + Z^{(2)}_{\Lambda_b} E_{\Lambda_b} ) e^{-m_{\Lambda^*}(t-t')} e^{-E_{\Lambda_b} t'} \cr
 &\hspace{3ex}+ \text{(excited-state contributions)}, \\ \nonumber \\
 {S_+^A(\mathbf{p},t,t^\prime)}&=\frac{3 E_{\Lambda_b} m_{\Lambda_b}}{ (E_{\Lambda_b}-m_{\Lambda_b})^{3/2} (E_{\Lambda_b}+m_{\Lambda_b})^{1/2} (m_{\Lambda_b}-m_{\Lambda^*})}\:{\mathscr{S}_+^A(\mathbf{p},t,t^\prime)}\cr
 &=g_+  \: Z_{\Lambda^*} ( Z^{(1)}_{\Lambda_b} m_{\Lambda_b} + Z^{(2)}_{\Lambda_b} E_{\Lambda_b} ) e^{-m_{\Lambda^*}(t-t')} e^{-E_{\Lambda_b} t'} \cr
 &\hspace{3ex}+ \text{(excited-state contributions)}, \\ \nonumber \\
  {S_\perp^A(\mathbf{p},t,t^\prime)}&=-\frac{3 E_{\Lambda_b} m_{\Lambda_b}}{ 2(E_{\Lambda_b}-m_{\Lambda_b})^2 (E_{\Lambda_b}+m_{\Lambda_b})}\:{\mathscr{S}_\perp^A(\mathbf{p},t,t^\prime)}\cr
 &=g_{\perp}  \: Z_{\Lambda^*} ( Z^{(1)}_{\Lambda_b} m_{\Lambda_b} + Z^{(2)}_{\Lambda_b} E_{\Lambda_b} ) e^{-m_{\Lambda^*}(t-t')} e^{-E_{\Lambda_b} t'} \cr
 &\hspace{3ex}+ \text{(excited-state contributions)}, \\ \nonumber \\ 
  {S_{\perp^\prime}^A(\mathbf{p},t,t^\prime)}&=-\frac{ E_{\Lambda_b} m_{\Lambda_b}}{ 2(E_{\Lambda_b}-m_{\Lambda_b})^2 (E_{\Lambda_b}+m_{\Lambda_b})}\:{\mathscr{S}_{\perp^\prime}^A(\mathbf{p},t,t^\prime)}\cr
 &=g_{\perp^\prime}  \: Z_{\Lambda^*} ( Z^{(1)}_{\Lambda_b} m_{\Lambda_b} + Z^{(2)}_{\Lambda_b} E_{\Lambda_b} ) e^{-m_{\Lambda^*}(t-t')} e^{-E_{\Lambda_b} t'} \cr
 &\hspace{3ex}+ \text{(excited-state contributions)},\\  \nonumber \\ 
 {S^{TV}_+(\mathbf{p},t,t^\prime)}&=-\frac{3 E_{\Lambda_b} m_{\Lambda_b}}{ (E_{\Lambda_b}-m_{\Lambda_b})^{1/2} (E_{\Lambda_b}+m_{\Lambda_b})^{3/2} q^2}\:{\mathscr{S}_+^{TV}(\mathbf{p},t,t^\prime)}\cr
 &=h_+  \: Z_{\Lambda^*} ( Z^{(1)}_{\Lambda_b} m_{\Lambda_b} + Z^{(2)}_{\Lambda_b} E_{\Lambda_b} ) e^{-m_{\Lambda^*}(t-t')} e^{-E_{\Lambda_b} t'} \cr
 &\hspace{3ex}+ \text{(excited-state contributions)}, \\ \nonumber \\
  {S^{TV}_\perp(\mathbf{p},t,t^\prime)}&=-\frac{3 E_{\Lambda_b} m_{\Lambda_b}}{ 2(E_{\Lambda_b}-m_{\Lambda_b}) (E_{\Lambda_b}+m_{\Lambda_b})^2 (m_{\Lambda_b}+m_{\Lambda^*}) }\:{\mathscr{S}^{TV}_\perp(\mathbf{p},t,t^\prime)}\cr
 &=h_{\perp}  \: Z_{\Lambda^*} ( Z^{(1)}_{\Lambda_b} m_{\Lambda_b} + Z^{(2)}_{\Lambda_b} E_{\Lambda_b} ) e^{-m_{\Lambda^*}(t-t')} e^{-E_{\Lambda_b} t'} \cr
 &\hspace{3ex}+ \text{(excited-state contributions)},
 \end{align}
 
 \begin{align}
  {S^{TV}_{\perp^\prime}(\mathbf{p},t,t^\prime)}&=-\frac{ E_{\Lambda_b} m_{\Lambda_b}}{ 2(E_{\Lambda_b}-m_{\Lambda_b}) (E_{\Lambda_b}+m_{\Lambda_b})^2 (m_{\Lambda_b}+m_{\Lambda^*})}\:{\mathscr{S}^{TV}_{\perp^\prime}(\mathbf{p},t,t^\prime)}\cr
 &=h_{\perp^\prime}  \: Z_{\Lambda^*} ( Z^{(1)}_{\Lambda_b} m_{\Lambda_b} + Z^{(2)}_{\Lambda_b} E_{\Lambda_b} ) e^{-m_{\Lambda^*}(t-t')} e^{-E_{\Lambda_b} t'} \cr
 &\hspace{3ex}+ \text{(excited-state contributions)},  \\ \nonumber \\
 {S^{TA}_+(\mathbf{p},t,t^\prime)}&=\frac{3 E_{\Lambda_b} m_{\Lambda_b}}{ (E_{\Lambda_b}+m_{\Lambda_b})^{1/2} (E_{\Lambda_b}-m_{\Lambda_b})^{3/2} q^2}\:{\mathscr{S}_+^{TA}(\mathbf{p},t,t^\prime)}\cr
 &=\widetilde{h}_+  \: Z_{\Lambda^*} ( Z^{(1)}_{\Lambda_b} m_{\Lambda_b} + Z^{(2)}_{\Lambda_b} E_{\Lambda_b} ) e^{-m_{\Lambda^*}(t-t')} e^{-E_{\Lambda_b} t'} \cr
 &\hspace{3ex}+ \text{(excited-state contributions)}, \\ \nonumber \\
  {S^{TA}_\perp(\mathbf{p},t,t^\prime)}&=-\frac{3 E_{\Lambda_b} m_{\Lambda_b}}{ 2(E_{\Lambda_b}+m_{\Lambda_b}) (E_{\Lambda_b}-m_{\Lambda_b})^2 (m_{\Lambda_b}-m_{\Lambda^*}) }\:{\mathscr{S}^{TA}_\perp(\mathbf{p},t,t^\prime)}\cr
 &=\widetilde{h}_{\perp}  \: Z_{\Lambda^*} ( Z^{(1)}_{\Lambda_b} m_{\Lambda_b} + Z^{(2)}_{\Lambda_b} E_{\Lambda_b} ) e^{-m_{\Lambda^*}(t-t')} e^{-E_{\Lambda_b} t'} \cr
 &\hspace{3ex}+ \text{(excited-state contributions)}, \\ \nonumber \\ 
  {S^{TA}_{\perp^\prime}(\mathbf{p},t,t^\prime)}&=-\frac{ E_{\Lambda_b} m_{\Lambda_b}}{ 2(E_{\Lambda_b}+m_{\Lambda_b}) (E_{\Lambda_b}-m_{\Lambda_b})^2 (m_{\Lambda_b}-m_{\Lambda^*})}\:{\mathscr{S}^{TA}_{\perp^\prime}(\mathbf{p},t,t^\prime)}\cr
 &=\widetilde{h}_{\perp^\prime}  \: Z_{\Lambda^*} ( Z^{(1)}_{\Lambda_b} m_{\Lambda_b} + Z^{(2)}_{\Lambda_b} E_{\Lambda_b} ) e^{-m_{\Lambda^*}(t-t')} e^{-E_{\Lambda_b} t'} \cr
 &\hspace{3ex}+ \text{(excited-state contributions)}.
\end{align} 
The excited-state contributions decay exponentially faster than the ground-state contributions. The unwanted factors of $Z_{\Lambda^*} ( Z^{(1)}_{\Lambda_b} m_{\Lambda_b} + Z^{(2)}_{\Lambda_b} E_{\Lambda_b} ) e^{-m_{\Lambda^*}(t-t')} e^{-E_{\Lambda_b} t'}$ will cancel in ratios of the above quantities at large times.

\subsection{Results for the form factors with relative signs preserved}

\begin{figure}
\flushleft

\includegraphics[width=0.245\linewidth,valign=t]{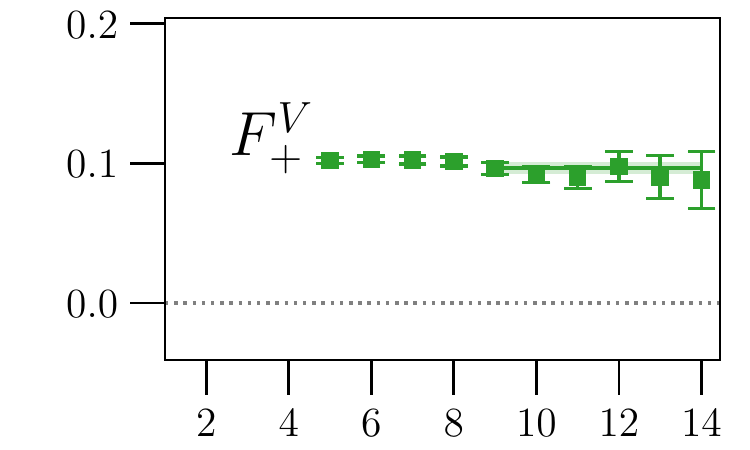} \includegraphics[width=0.245\linewidth,valign=t]{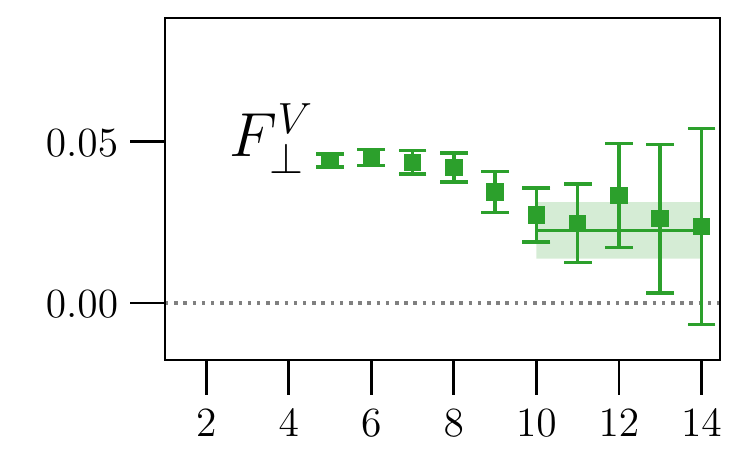} \includegraphics[width=0.245\linewidth,valign=t]{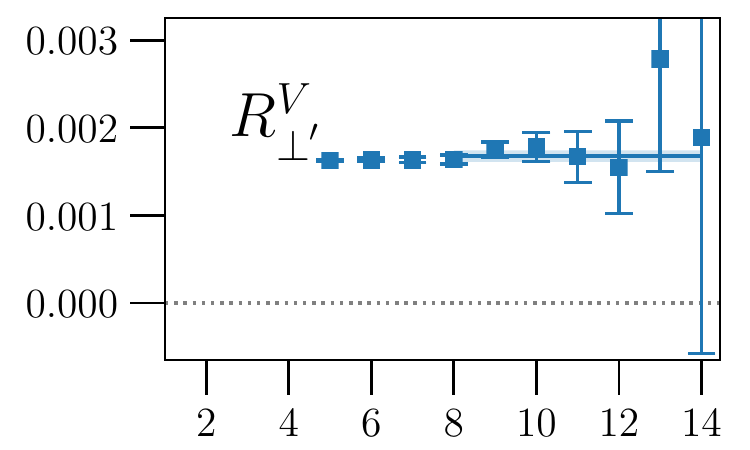} \includegraphics[width=0.245\linewidth,valign=t]{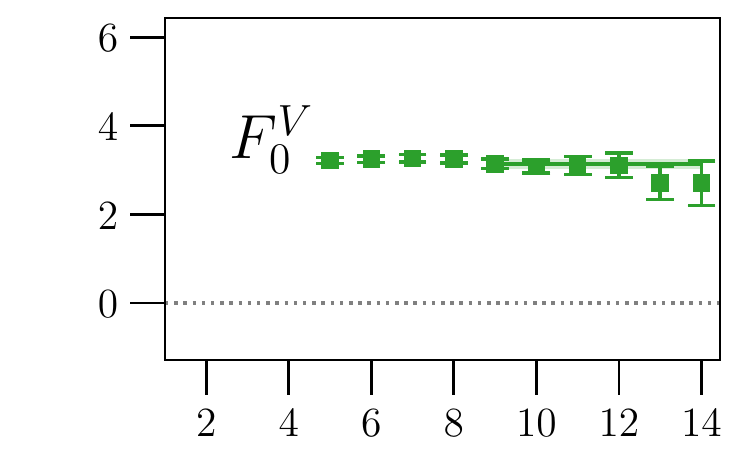}

\includegraphics[width=0.245\linewidth,valign=t]{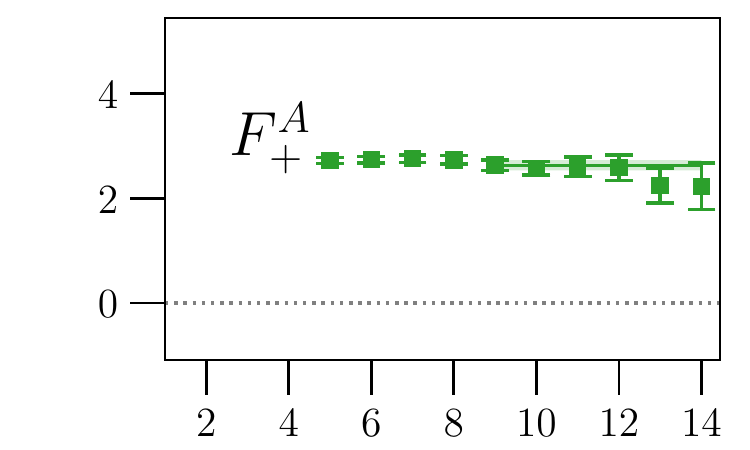} \includegraphics[width=0.245\linewidth,valign=t]{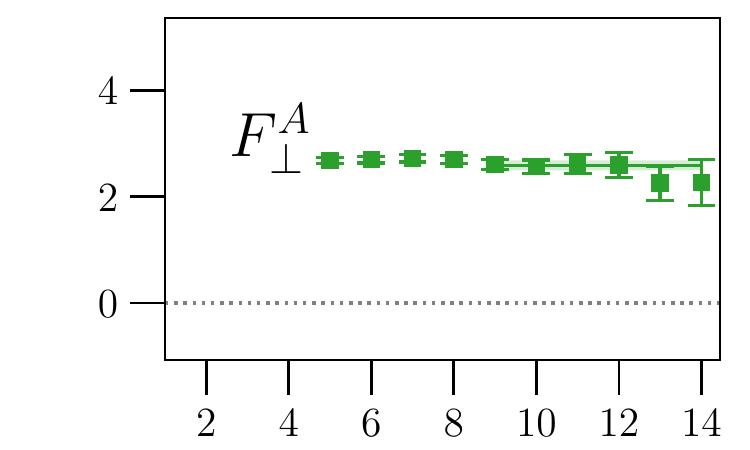} \includegraphics[width=0.245\linewidth,valign=t]{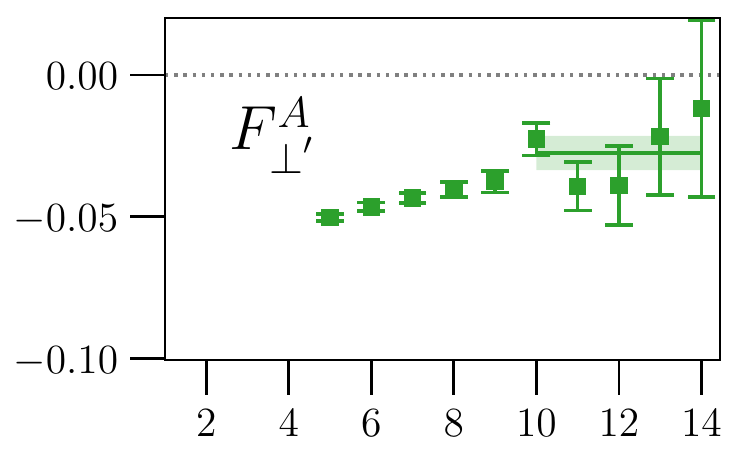} \includegraphics[width=0.245\linewidth,valign=t]{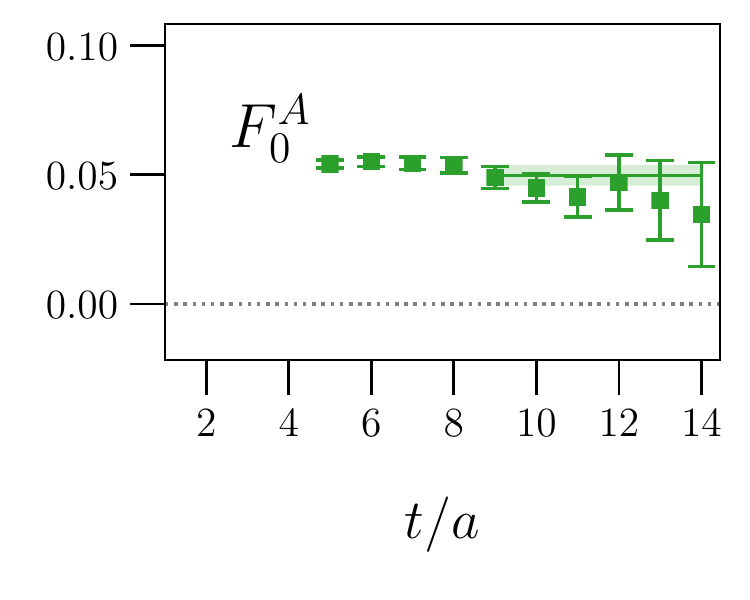}

\vspace{-6.0ex}

\includegraphics[width=0.245\linewidth,valign=t]{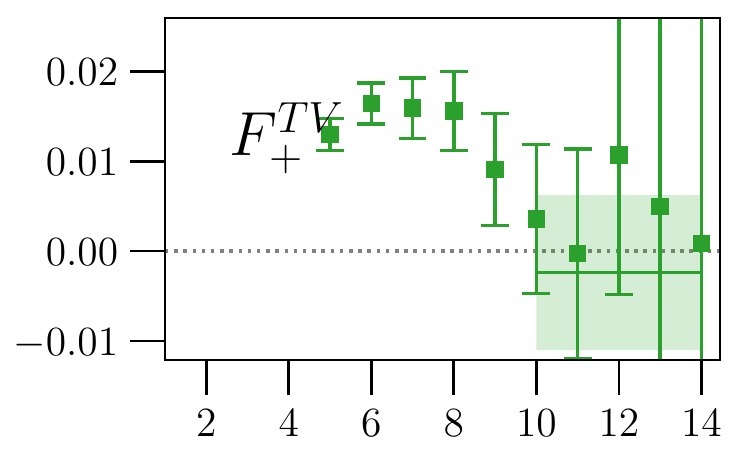} \includegraphics[width=0.245\linewidth,valign=t]{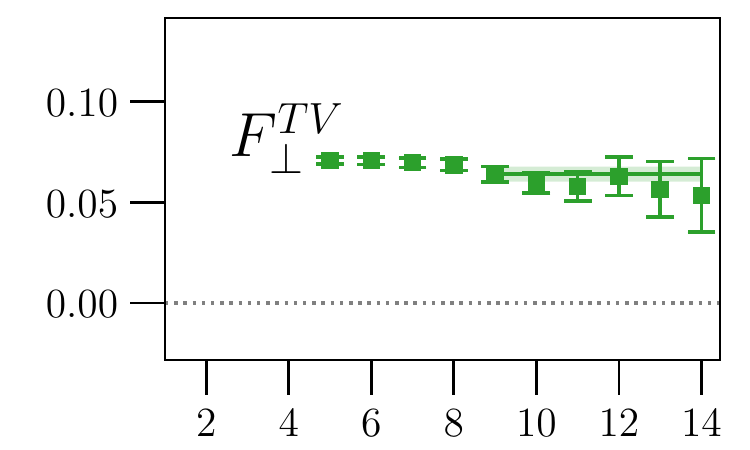} \includegraphics[width=0.245\linewidth,valign=t]{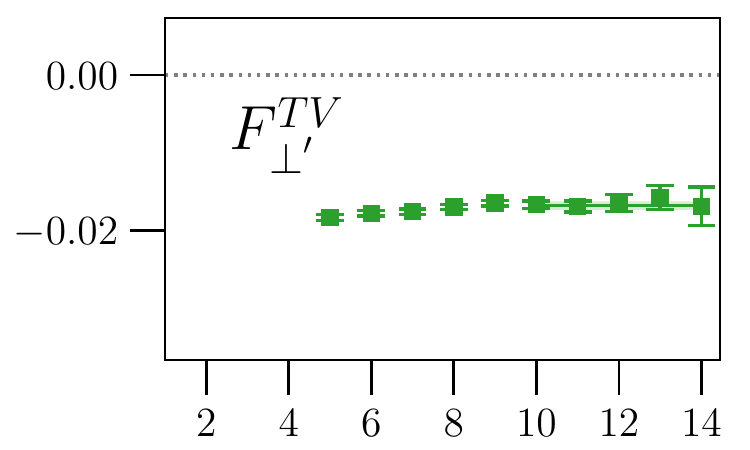}

\includegraphics[width=0.245\linewidth,valign=t]{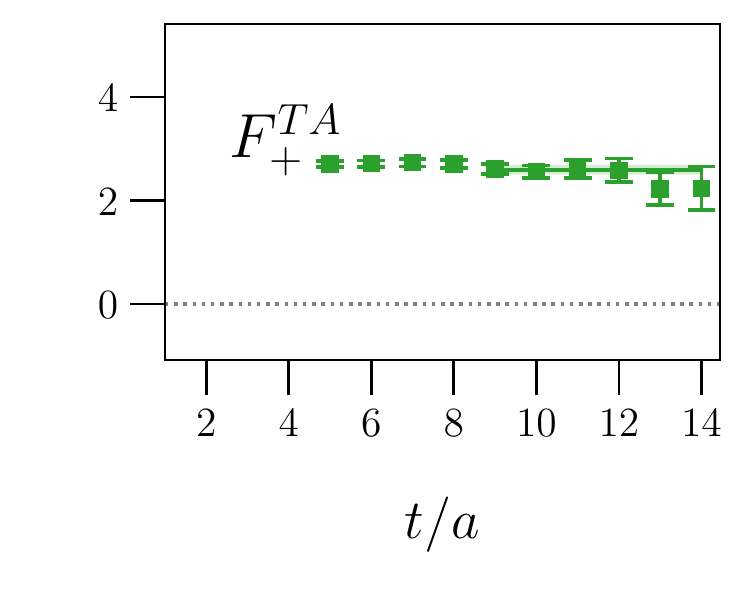} \includegraphics[width=0.245\linewidth,valign=t]{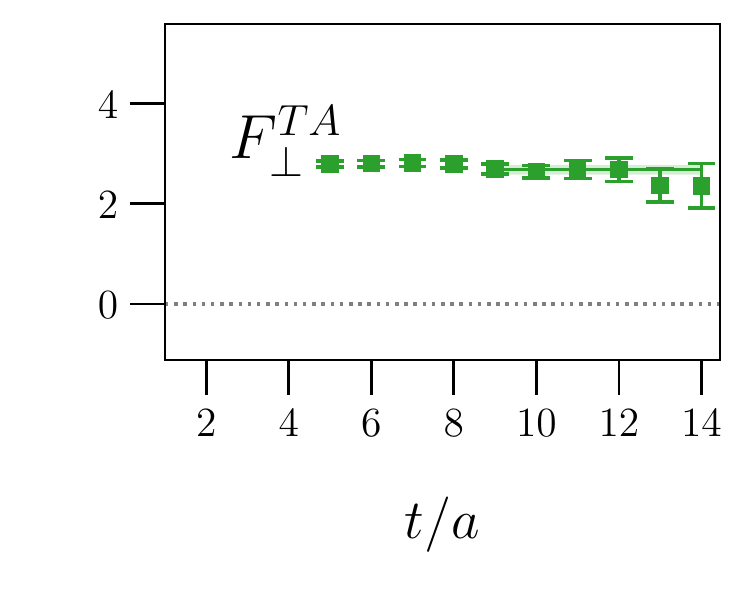} \includegraphics[width=0.245\linewidth,valign=t]{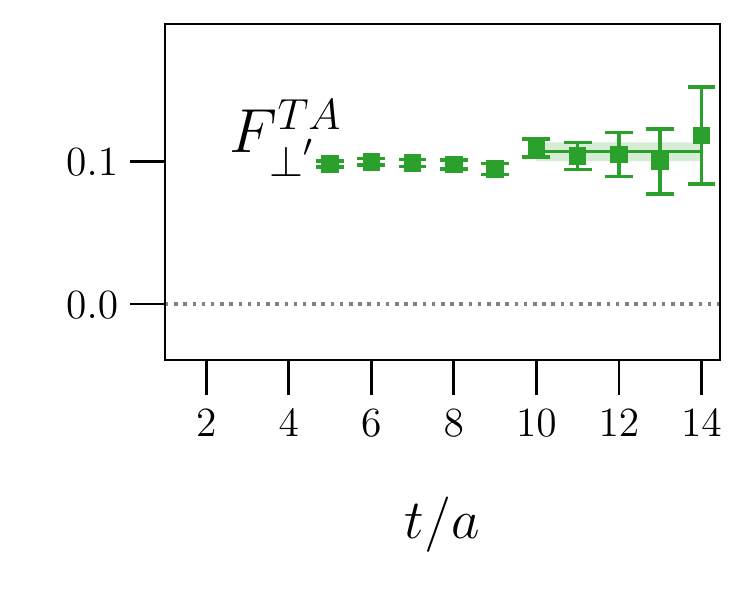}

\vspace{-1ex}
 \caption{\label{fig:ratios}Numerical results for the quantities $F^X_\lambda(\mathbf{p},t)$, defined in Eq.~(\protect\ref{eq:FX}), as a function of the source-sink separation, for $\mathbf{p}=(0,0,3)\frac{2\pi}{L}$ and for the F004 ensemble. Also shown is $R_{\perp^\prime}^V(\mathbf{p}, t)$, which is used to extract the square of the reference form factor $f_{\perp^\prime}$. The horizontal lines indicate the ranges and extracted values of constant fits.}
\end{figure}

The fourteen form factors with relative sign information preserved can now be obtained by extracting the magnitude of a single reference form factor as in Sec.~\ref{sec:FFsquares}, and multiplying with ratios of the projected three-point functions $S^X_\lambda(\mathbf{p},t,t^\prime)$. We choose $f_{\perp^\prime}$ to
be the reference form factor because the results for the corresponding $R_{\perp^\prime}^V$ show good plateaus and reasonably small statistical uncertainties (see the third plot from the left in the top row of Fig.~\ref{fig:ratios}). We again set $t^\prime=t/2$, and define the functions
\begin{align}
F^X_\lambda(\mathbf{p},t)=\frac{S^X_\lambda(\mathbf{p},t,t/2)}{S^V_{\perp^\prime}(\mathbf{p},t,t/2)}\sqrt{R_{\perp^\prime}^V(\mathbf{p})}, \label{eq:FX}
\end{align}
where $R_{\perp^\prime}^V(\mathbf{p})$ denotes the result of a constant fit to $R_{\perp^\prime}^V(\mathbf{p}, t)$ in the region of ground-state saturation. The functions
$F^X_\lambda(\mathbf{p},t)$ are equal to the individual helicity form factors up to excited-state contamination that decays exponentially with $t$. We perform constant fits to $F^X_\lambda(\mathbf{p},t)$ in the plateau regions, requiring good quality-of-fit and stability under variations of the starting time. Plots of $F^X_\lambda(\mathbf{p},t)$ and the associated fits for one ensemble and one momentum are shown in Fig.~\ref{fig:ratios}. All fit results are listed in Table \ref{tab:FFvalues}. The uncertainties were computed using statistical bootstrap.

\begin{table}
 \begin{tabular}{cccllllll}
  \hline\hline
  Form factor                & & $|\mathbf{p}|/(2\pi/L)$ & & \hspace{2ex}C01 & & \hspace{2ex}C005 & & \hspace{2ex}F004 \\
  \hline
  $f_0$                             &&   2   &&   $\wm    3.77(18)$   &&   $\wm    3.53(24)$   &&   $\wm    3.36(14)$\\
                                    &&   3   &&   $\wm    3.38(14)$   &&   $\wm    3.15(20)$   &&   $\wm    3.14(11)$\\  
  $f_+$                             &&   2   &&   $\wm  0.0773(40)$   &&   $\wm  0.0714(55)$   &&   $\wm  0.0698(36)$\\
                                    &&   3   &&   $\wm  0.1040(49)$   &&   $\wm  0.0949(71)$   &&   $\wm  0.0965(43)$\\
  $f_{\perp}$                       &&   2   &&   $\wm   0.002(10)$   &&   $   -  0.017(13)$   &&   $   - 0.0204(81)$\\
                                    &&   3   &&   $\wm   0.048(10)$   &&   $\wm   0.018(14)$   &&   $\wm  0.0225(87)$\\
  $f_{\perp^{\prime}}$              &&   2   &&   $\wm 0.04433(73)$   &&   $\wm  0.0434(16)$   &&   $\wm 0.04399(67)$\\
                                    &&   3   &&   $\wm 0.04051(89)$   &&   $\wm  0.0401(19)$   &&   $\wm 0.04093(81)$\\
  $g_0$                             &&   2   &&   $\wm  0.0273(40)$   &&   $\wm  0.0250(50)$   &&   $\wm  0.0224(35)$\\
                                    &&   3   &&   $\wm  0.0559(47)$   &&   $\wm  0.0508(61)$   &&   $\wm  0.0498(40)$\\                                    
  $g_+$                             &&   2   &&   $\wm    3.17(17)$   &&   $\wm    2.95(22)$   &&   $\wm    2.82(13)$\\
                                    &&   3   &&   $\wm    2.85(13)$   &&   $\wm    2.65(18)$   &&   $\wm    2.63(10)$\\
  $g_{\perp}$                       &&   2   &&   $\wm    3.12(16)$   &&   $\wm    2.91(21)$   &&   $\wm    2.76(13)$\\
                                    &&   3   &&   $\wm    2.80(12)$   &&   $\wm    2.61(17)$   &&   $\wm   2.589(95)$\\
  $g_{\perp^{\prime}}$              &&   2   &&   $   -  0.029(14)$   &&   $   -  0.052(21)$   &&   $   - 0.0261(86)$\\
                                    &&   3   &&   $   -  0.025(10)$   &&   $   -  0.040(14)$   &&   $   - 0.0275(60)$\\
  $h_+$                             &&   2   &&   $-  0.0162(95)$   &&   $-   0.034(13)$   &&   $-  0.0436(80)$\\
                                    &&   3   &&   $\wm  0.028(10)$   &&   $-   0.000(14)$   &&   $-  0.0024(86)$\\
  $h_{\perp}$                       &&   2   &&   $\wm   0.0440(36)$   &&   $\wm 0.0384(47)$   &&   $\wm 0.0388(32)$\\
                                    &&   3   &&   $\wm   0.0701(44)$   &&   $\wm    0.0616(59)$   &&   $\wm   0.0640(37)$\\
  $h_{\perp^{\prime}}$              &&   2   &&   $- 0.01582(73)$   &&   $-  0.0155(12)$   &&   $- 0.01738(47)$\\
                                    &&   3   &&   $- 0.01495(82)$   &&   $-  0.0144(13)$   &&   $- 0.01684(55)$\\
  $\widetilde{h}_+$                 &&   2   &&   $   \wm   3.15(16)$   &&   $   \wm   2.91(21)$   &&   $   \wm   2.78(12)$\\
                                    &&   3   &&   $   \wm   2.82(12)$   &&   $   \wm   2.61(17)$   &&   $   \wm  2.593(93)$\\
  $\widetilde{h}_{\perp}$           &&   2   &&   $   \wm   3.22(16)$   &&   $   \wm   3.01(21)$   &&   $   \wm   2.86(13)$\\
                                    &&   3   &&   $   \wm   2.89(12)$   &&   $   \wm   2.70(18)$   &&   $   \wm   2.68(10)$\\
  $\widetilde{h}_{\perp^{\prime}}$  &&   2   &&   $   \wm  0.098(14)$   &&   $   \wm  0.087(22)$   &&   $   \wm 0.1183(83)$\\
                                    &&   3   &&   $   \wm  0.091(11)$   &&   $   \wm  0.079(16)$   &&   $   \wm 0.1067(66)$\\
  \hline\hline
 \end{tabular}
\caption{\label{tab:FFvalues}Values of the form factors extracted from constant fits of $F^X_\lambda(\mathbf{p},t)$ in the plateau regions,  for each ensemble and for the two different $\Lambda_b$ momenta.}
\end{table}

\FloatBarrier
\section{Chiral and continuum extrapolations of the form factors}
\label{sec:FFextrap}
\FloatBarrier

The final step in the analysis of the form factors is to fit suitable functions describing the dependence on the kinematics, the light-quark mass (or, equivalently, $m_\pi^2$), and the lattice spacing to the results given in Table \ref{tab:FFvalues}. Given that we have data for only two different momenta that correspond to values of $q^2$ near the kinematic endpoint, we describe the kinematic dependence of each form factor by a linear function of the dimensionless variable 
\begin{equation}
w(q^2)=v\cdot v^\prime=\frac{m_{\Lambda_b}^2+m_{\Lambda^*}^2-q^2}{2m_{\Lambda_b}m_{\Lambda^*}}.
\end{equation}
We expect this description to be accurate only in the high-$q^2$ region. To allow for dependence on the light-quark mass and lattice spacing, we use the model
\begin{equation}\label{eq:extrapolation}
f(q^2)=F^{f}\left[1+C^{f}\frac{m_{\pi}^2-m_{\pi,\rm phys}^2}{(4\pi f_{\pi})^2}+D^{f}a^2\Lambda^2\right]+A^{f}\left[1+\tilde{C}^{f}\frac{m_{\pi}^2-m_{\pi,\rm phys}^2}{(4\pi f_{\pi})^2}+\tilde{D}^{f}a^2\Lambda^2\right](w-1), \\
\end{equation}
with independent fit parameters $F^f$, $A^f$, $C^f$, $D^f$, $\tilde{C}^f$, and $\tilde{D}^f$ for each form factor $f$. Here, we introduced $f_{\pi}=132\,\text{MeV}$ and $\Lambda=300\,\text{MeV}$ to make all parameters dimensionless. In the physical limit $m_\pi=m_{\pi,\rm phys}=135\,\text{MeV}$, $a=0$, the fit functions reduce to the form
\begin{equation}\label{eq:physicalFF}
f(q^2)=F^{f}+A^{f}(w-1),
\end{equation}
which only depend on the parameters $F^f$ and $A^f$. The model (\ref{eq:extrapolation}) can be thought of as expansions of both the zero-recoil form factors $F^f$ and the slopes $A^f$ in terms of the light-quark mass and the square of the lattice spacing. The limited number of data points made it necessary to constrain the size of the coefficients $C^f$, $D^f$, $\tilde{C}^f$, and $\tilde{D}^f$ to be not unnaturally large. To this end, we introduced Gaussian priors for $C^f$, $D^f$, $\tilde{C}^f$, and $\tilde{D}^f$ with central values equal to 0 and widths equal to 10.

\begin{figure}
 \centerline{\includegraphics[width=0.6\linewidth]{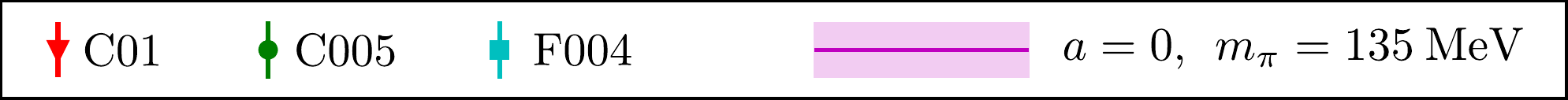}}
 
 \vspace{1ex}

 \includegraphics[width=0.47\linewidth]{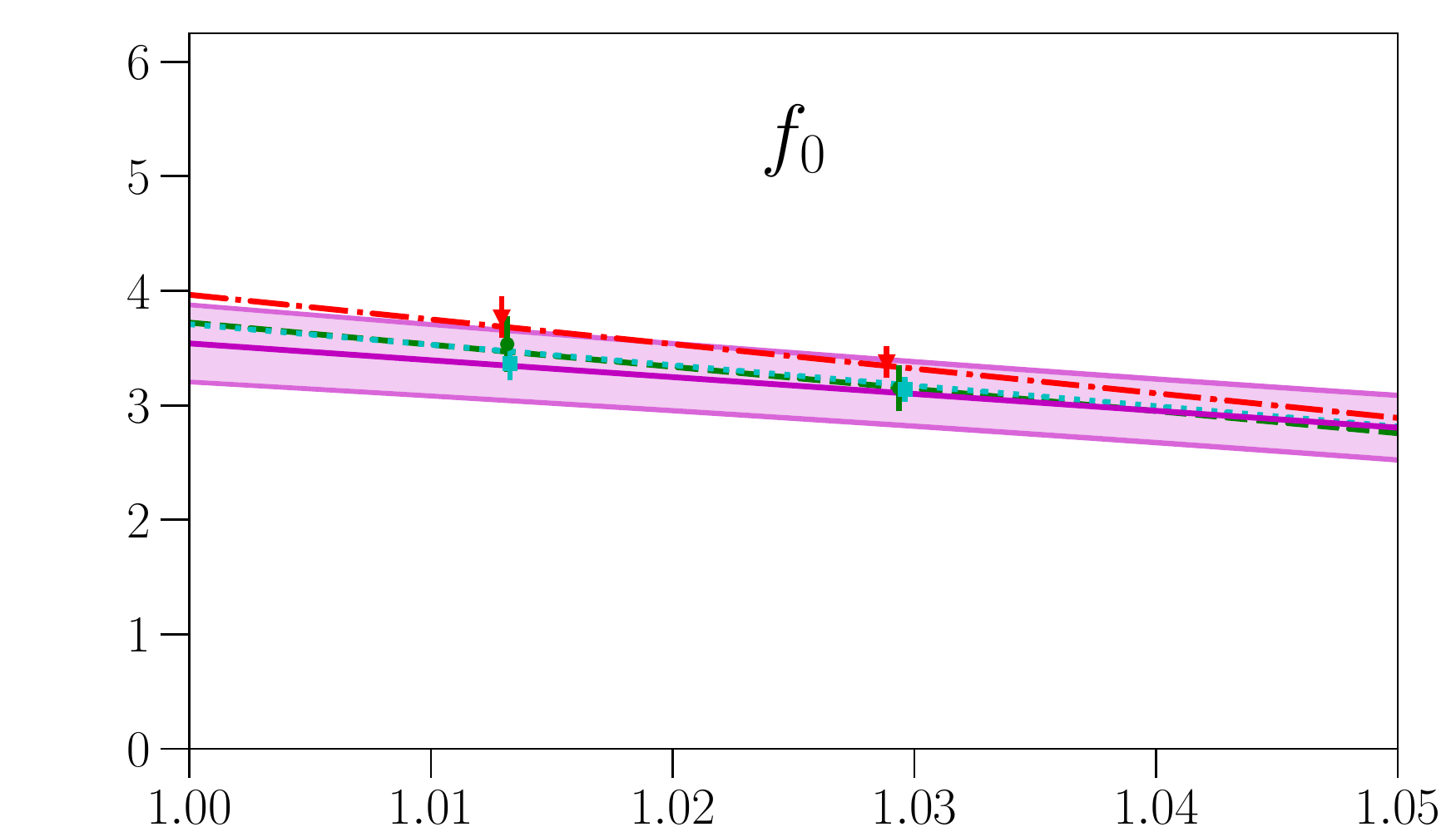} \hfill \includegraphics[width=0.47\linewidth]{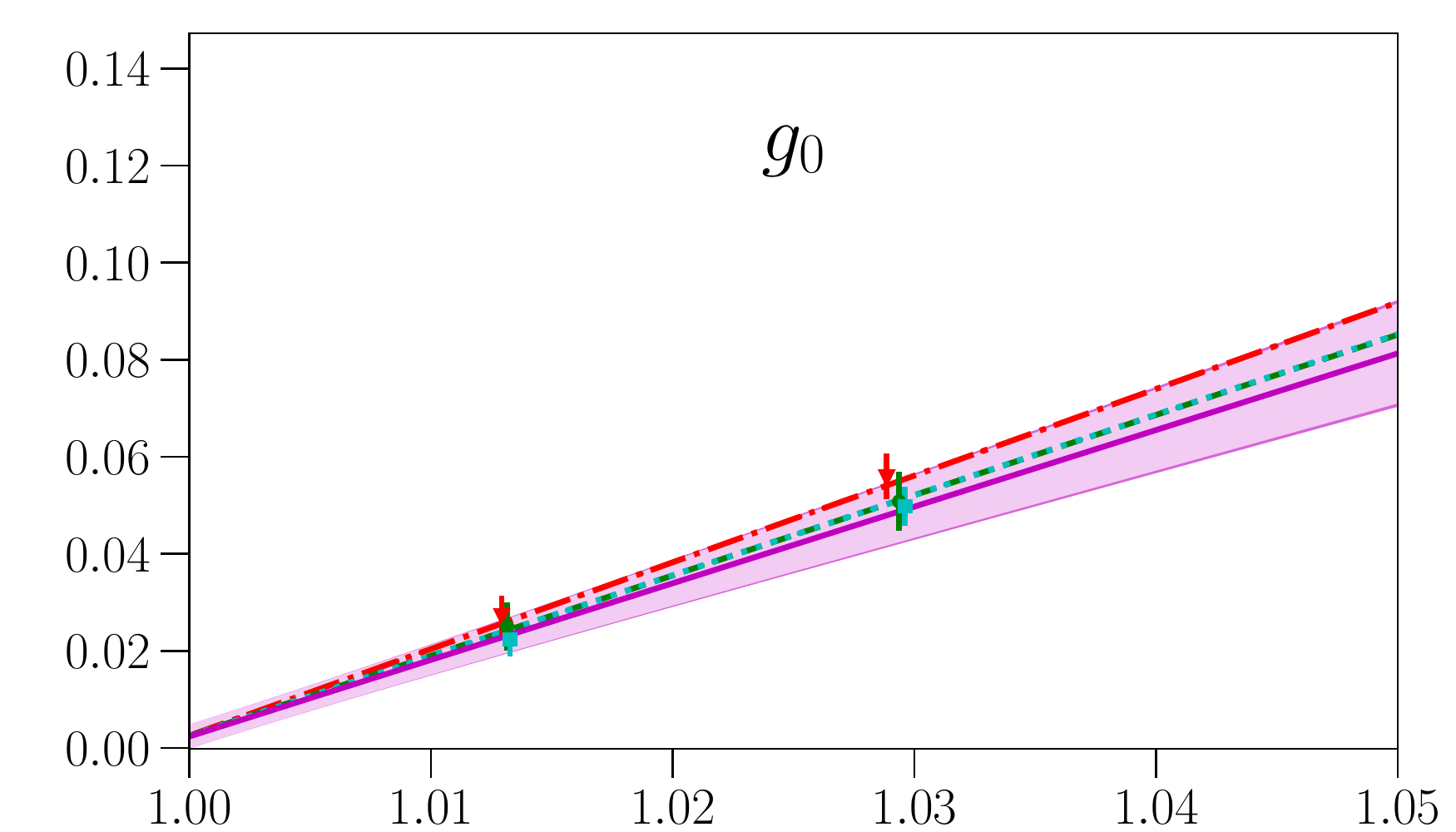} \\
 \includegraphics[width=0.47\linewidth]{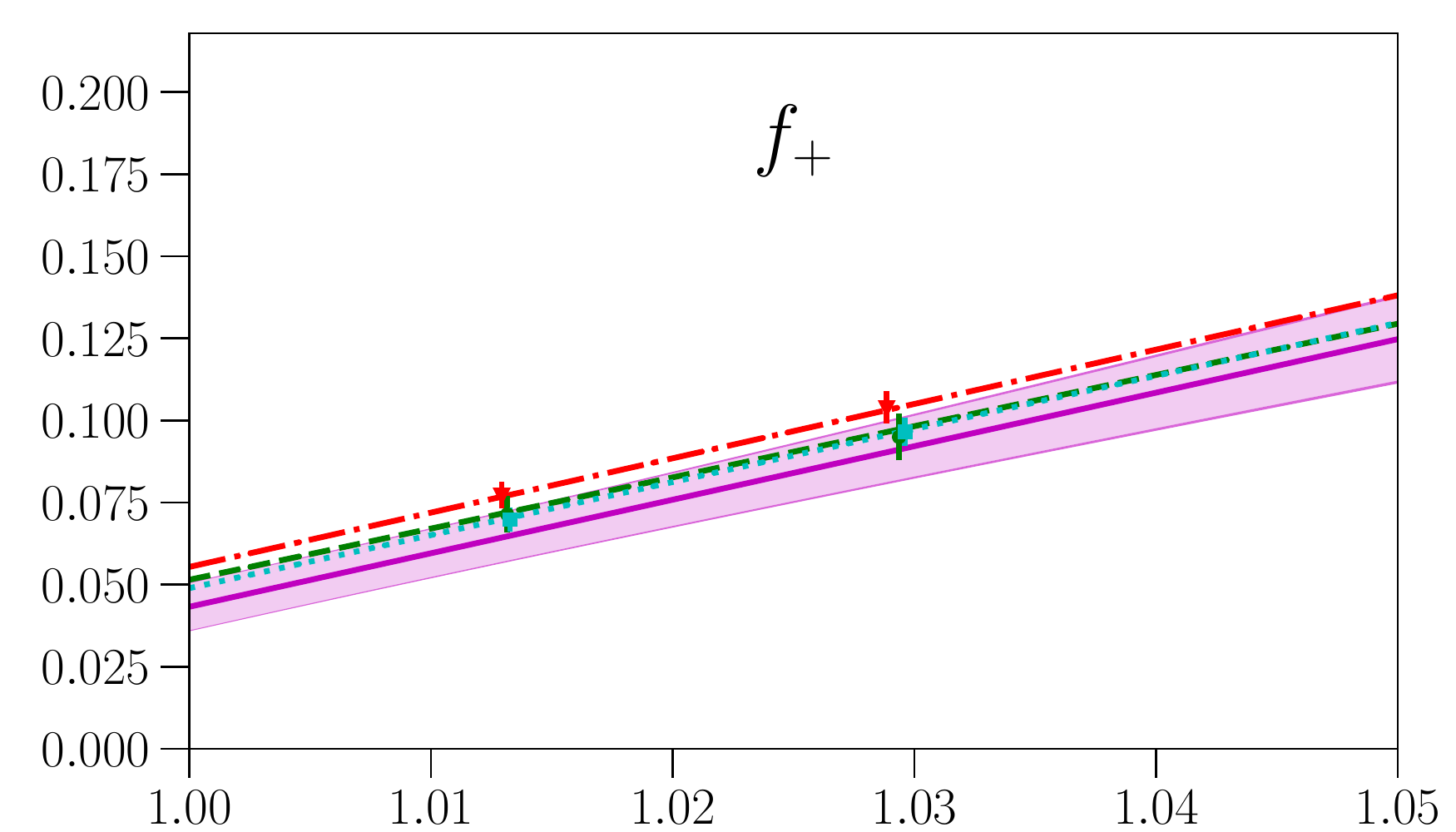} \hfill \includegraphics[width=0.47\linewidth]{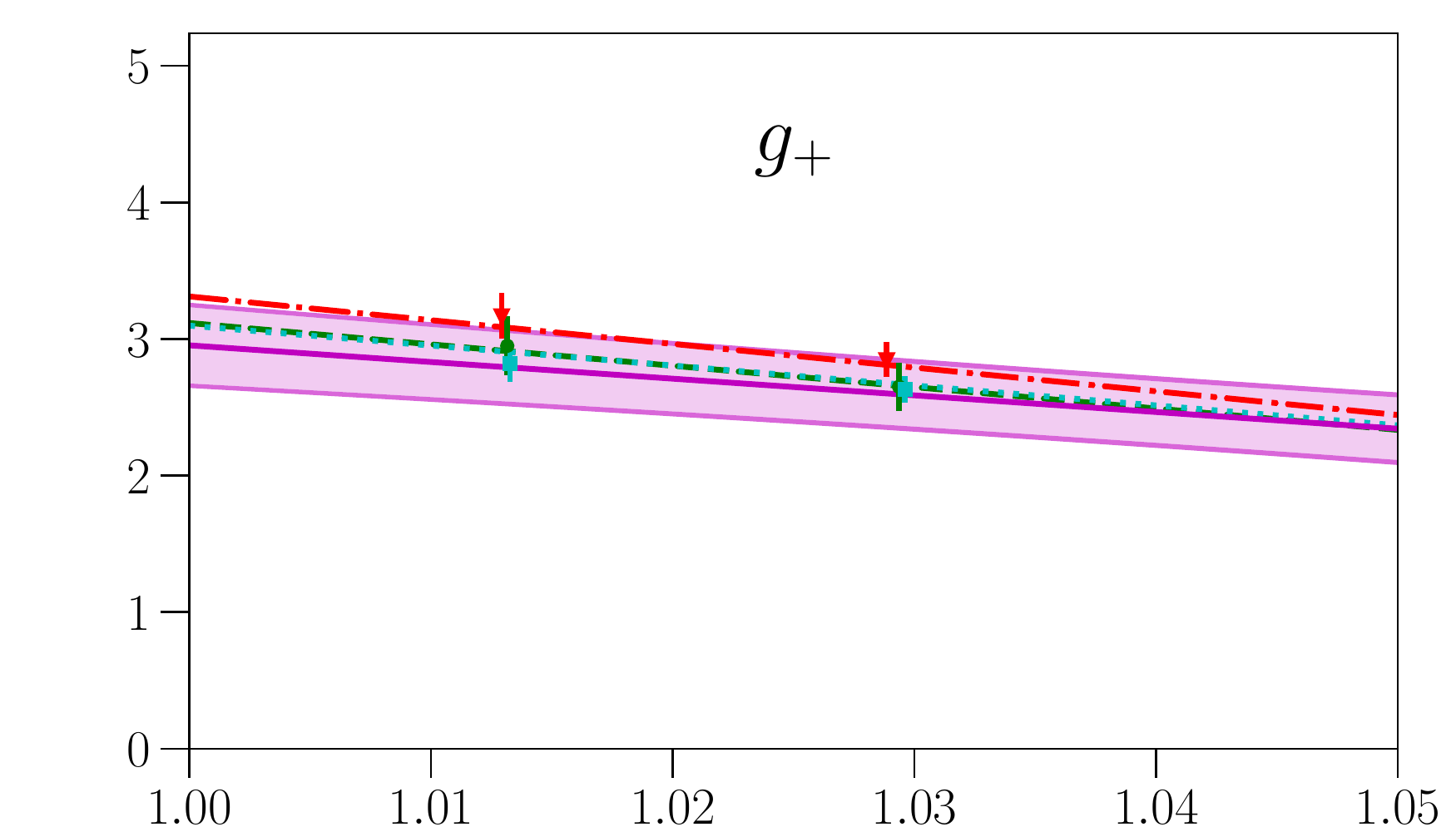} \\
 \includegraphics[width=0.47\linewidth]{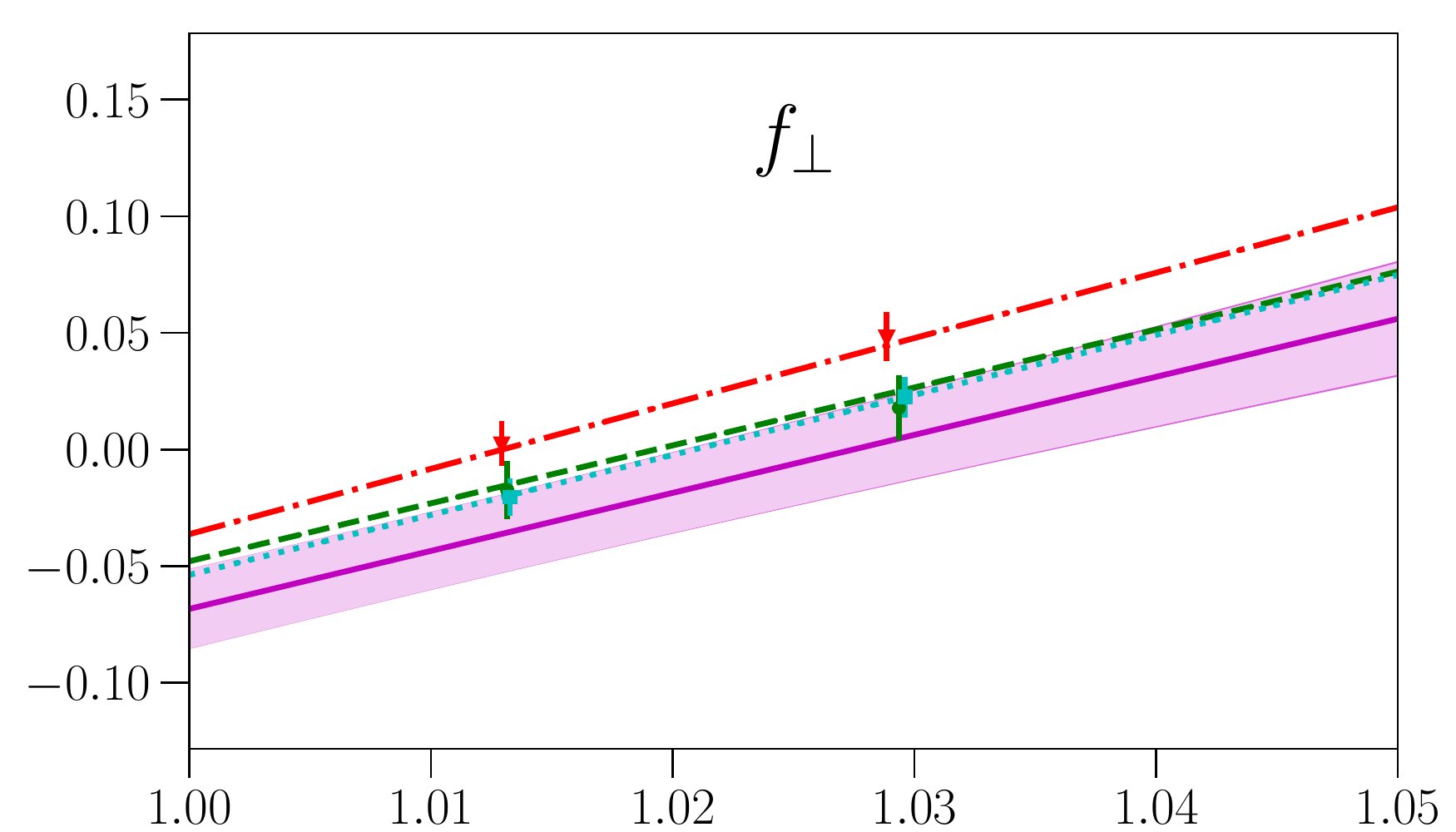} \hfill \includegraphics[width=0.47\linewidth]{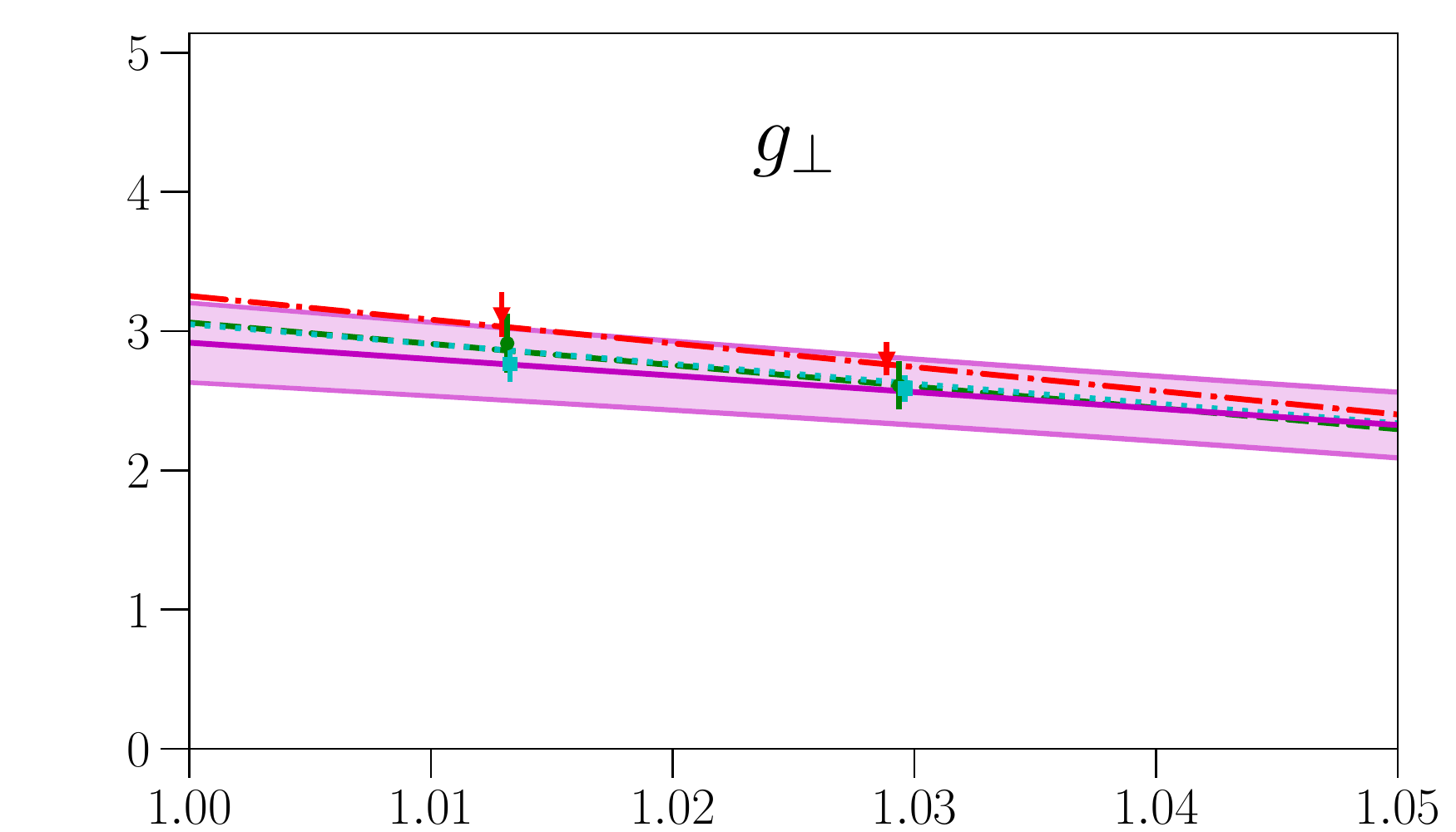} \\
 \includegraphics[width=0.47\linewidth]{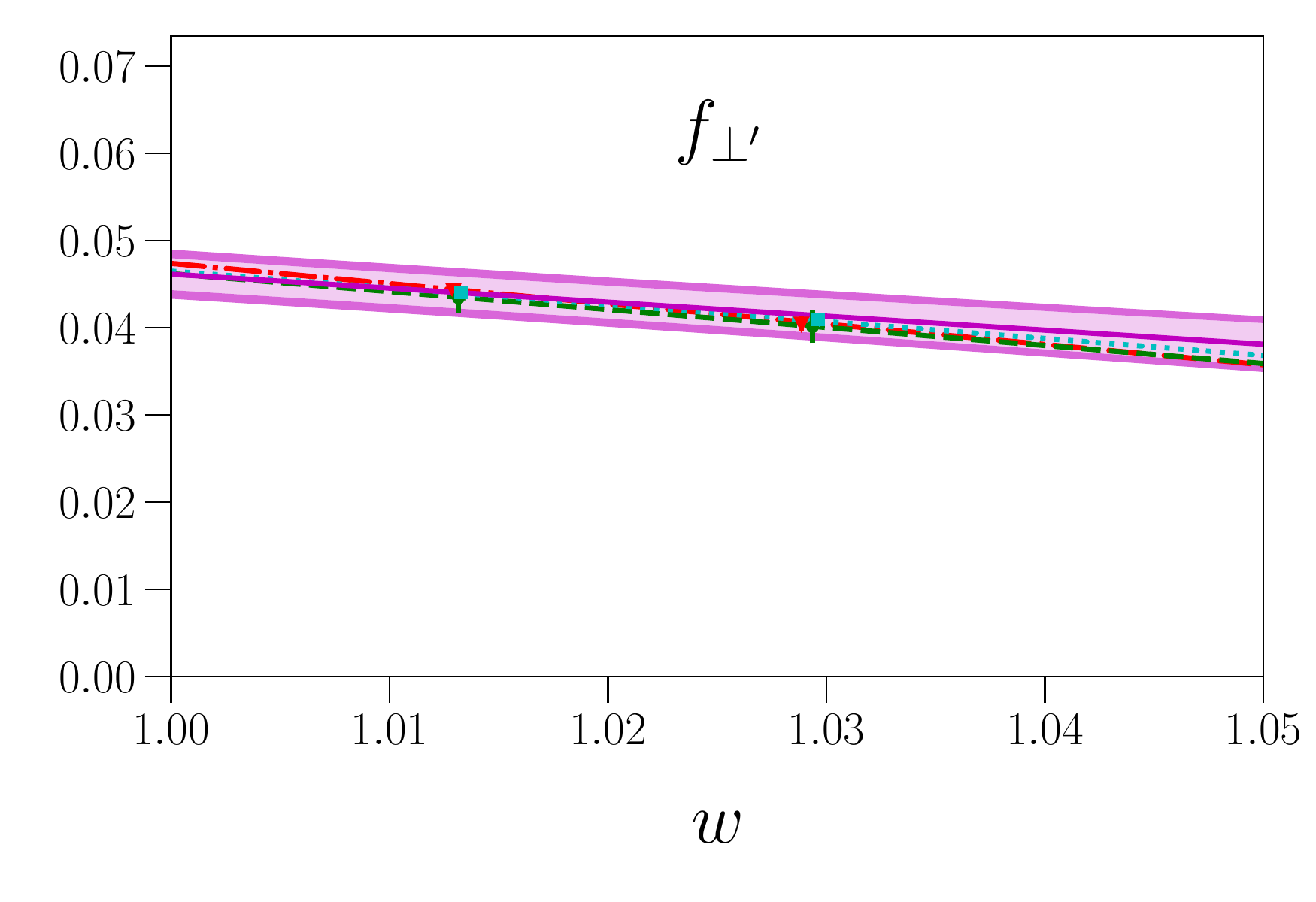} \hfill \includegraphics[width=0.47\linewidth]{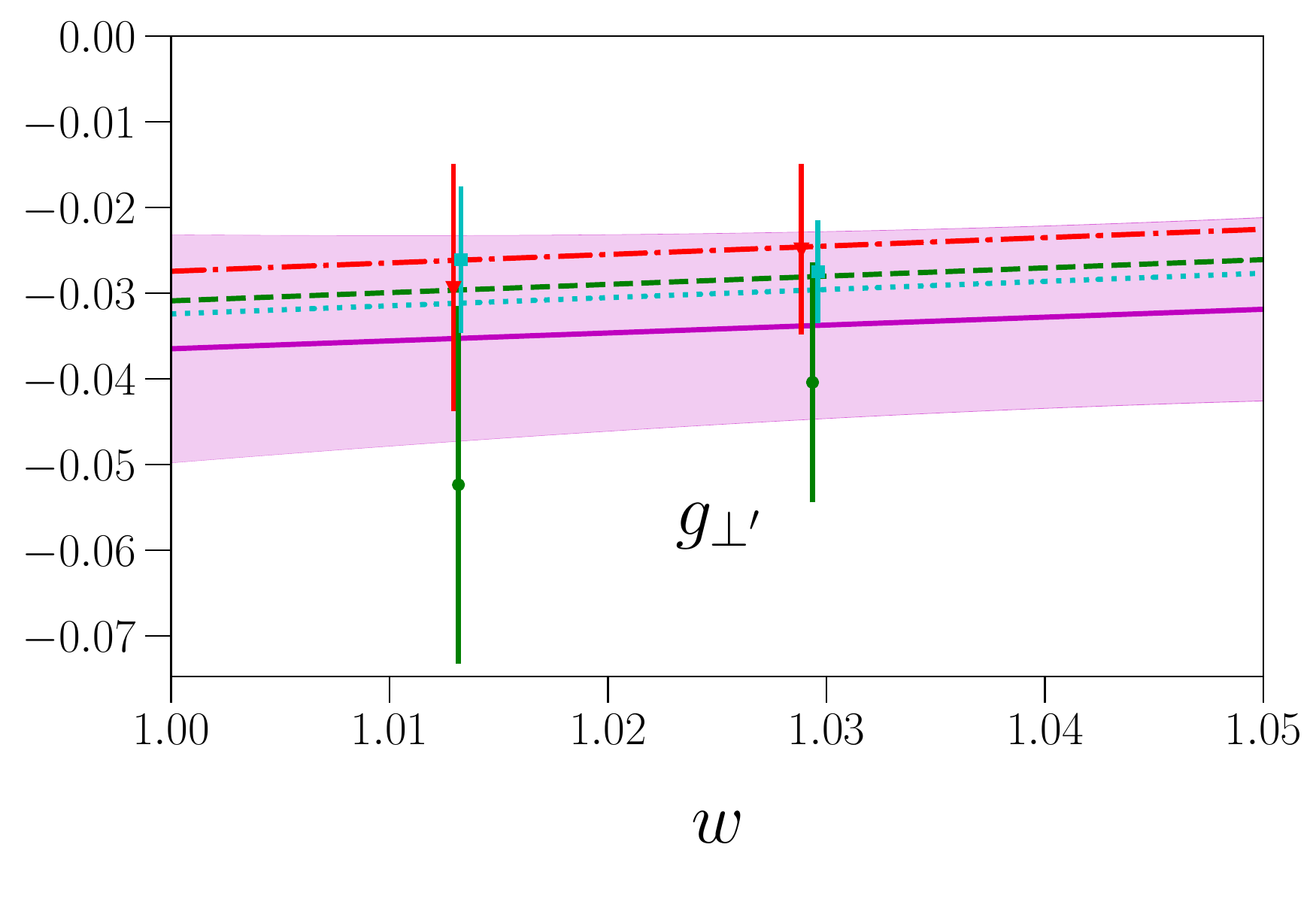} 
 
 \caption{\label{fig:FFextrap1}Chiral and continuum extrapolations of the vector and axial vector form factors. The  solid magenta curves show the form factors in the physical limit, with inner light magenta bands indicating the $1\sigma$ statistical uncertainties and outer dark magenta bands indicating the quadrature sums of statistical and estimated systematic uncertainties. The dashed-dotted, dashed, and dotted curves show the fit models evaluated at the pion masses and lattice spacings of the individual data sets C01, C005, and F004, respectively, where the uncertainty bands are omitted for clarity.}
\end{figure}

\begin{figure}
 \centerline{\includegraphics[width=0.6\linewidth]{figures/legend_datasets.pdf}}
 
 \vspace{1ex}

 \includegraphics[width=0.47\linewidth]{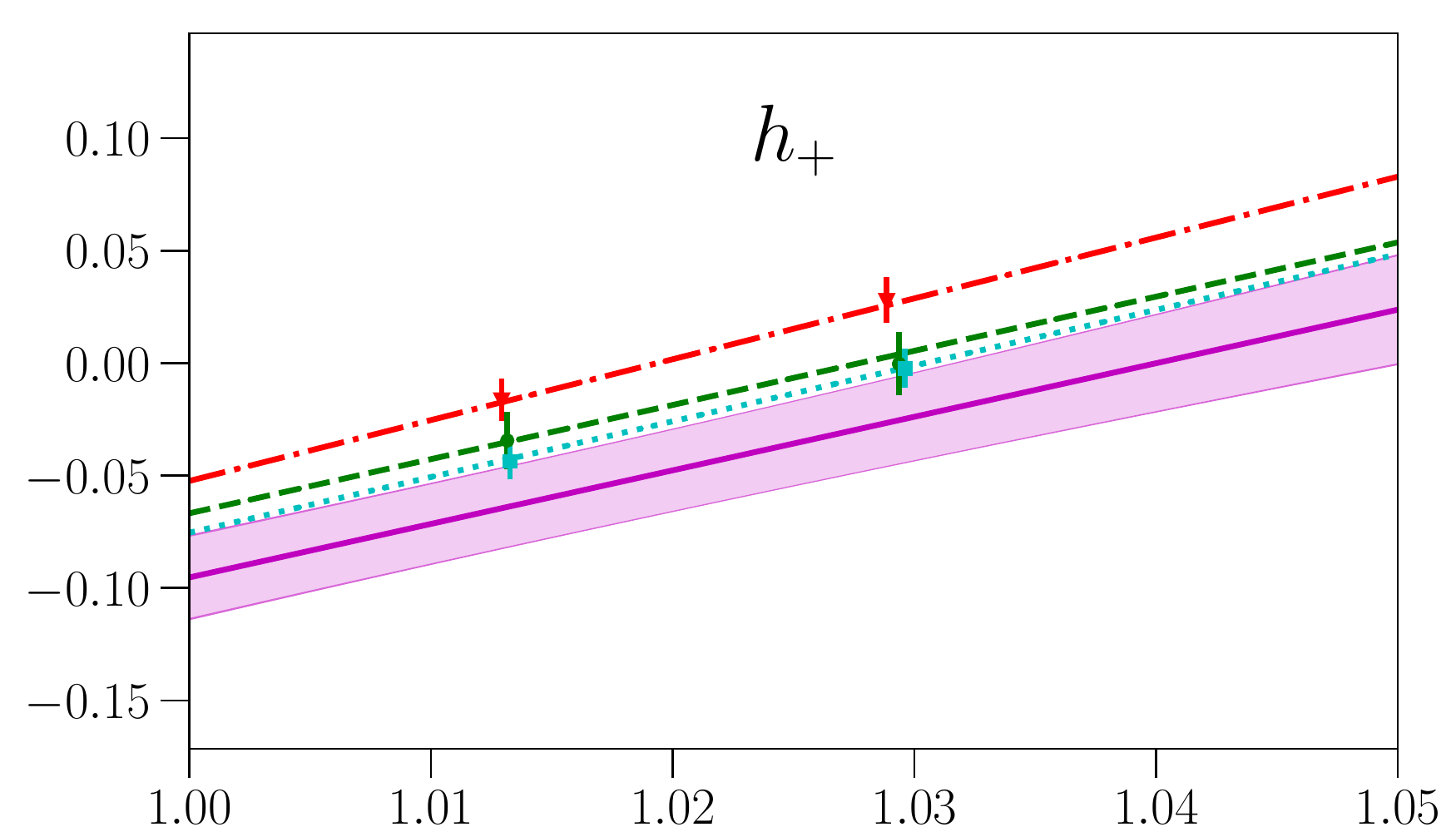} \hfill \includegraphics[width=0.47\linewidth]{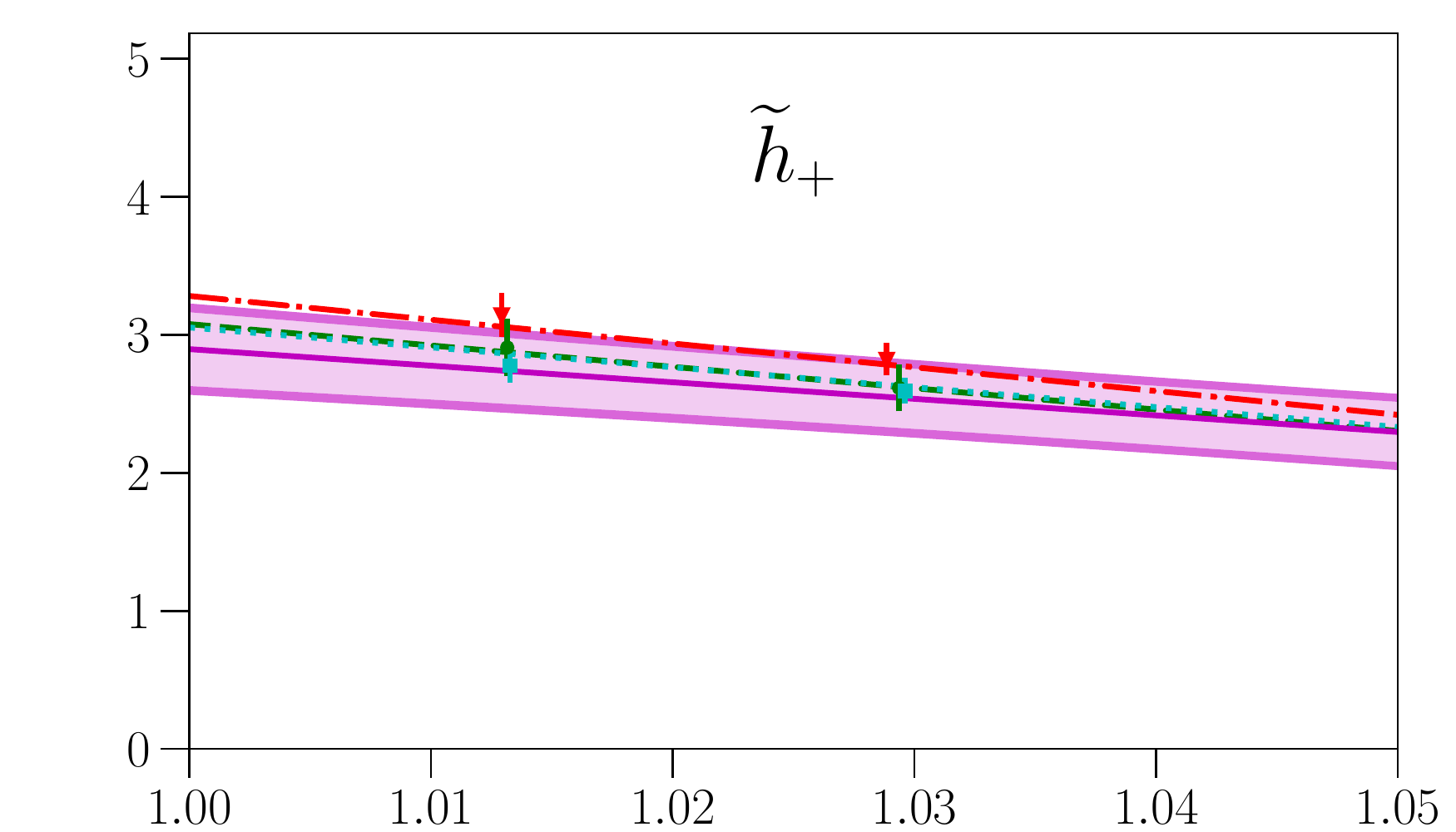} \\
 \includegraphics[width=0.47\linewidth]{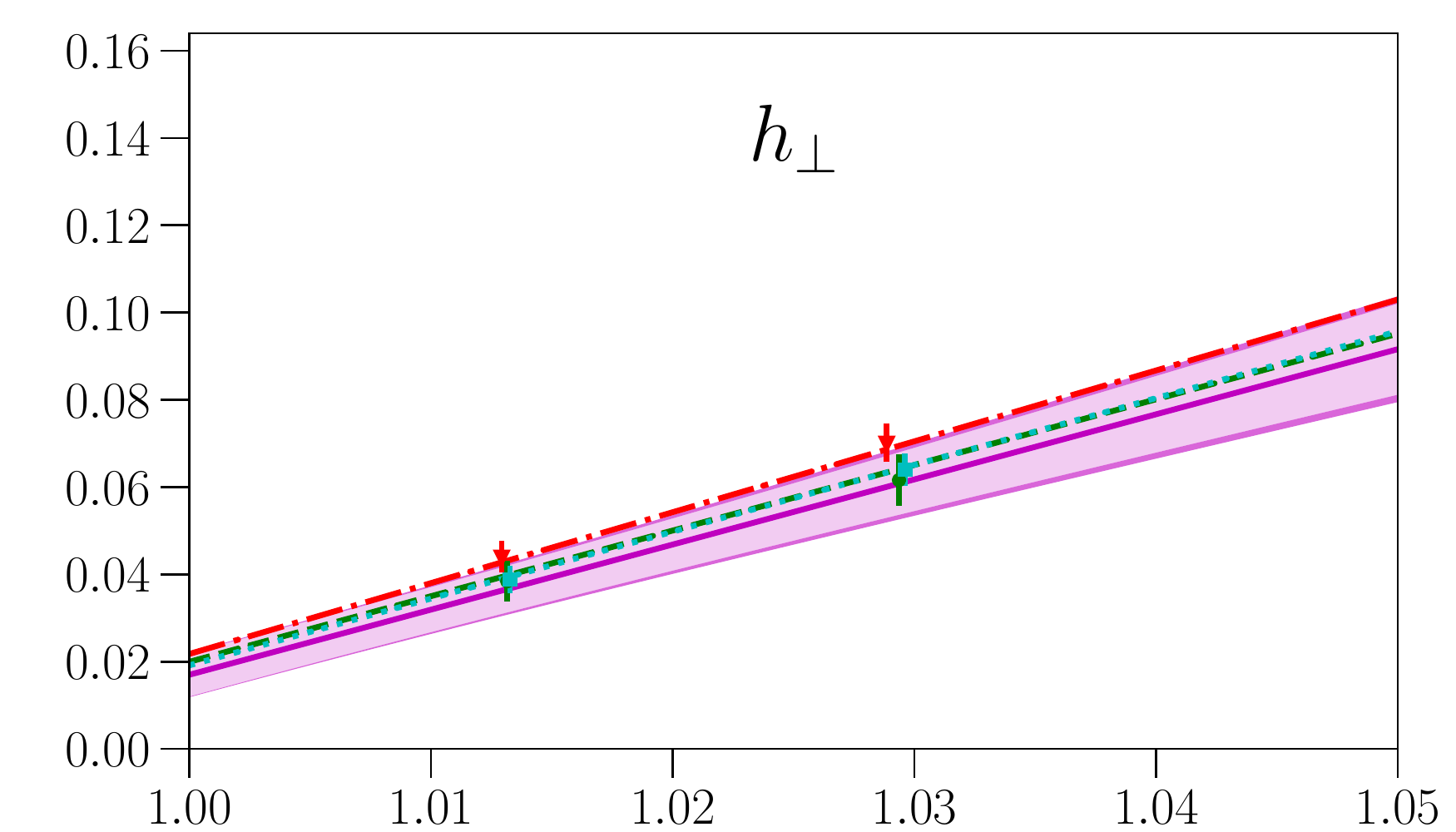} \hfill \includegraphics[width=0.47\linewidth]{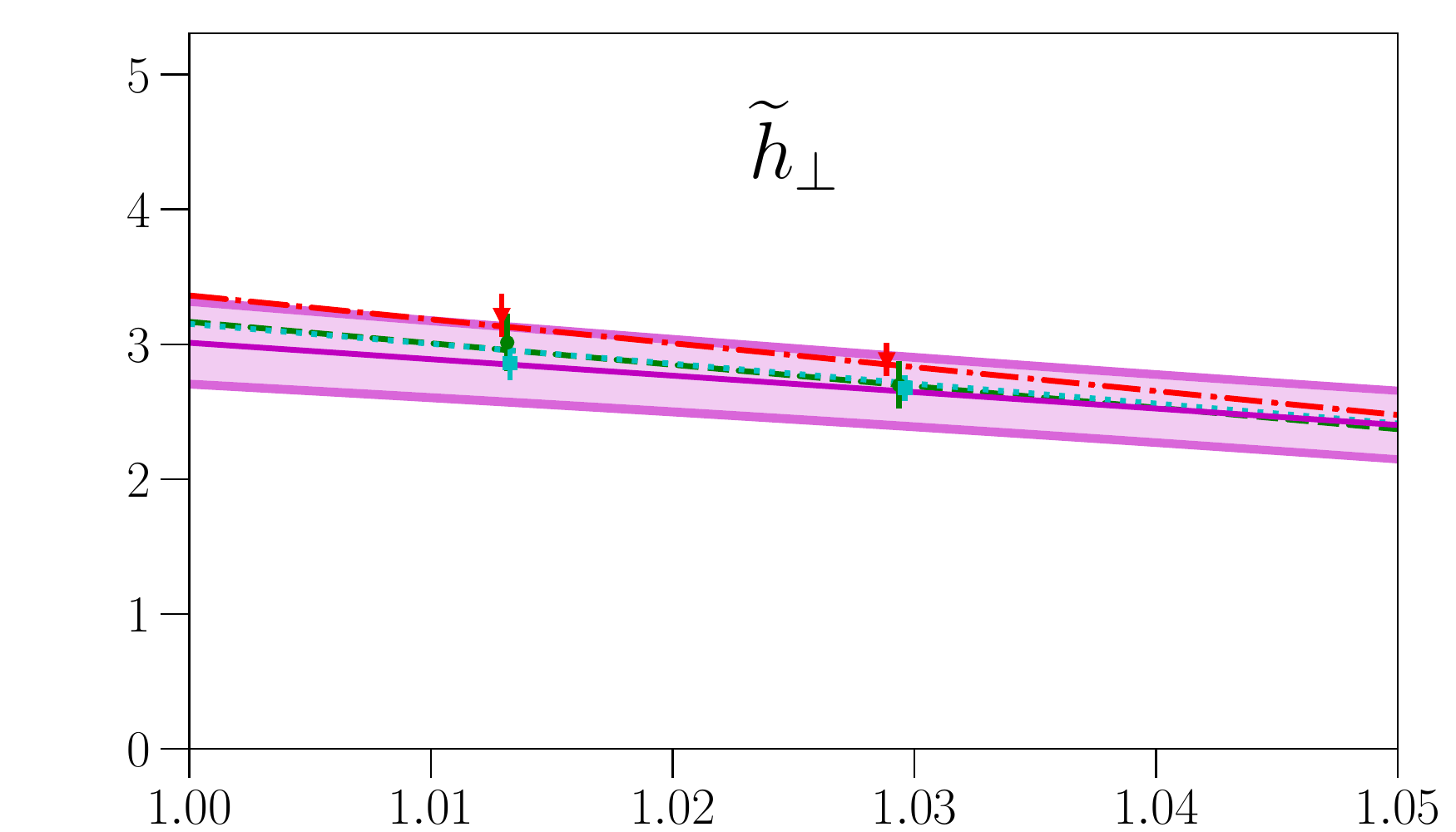} \\
 \includegraphics[width=0.47\linewidth]{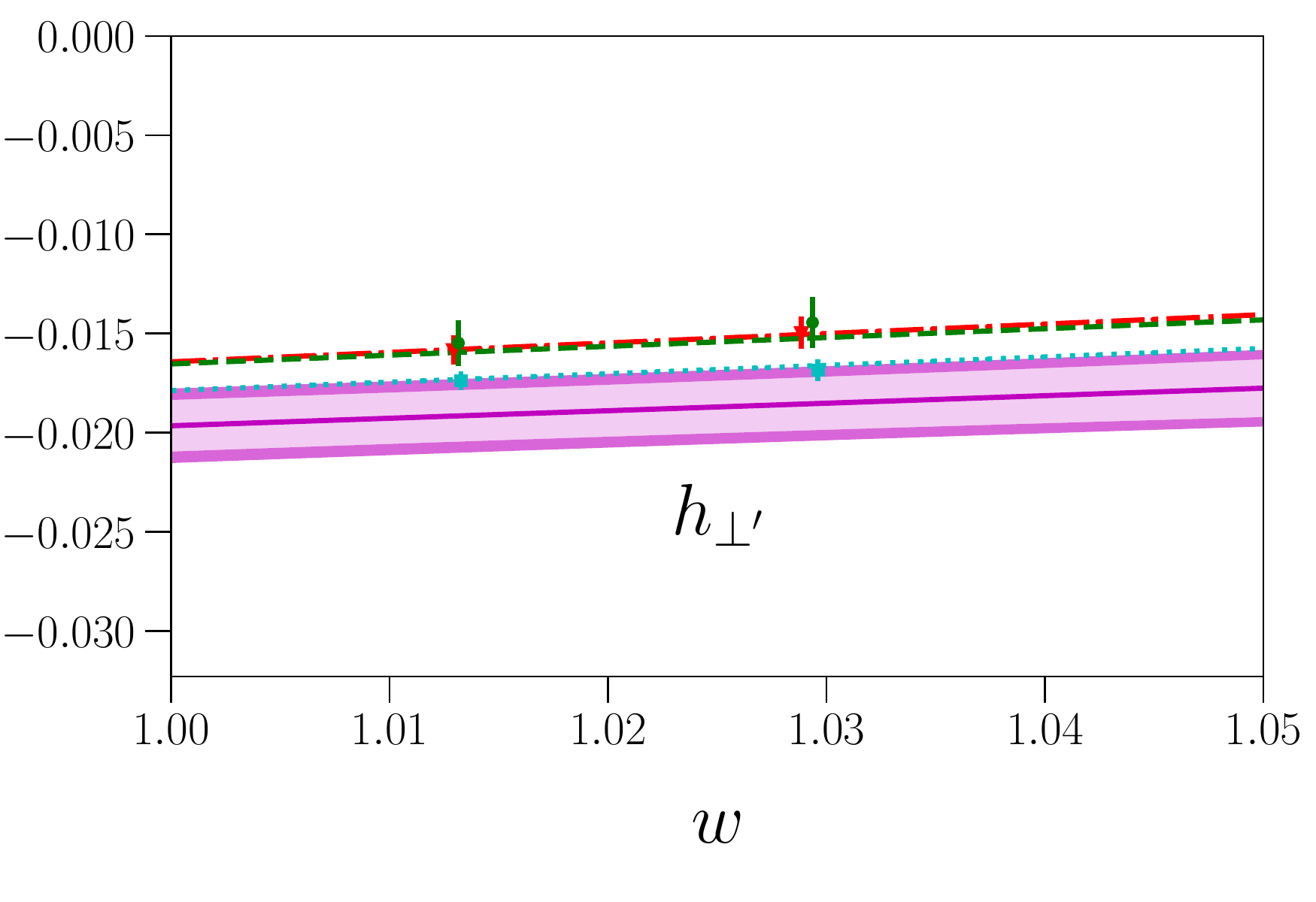} \hfill \includegraphics[width=0.47\linewidth]{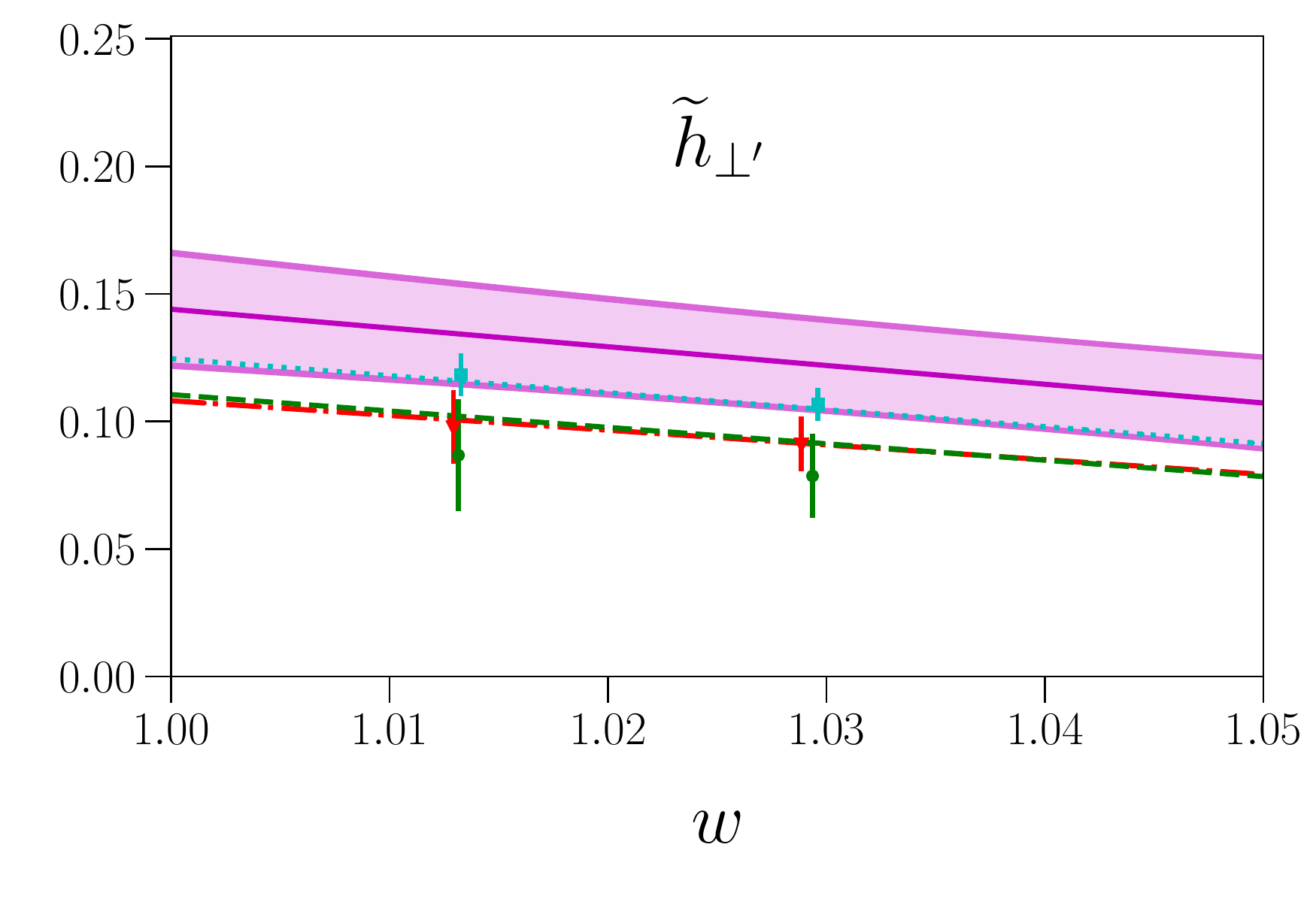} 
 
 \caption{\label{fig:FFextrap2}Like Fig.~\protect\ref{fig:FFextrap1}, but for the tensor form factors.}
\end{figure}

\begin{table}
\begin{tabular}{llllll}
\hline\hline
$f$ & & \hspace{4ex}$F^f$ & & \hspace{4ex}$A^f$ \\
\hline
${f_0}$                      && $\wm 3.54(29)$  && $-14.7(3.3)$ \\ 
${f_+}$                      && $\wm 0.0432(64)$  && $\nb\wm 1.63(19)$ \\ 
${f_\perp}$                  && $-0.068(18)$  &&  $\nb\wm 2.49(35)$ \\ 
${f_{\perp^\prime}}$         && $\wm 0.0461(18)$  && \nb$-0.161(27)$ \\ 
${g_0}$                      && $\wm 0.0024(38)$  && \nb$\wm 1.58(17)$ \\ 
${g_+}$                      && $\wm 2.95(25)$  && $-12.2(2.9)$ \\ 
${g_\perp}$                  && $\wm 2.92(24)$  && $-11.8(2.8)$ \\ 
${g_{\perp^\prime}}$         && $-0.037(14)$  && \nb$\wm 0.09(25)$ \\ 
${h_+}$                      && $-0.095(19)$  &&  \nb$\wm 2.38(32)$ \\ 
${h_\perp}$                  && $\wm 0.0170(43)$  &&  \nb$\wm 1.49(16)$ \\ 
${h_{\perp^\prime}}$         && $- 0.0196(13)$  && \nb$\wm 0.038(11)$ \\ 
${\tilde{h}_+}$              && $\wm 2.90(24)$  && $-12.0(2.9)$ \\ 
${\tilde{h}_\perp}$          && $\wm 3.01(25)$  && $-12.2(2.8)$ \\ 
${\tilde{h}_{\perp^\prime}}$ && $\wm 0.144(24)$  && \nb$-0.74(37)$ \\ 
\hline\hline 
\end{tabular}
\caption{\label{tab:FFparams} The nominal fit parameters describing the form factors in the physical limit. The parametrizations, which are accurate only in the high-$q^2$ region, are given by $f=F^{f}+A^{f}(w-1)$, where $w=v\cdot v^\prime=(m_{\Lambda_b}^2+m_{\Lambda^*}^2-q^2)/(2m_{\Lambda_b}m_{\Lambda^*})$. The $28\times 28$ covariance matrix is available as an ancillary file. The uncertainties given here are statistical only; see the main text for a discussion of systematic uncertainties.}
\end{table}

\begin{table}
\begin{tabular}{lllll}
  \hline\hline
  $f$                & & \hspace{4ex}$F^f_{\rm HO}$ & & \hspace{4ex}$A^f_{\rm HO}$ \\
  \hline
  $f_0$                    &&   $\wm    3.55(33)$   &&   $   -   14.6(3.3)$\\  
  $f_+$                    &&   $\wm  0.0433(70)$   &&   $\wm    1.64(20)$\\
  $f_{\perp}$              &&   $   -  0.068(19)$   &&   $\wm    2.51(38)$\\
  $f_{\perp^{\prime}}$     &&   $\wm  0.0462(27)$   &&   $   -  0.161(28)$\\
  $g_0$                    &&   $\wm  0.0024(39)$   &&   $\wm    1.59(18)$\\
  $g_+$                    &&   $\wm    2.95(29)$   &&   $   -   12.1(2.9)$\\  
  $g_{\perp}$              &&   $\wm    2.92(28)$   &&   $   -   11.7(2.8)$\\
  $g_{\perp^{\prime}}$     &&   $   -  0.037(14)$   &&   $\wm    0.09(25)$\\
  $h_+$                    &&   $   -  0.095(20)$   &&   $\wm    2.39(37)$\\
  $h_{\perp}$              &&   $\wm  0.0169(47)$   &&   $\wm    1.50(20)$\\
  $h_{\perp^{\prime}}$     &&   $   - 0.0197(19)$   &&   $\wm   0.038(11)$\\
  $\widetilde{h}_+$        &&   $\wm    2.90(32)$   &&   $   -   11.9(2.9)$\\
  $\widetilde{h}_{\perp}$  &&   $\wm    3.01(33)$   &&   $   -   12.1(2.9)$\\
  $\widetilde{h}_{\perp^{\prime}}$  &&   $\wm   0.145(27)$   &&   $   -   0.74(37)$\\
  \hline\hline
 \end{tabular}
\caption{\label{tab:FFparams_HO} Form-factor parameters obtained from fits including higher-order terms. These parameters are used only to estimate systematic uncertainties as explained in the main text. The $28\times 28$ covariance matrix is available as an ancillary file.}
\end{table}

Our results for the physical-limit parameters $F^f$ and $A^f$ are given in Table \ref{tab:FFparams}. The full $28\times 28$ covariance matrix of the parameters for all fourteen form factors is available as an ancillary file. The form factors in the physical
limit are plotted in Figs.~\ref{fig:FFextrap1} and \ref{fig:FFextrap2}. The
dashed-dotted, dashed, and dotted curves show the fit models evaluated at the pion masses and lattice spacings of the individual data sets C01, C005, and F004, respectively, where the uncertainty bands are omitted for clarity. We see that the data are well described by the model. The results for the parameters $C^f$, $D^f$, $\tilde{C}^f$, and $\tilde{D}^f$ are in fact consistent with zero within the statistical uncertainties. To report the values of $\chi^2/{\rm d.o.f.}$ of the fits, we need to make a choice for the number of parameters to be subtracted from the number of data points to obtain the number of degrees of freedom. If we count $F^f$, $A^f$, $C^f$, and $D^f$ as parameters that are primarily constrained by the data, then ${\rm d.o.f.}=6-4=2$. With this choice, the fits have $\chi^2/{\rm d.o.f.}$ in the range from approximately 0.3 to 1.2.

To estimates systematic uncertainties associated with the chiral and continuum extrapolation, we additionally performed ``higher-order'' fits using the model
\begin{eqnarray}
\nonumber f_{\rm HO}(q^2)&=&F_{\rm HO}^{f}\left[1+C_{\rm HO}^{f}\frac{m_{\pi}^2-m_{\pi,\rm phys}^2}{(4\pi f_{\pi})^2}+H_{\rm HO}^{f}\frac{m_{\pi}^3-m_{\pi,\rm phys}^3}{(4\pi f_{\pi})^3}+D_{\rm HO}^{f}a^2\Lambda^2+E_{\rm HO}^{f}a \Lambda +G_{\rm HO}^{f}a^3\Lambda^3 \right] \\
 &&+A_{\rm HO}^{f}\left[1+\tilde{C}_{\rm HO}^{f}\frac{m_{\pi}^2-m_{\pi,\rm phys}^2}{(4\pi f_{\pi})^2}+\tilde{H}_{\rm HO}^{f}\frac{m_{\pi}^3-m_{\pi,\rm phys}^3}{(4\pi f_{\pi})^3}+\tilde{D}_{\rm HO}^{f}a^2\Lambda^2+\tilde{E}_{\rm HO}^{f}a \Lambda +\tilde{G}_{\rm HO}^{f}a^3\Lambda^3\right](w-1), \hspace{6ex} \label{eq:extrapolationHO} 
\end{eqnarray}
using Gaussian priors for the parameters $C_{\rm HO}^f$, $H_{\rm HO}^f$, $D_{\rm HO}^f$, $G_{\rm HO}^f$, $\tilde{C}_{\rm HO}^f$,  $\tilde{H}_{\rm HO}^f$, $\tilde{D}_{\rm HO}^f$, $\tilde{G}_{\rm HO}^f$ with central values equal to 0 and widths equal to 10. The terms with coefficients $E_{\rm HO}^{f}$ and $\tilde{E}_{\rm HO}^{f}$ are meant to describe the effects of the incomplete $\mathcal{O}(a)$ improvement of the heavy-light currents using only the $d_1$ correction term in Eq.~(\ref{eq:improvedcurrent}) and with $d_1$ evaluated at mean-field-improved tree level. In Ref.~\cite{Detmold:2015aaa}, results for the $\Lambda_b \to p$ form factors (using the same actions and lattice spacings) using the incomplete ($d_1$ only) and full operator bases for the $\mathcal{O}(a)$ improvement were compared, albeit with all coefficients evaluated at one loop (the coefficients equivalent to $d_1$ are denoted as $c_\Gamma^R$ in Ref.~\cite{Detmold:2015aaa}). The results were found to differ only by less than 0.3\%. The one-loop and tree-level values of $d_1$ differ only by approximately $0.02$, but we also expect larger $\mathcal{O}(a)$ effects associated with the use of nonzero $\Lambda_b$ momentum. We therefore conservatively allow for the effect of the missing radiative corrections to the $\mathcal{O}(a)$ improvement to be as large as 5 percent at the coarse lattice spacing. This translates to setting the prior widths of the parameters  $E_{\rm HO}^{f}$ and $\tilde{E}_{\rm HO}^{f}$ to 0.3.

In the higher-order fits, we also incorporate the systematic uncertainties associated with the residual matching factors $\rho_\Gamma$, as well as scale-setting and isospin-symmetry-breaking/QED effects. The residual matching factors were computed at one loop for the vector and axial-vector currents, and the size of the missing higher-order corrections was estimated to be below 0.07\% in Ref.~\cite{Detmold:2015aaa}, a result of the smallness of the one-loop corrections (this is the benefit of the ``mostly nonperturbative'' method). Nevertheless, because we improved the tuning of the $b$-quark action parameters here without recomputing the one-loop corrections to the current matching factors, we allow for matching uncertainties in the vector and axial vector form factors as large as 2\%. For the tensor form factors, we estimate the size of the missing one-loop corrections to the residual matching factors to be 5.316\% at $\mu=m_b$ as discussed in Sec.~\ref{sec:threept}. The neglected effects from $m_u-m_d\neq 0$ and QED in the form factors are estimated to be approximately $1\%$. The current-matching and isospin-breaking/QED uncertainties were included in the higher-order fits by multiplying each form factor with Gaussian random distributions of central value 1 and width corresponding to the estimated uncertainty. These distributions were taken to be correlated within each of the groups  $\{f_0, f_+, f_\perp, f_{\perp^\prime}\}$,   $\{g_0, g_+, g_\perp, g_{\perp^\prime}\}$,  $\{h_+, h_\perp, h_{\perp^\prime}\}$, $\{\tilde{h}_+, \tilde{h}_\perp, \tilde{h}_{\perp^\prime}\}$, but uncorrelated across different groups. The scale-setting uncertainties were incorporated by promoting the lattice spacings to fit parameters, constrained to have the known values and uncertainties.

In the physical limit, the higher-order fit functions again reduce to the form as in Eq.~(\ref{eq:physicalFF}), with $F^f$ and $A^f$ replaced by $F^f_{\rm HO}$ and $A^f_{\rm HO}$. The results for these parameters are given in Table \ref{tab:FFparams_HO}, and the corresponding covariance matrix is available as another ancillary file. As in Refs.~\cite{Detmold:2015aaa,Detmold:2016pkz}, we evaluate the systematic form-factor uncertainty of any observable $O$ through
\begin{equation}
 \sigma_{O,{\rm syst}} = {\rm max}\left( |O_{\rm HO}-O|,\: \sqrt{|\sigma_{O,{\rm HO}}^2-\sigma_O^2|}  \right), \label{eq:sigmasyst}
\end{equation}
where $O$, $\sigma_O$ denote the central value and uncertainty obtained using the parameter values and covariance matrix of the nominal fit and $O_{\rm HO}$, $\sigma_{O,{\rm HO}}^2$ denote the central value and uncertainty obtained using the parameter values and covariance matrix of the higher-order fit. The systematic and statistical uncertainties are then added in quadrature to obtain the total uncertainties. The total uncertainties of the form factors themselves are shown with the dark-magenta bands in Figs.~\ref{fig:FFextrap1} and \ref{fig:FFextrap2}. For some of the form factors, the statistical uncertainties are so large that adding the systematic uncertainties does not visibly increase the width of the band. When applying Eq.~(\ref{eq:sigmasyst}) to the $\Lambda_b \to \Lambda^*(1520)\ell^+\ell^-$ differential branching fraction in the region $q^2\geq 16\:{\rm GeV}^2$, we find that the systematic uncertainties in the form factors contribute an uncertainty ranging from 9.7 to 11.4 percent in $d\mathcal{B}/dq^2$. Because $d\mathcal{B}/dq^2$ depends quadratically on the form factors, this corresponds to an effective form-factor systematic uncertainty in the range from 4.9 to 5.7 percent.

Finally, note that our estimates of systematic uncertainties do not account for errors introduced by performing the data analysis as if the $\Lambda^*(1520)$ is a stable hadron. We expect these errors to be small, given the narrow width of the $\Lambda^*(1520)$ and our restriction to the rest frame. A more rigorous determination of $\Lambda_b \to \Lambda^*(1520)$ form factors that treats the $\Lambda^*(1520)$ as an unstable resonance in coupled-channel $p$-$K$, $\Sigma$-$\pi$ scattering may be possible using the finite-volume formalism of Refs.~\cite{Briceno:2014uqa,Briceno:2015csa}, but this is far beyond the scope of the present work. In the absence of such an analysis, we also cannot reliably estimate finite-volume effects in the form factors, although we note that $m_\pi L>4$ for all ensembles used here.

\FloatBarrier
\section{\texorpdfstring{$\bm{\Lambda_b \to \Lambda^*(1520)\ell^+\ell^-}$}{Lambdab to Lambda*(1520)mu+mu-} observables}
\label{sec:observables}
\FloatBarrier

To calculate the $\Lambda_b \to \Lambda^*(1520)\ell^+\ell^-$ observables, we employ the usual operator-product expansion that allows us to express the decay amplitude in terms of local hadronic matrix elements \cite{Beylich:2011aq}. For the differential decay rate in the Standard Model, we find
\begin{equation}
\frac{\mathrm{d}\Gamma}{\mathrm{d}q^2}=  \frac{G_{F}^2\alpha_{\rm em}^2}{3\cdot 2^{10} \pi^5 m^3_{\Lambda_b}} \left|V_{tb}V_{ts}^{*}\right|^2 \upsilon \sqrt{s_+s_-}  \left[A_1 \left(2 m_{\ell}^2+q^2\right)+A_2 q^2 \upsilon^2+6 A_t m_{\ell}^2 \right],
\end{equation}
where $ \upsilon=\sqrt{1-4 m_{\ell}^2/q^2}$, and the quantities $A_1$, $A_2$, and $A_t$ are given by
\begin{align}
\nonumber A_1&=\left| H_1\left(-1,\frac{1}{2},\frac{3}{2}\right)\right|^2+\left| H_1\left(-1,-\frac{1}{2},\frac{1}{2}\right)\right|^2+\left| H_1\left(0,\frac{1}{2},\frac{1}{2}\right)\right|^2\\
&+\left| H_1\left(0,-\frac{1}{2},-\frac{1}{2}\right)\right|^2+\left| H_1\left(1,\frac{1}{2},-\frac{1}{2}\right)\right|^2+\left| H_1\left(1,-\frac{1}{2},-\frac{3}{2}\right)\right|^2,\\
\nonumber A_2&=\left| H_2\left(-1,\frac{1}{2},\frac{3}{2}\right)\right|^2+\left| H_2\left(-1,-\frac{1}{2},\frac{1}{2}\right)\right|^2+\left| H_2\left(0,\frac{1}{2},\frac{1}{2}\right)\right|^2\\
&+\left| H_2\left(0,-\frac{1}{2},-\frac{1}{2}\right)\right|^2+\left| H_2\left(1,\frac{1}{2},-\frac{1}{2}\right)\right|^2+\left| H_2\left(1,-\frac{1}{2},-\frac{3}{2}\right)\right|^2,\\
A_t&=\left| H_2\left(t,\frac{1}{2},\frac{1}{2}\right)\right|^2+\left| H_2\left(t,-\frac{1}{2},-\frac{1}{2}\right)\right|^2.
\end{align}
Here, $H_1$ and $H_2$ are linear combinations of hadronic helicity amplitudes with the appropriate Wilson coefficients:
\begin{eqnarray}
H_{1}&=&-\frac{2 m_{b}}{q^{2}}C^{\rm eff}_{7}(q^2)\left(H_T+H_{T5}\right)+C^{\rm eff}_{9}(q^2)\left(H_V-H_A\right),
\label{eq:h1} \\
H_{2}&=&C_{10}\left(H_V-H_A\right).
\label{eq:h2}
\end{eqnarray}
In terms of the form factors, the helicity amplitudes (in our sign conventions) for the vector, axial-vector, and tensor currents are equal to
\begin{align}
H_V\left(t,\frac{1}{2},\frac{1}{2}\right)=H_V\left(t,-\frac{1}{2},-\frac{1}{2}\right)&=-f_0\frac{ (m_{\Lambda_b}-m_{\Lambda^*}) \sqrt{s_-}}{\sqrt{6\, q^2}}, \\
H_V\left(0,\frac{1}{2},\frac{1}{2}\right)=H_V\left(0,-\frac{1}{2},-\frac{1}{2}\right)&=-f_{+}\frac{ (m_{\Lambda_b}+m_{\Lambda^*}) \sqrt{s_+}}{\sqrt{6\, q^2}}, \\
H_V\left(1,\frac{1}{2},-\frac{1}{2}\right)=-H_V\left(-1,-\frac{1}{2},\frac{1}{2}\right)&=-f_{\perp}\frac{ \sqrt{s_+}}{\sqrt{3}}, \\
H_V\left(1,-\frac{1}{2},-\frac{3}{2}\right)=H_V\left(-1,\frac{1}{2},\frac{3}{2}\right)&=f_{\perp^\prime} \sqrt{s_+},
\end{align}
\begin{align}
H_A\left(t,\frac{1}{2},\frac{1}{2}\right)=-H_A\left(t,-\frac{1}{2},-\frac{1}{2}\right)&=g_0\frac{ (m_{\Lambda_b}+m_{\Lambda^*}) \sqrt{s_+}}{\sqrt{6\, q^2}}, \\
H_A\left(0,\frac{1}{2},\frac{1}{2}\right)=-H_A\left(0,-\frac{1}{2},-\frac{1}{2}\right)&=g_{+}\frac{ (m_{\Lambda_b}-m_{\Lambda}^*) \sqrt{s_-}}{\sqrt{6\, q^2}}, \\
H_A\left(1,\frac{1}{2},-\frac{1}{2}\right)=-H_A\left(-1,-\frac{1}{2},\frac{1}{2}\right)&=-g_{\perp} \frac{\sqrt{s_-}}{\sqrt{3}}, \\
H_A\left(1,-\frac{1}{2},-\frac{3}{2}\right)=-H_A\left(-1,\frac{1}{2},\frac{3}{2}\right)&=-g_{\perp^\prime} \sqrt{s_-},
\end{align}
and
\begin{align}
H_T\left(t,\frac{1}{2},\frac{1}{2}\right)=H_T\left(t,-\frac{1}{2},-\frac{1}{2}\right)&=0, \\
H_T\left(0,\frac{1}{2},\frac{1}{2}\right)=H_T\left(0,-\frac{1}{2},-\frac{1}{2}\right)&=-h_{+}\frac{ (m_{\Lambda_b}+m_{\Lambda^*}) \sqrt{s_+}}{\sqrt{6\, q^2}}, \\
H_T\left(1,\frac{1}{2},-\frac{1}{2}\right)=H_T\left(-1,-\frac{1}{2},\frac{1}{2}\right)&=-h_{\perp}\frac{ \sqrt{s_+}}{\sqrt{3}}, \\
H_T\left(1,-\frac{1}{2},-\frac{3}{2}\right)=H_T\left(-1,\frac{1}{2},\frac{3}{2}\right)&=h_{\perp^\prime} \sqrt{s_+}, \\
H_{T5}\left(t,\frac{1}{2},\frac{1}{2}\right)=-H_{T5}\left(t,-\frac{1}{2},-\frac{1}{2}\right)&=0, \\
H_{T5}\left(0,\frac{1}{2},\frac{1}{2}\right)=-H_{T5}\left(0,-\frac{1}{2},-\frac{1}{2}\right)&=\widetilde{h}_{+}\frac{ (m_{\Lambda_b}-m_{\Lambda}^*) \sqrt{s_-}}{\sqrt{6\, q^2}}, \\
H_{T5}\left(1,\frac{1}{2},-\frac{1}{2}\right)=-H_{T5}\left(-1,-\frac{1}{2},\frac{1}{2}\right)&=-\widetilde{h}_{\perp}\frac{ \sqrt{s_-}}{\sqrt{3}}, \\
H_{T5}\left(1,-\frac{1}{2},-\frac{3}{2}\right)=-H_{T5}\left(-1,\frac{1}{2},\frac{3}{2}\right)&=-\widetilde{h}_{\perp^\prime} \sqrt{s_-}.
\end{align}
For the effective Wilson coefficients $C^{\rm eff}_{7}(q^2)$ and $C^{\rm eff}_{9}(q^2)$, we use the expressions given in Eqs.~(65) and (66) of Ref.~\cite{Detmold:2016pkz}. The Wilson coefficients $C_1$ through $C_{10}$, the strong and electromagnetic couplings, and the $b$ and $c$ quark masses are also evaluated as in Ref.~\cite{Detmold:2016pkz}. We take
\begin{equation}
 \left|V_{tb}V_{ts}^{*}\right| = 0.04120 \pm 0.00056 \label{eq:VtsVtb}
\end{equation}
from the Summer 2018 Standard-Model fit performed by the UTFit Collaboration \cite{UTfit}, and, to obtain $\mathrm{d}\mathcal{B}/\mathrm{d}q^2=\tau_{\Lambda_b}\mathrm{d}\Gamma/\mathrm{d}q^2$, the $\Lambda_b$ lifetime
\begin{equation}
 \tau_{\Lambda_b} = (1.471 \pm 0.009)\:\:{\rm ps} \label{eq:tauLb} 
\end{equation}
from the Review of Particle Physics \cite{Tanabashi:2018oca}.

The uncertainties estimated for the Standard-Model predictions shown below include the form-factor statistical and systematic uncertainties, the perturbative uncertainties, an estimate of quark-hadron duality violations (as in Ref.~\cite{Detmold:2016pkz}), and the parametric uncertainties from Eqs.~(\ref{eq:VtsVtb}), and (\ref{eq:tauLb}).

\begin{figure}
 \includegraphics[width=0.5\linewidth]{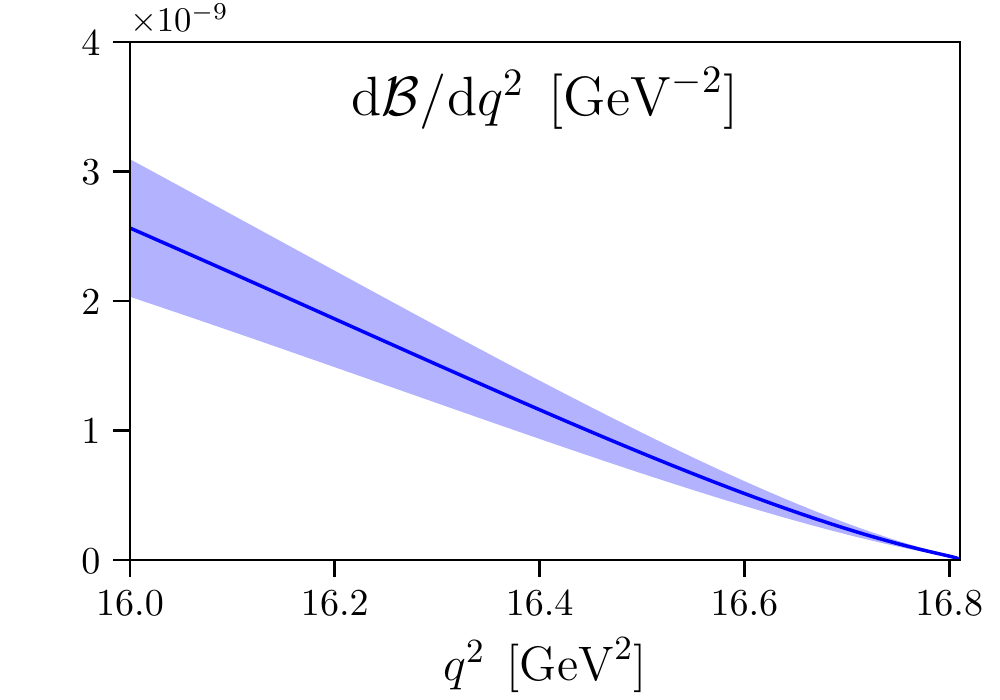}
 \caption{\label{fig:dBdqsqr}The $\Lambda_b \to \Lambda^*(1520)\ell^+\ell^-$ differential branching fraction in the high-$q^2$ region calculated in the Standard Model using our form factor results. Note that the factor of $\mathcal{B}(\Lambda^* \to p K^-)$ is not included here.}
\end{figure}

Our prediction for the differential branching fraction in the high-$q^2$ region is shown in Fig.~\ref{fig:dBdqsqr}. Here we have set $m_\ell=0$, which, in this kinetic region, is a good approximation for both electrons and muons. We only show results above $q^2=16\text{ GeV}^2$ because our lattice data only reach down to approximately $16.3\text{ GeV}^2$, and our parametrization of the $q^2$-dependence of the form factors is not expected to be reliable for lower $q^2$. In this kinematic region, our numerical results for $\mathrm{d}\mathcal{B}/\mathrm{d}q^2$ are approximately a factor of 2 lower than those obtained using the quark-model form factors of Ref.~\cite{Mott:2011cx}.

\begin{figure}
\includegraphics[width=0.47\linewidth]{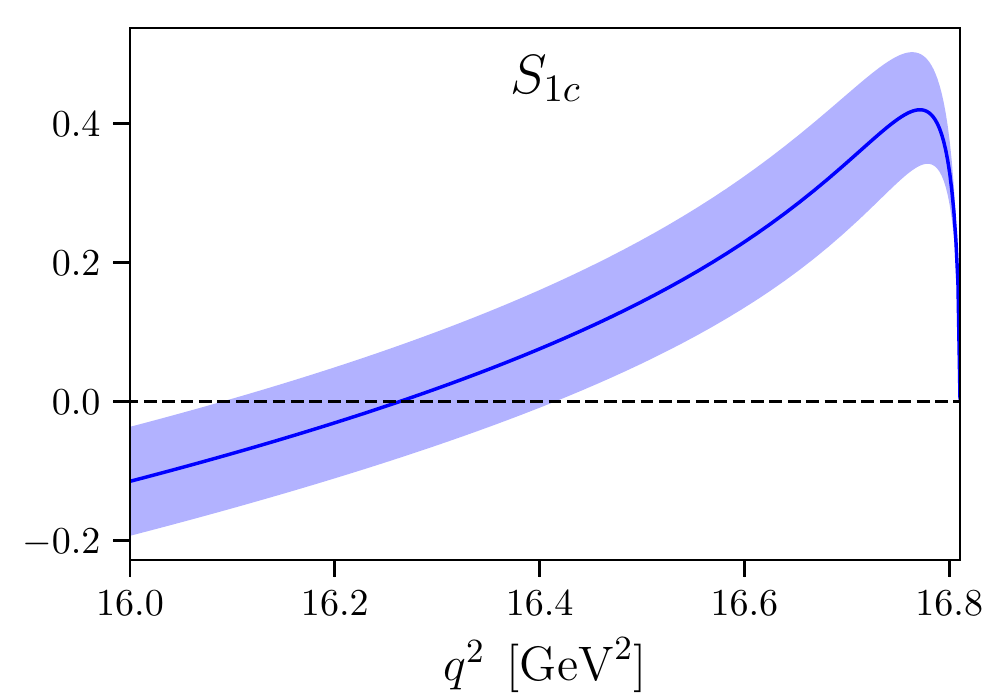} \hfill \includegraphics[width=0.47\linewidth]{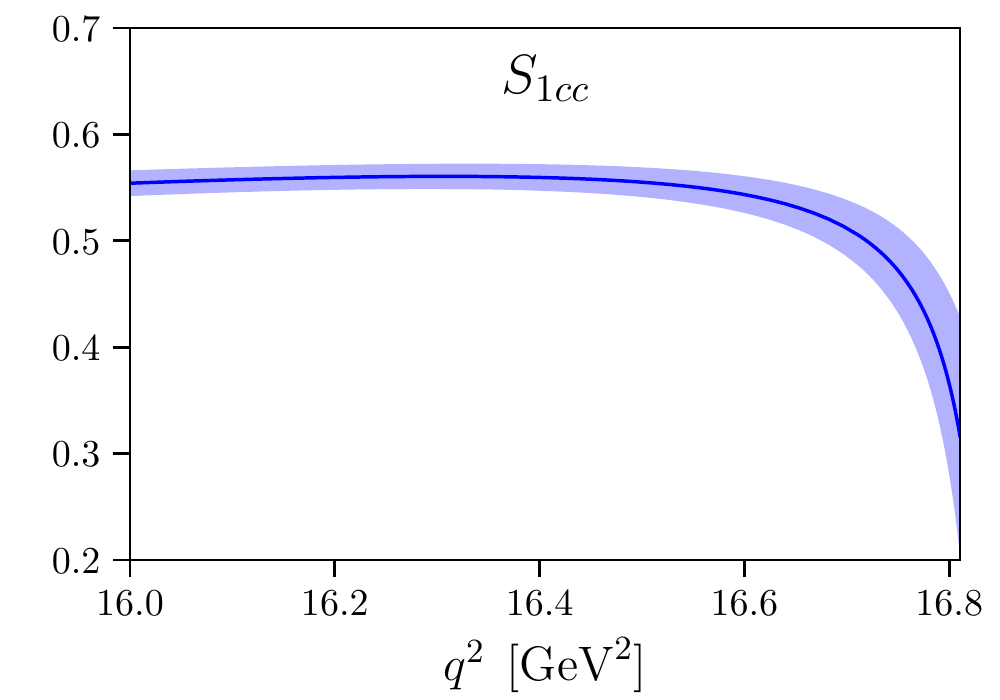}

\vspace{2ex}

 \includegraphics[width=0.47\linewidth]{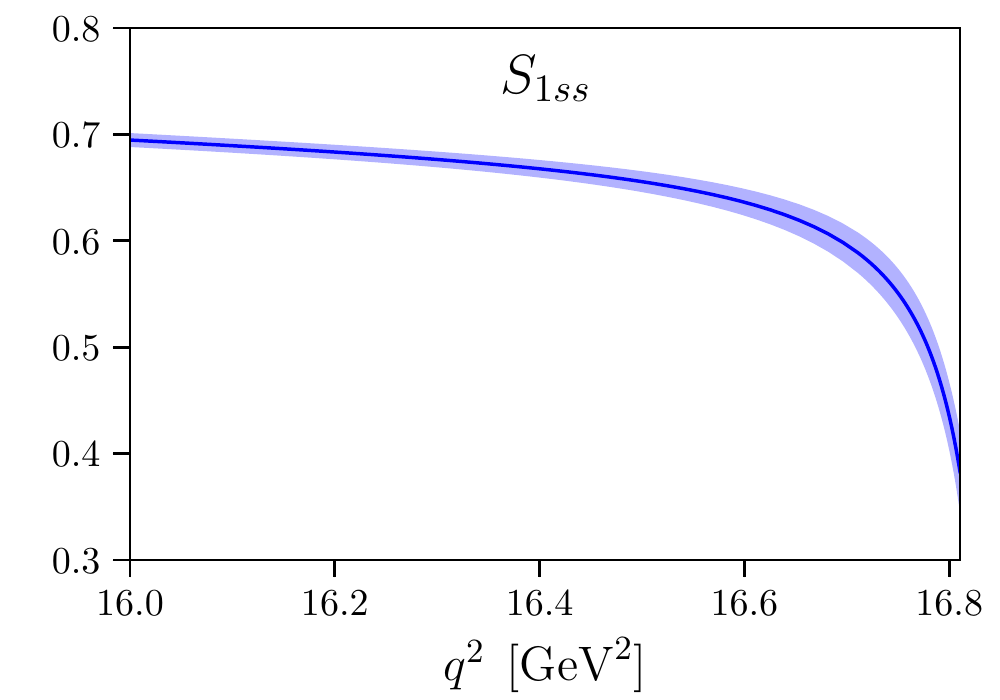} \hfill \includegraphics[width=0.47\linewidth]{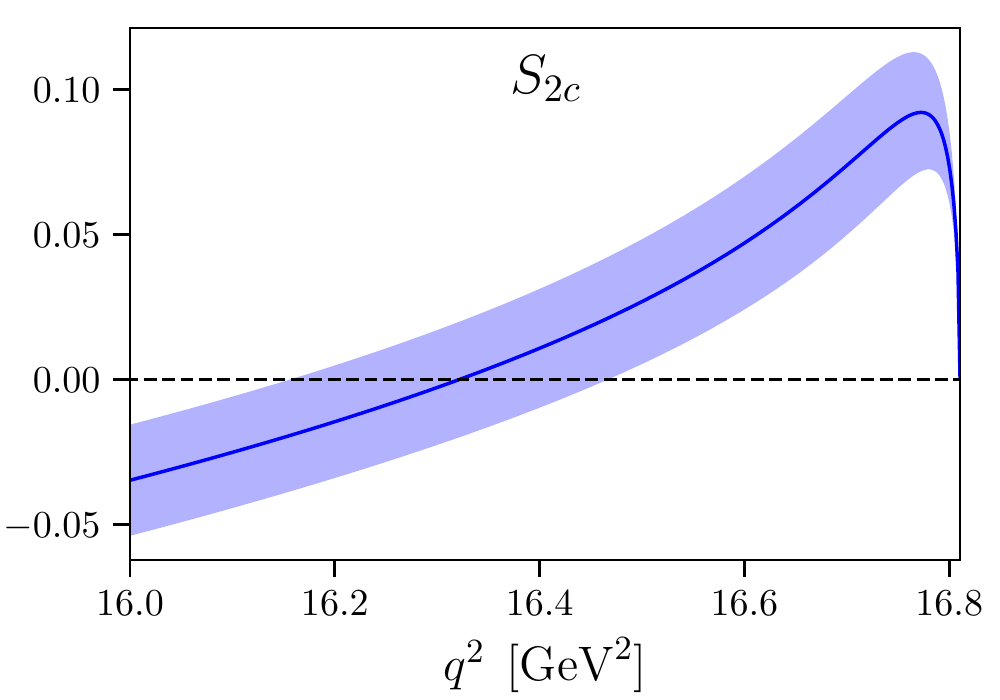} 

\vspace{2ex}
 
  \includegraphics[width=0.47\linewidth]{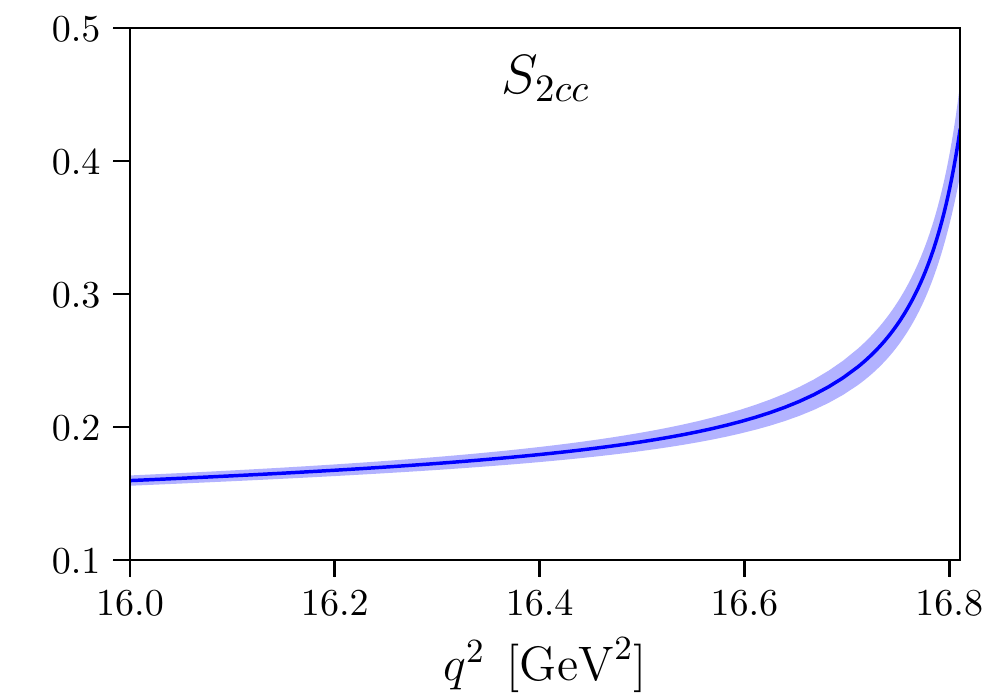} \hfill \includegraphics[width=0.47\linewidth]{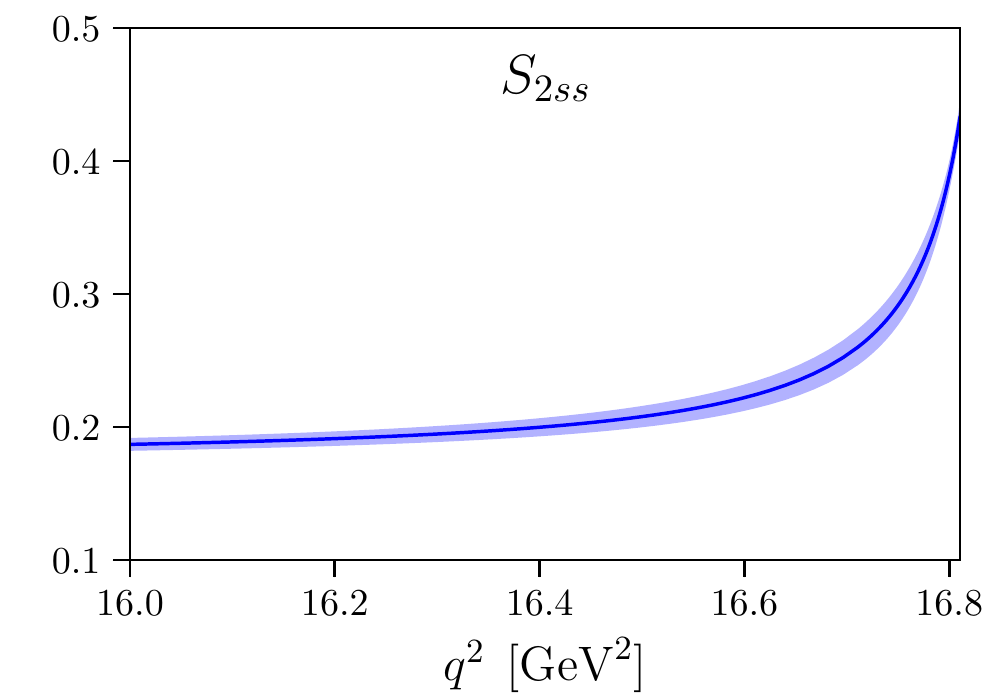} 
 \caption{\label{fig:angular1}The $\Lambda_b \to \Lambda^*(1520)(\to p K^-)\ell^+\ell^-$ angular observables $S_{1c}$, $S_{1cc}$, $S_{1ss}$, $S_{2c}$, $S_{2cc}$, and $S_{2ss}$ in the high-$q^2$ region calculated in the Standard Model using our form factor results.}
\end{figure}

\begin{figure}
\includegraphics[width=0.47\linewidth]{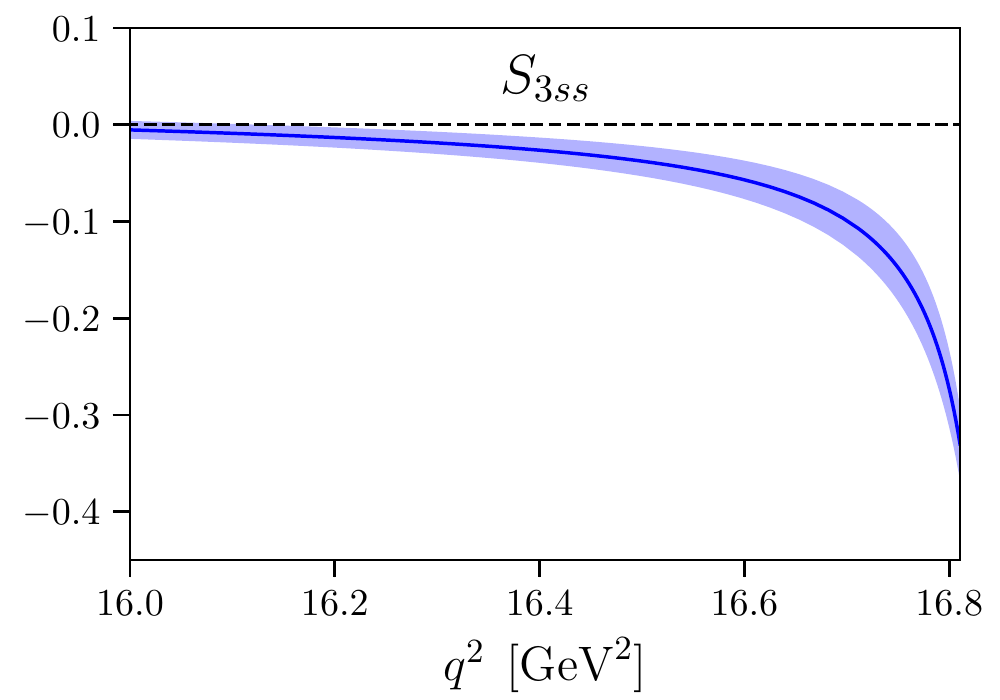} \hfill \includegraphics[width=0.47\linewidth]{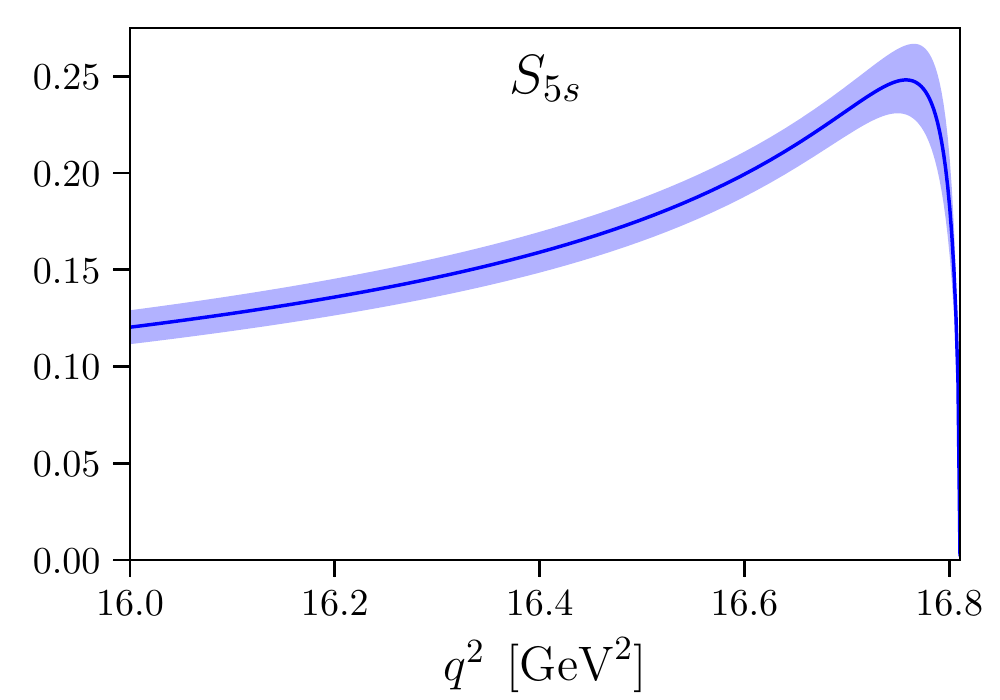} 

\vspace{2ex}

 \includegraphics[width=0.47\linewidth]{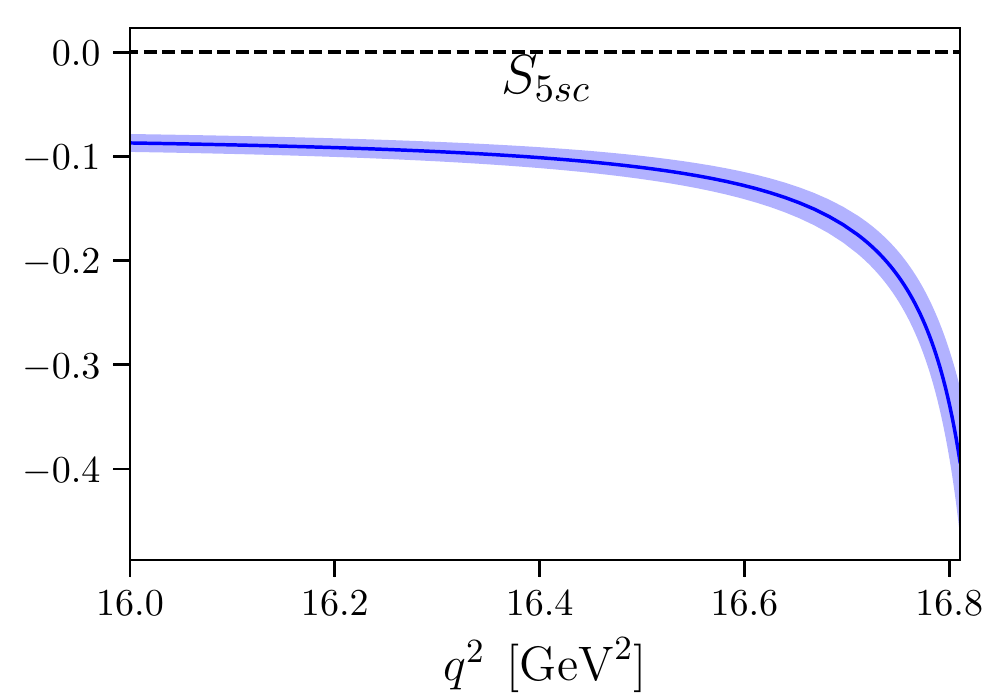} \\
 \caption{\label{fig:angular2}The $\Lambda_b \to \Lambda^*(1520)(\to p K^-)\ell^+\ell^-$ angular observables $S_{3ss}$, $S_{5s}$, and $S_{5sc}$ in the high-$q^2$ region calculated in the Standard Model using our form factor results.}
\end{figure}

\begin{figure}
 \includegraphics[width=0.47\linewidth]{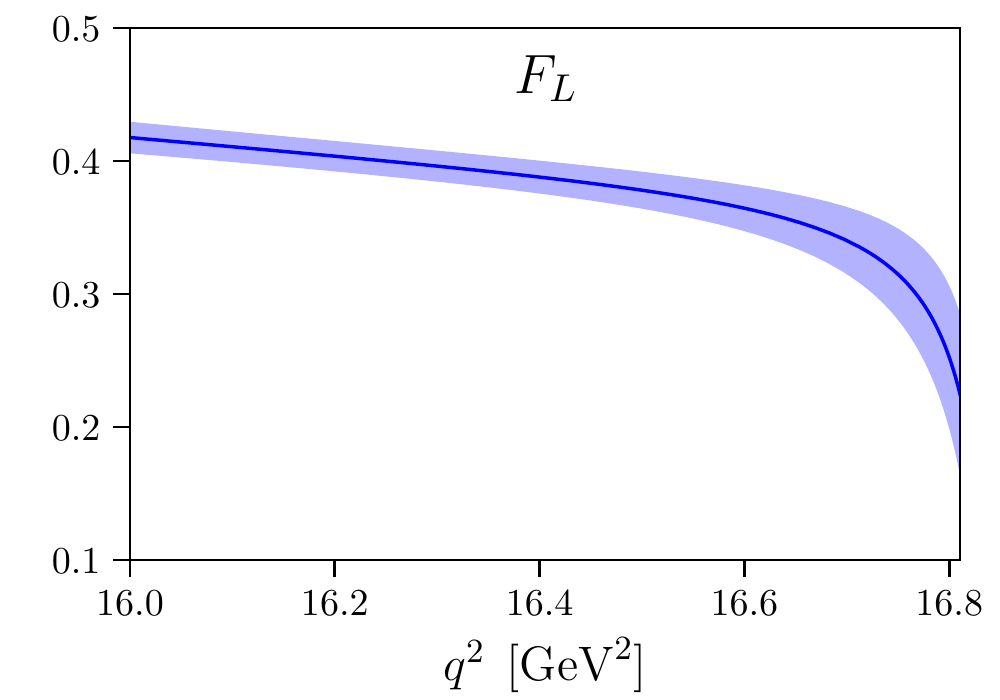} \hfill \includegraphics[width=0.47\linewidth]{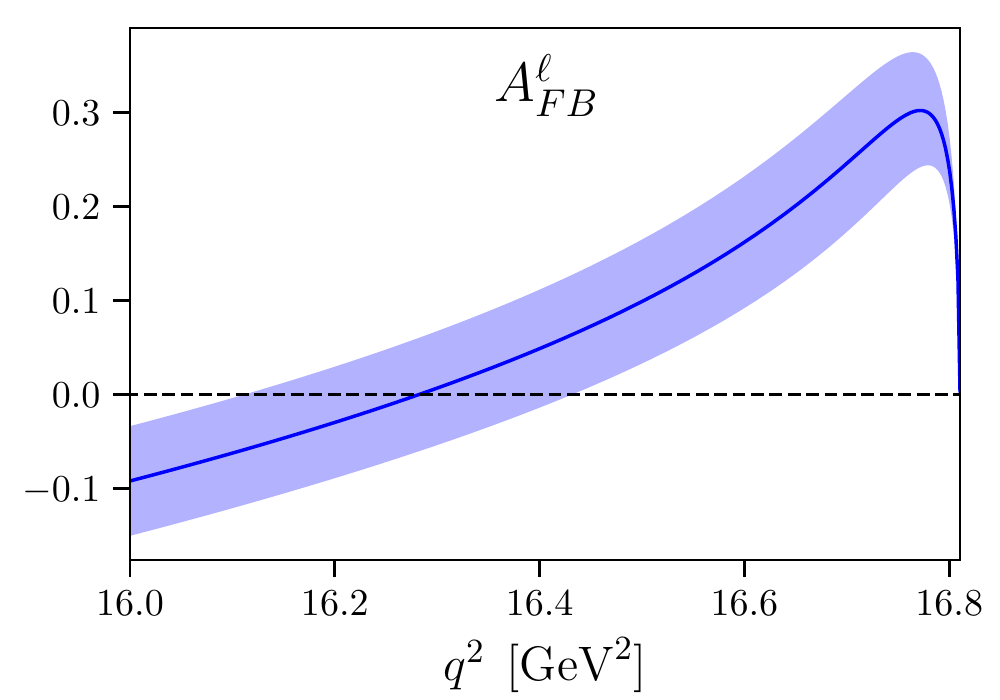}
 \caption{\label{fig:angular3}The $\Lambda_b \to \Lambda^*(1520)(\to p K^-)\ell^+\ell^-$ fraction of longitudinally polarized dileptons and the lepton-side forward-backward asymmetry in the high-$q^2$ region calculated in the Standard Model using our form factor results.}
\end{figure}

In the narrow-width approximation for the $\Lambda^*(1520)$ and for $m_\ell=0$, the $\Lambda_b \to \Lambda^*(1520)(\to p K^-)\ell^+\ell^-$ four-fold differential decay distribution in the Standard Model has the form 
\begin{eqnarray}
\nonumber \frac{\mathrm{d}^4\Gamma}{\mathrm{d}q^2\: \mathrm{d}\cos{\theta_\ell}\: \mathrm{d}\cos{\theta_{\Lambda^*}}\:\mathrm{d}\phi} &=& \frac{3}{8\pi}\Big[\cos^2\theta_{\Lambda^*} \left(L_{1c} \cos \theta_\ell+L_{1cc} \cos^2\theta_\ell+L_{1ss} \sin^2\theta_\ell\right)\\
\nonumber  &&\hspace{3ex}+ \sin^2\theta_{\Lambda^*} \left(L_{2c} \cos
   \theta_\ell+L_{2cc} \cos^2\theta_\ell+L_{2ss} \sin^2\theta_\ell\right)\\
\nonumber  &&\hspace{3ex}+ \sin^2\theta_{\Lambda^*} \left(L_{3ss} \sin^2\theta_\ell \cos^2
   \phi+L_{4ss} \sin^2\theta_\ell \sin \phi \cos
   \phi\right)\\
\nonumber    &&\hspace{3ex}+\sin \theta_{\Lambda^*} \cos \theta_{\Lambda^*} \cos \phi\: (L_{5s} \sin \theta_\ell+L_{5sc} \sin \theta_\ell \cos \theta_\ell)\\
    &&\hspace{3ex}+\sin \theta_{\Lambda^*} \cos \theta_{\Lambda^*}\sin \phi\: (L_{6s} \sin
   \theta_\ell+L_{6sc} \sin \theta_\ell \cos \theta_\ell) \Big], \label{eq:d4G}
\end{eqnarray}
where the angular coefficients $L_i$ are functions of $q^2$ only \cite{Descotes-Genon:2019dbw}.
The expressions for the $L_i$ in terms of form factors are given in Ref.~\cite{Descotes-Genon:2019dbw}, using a slightly different definition of the form factors that is related to ours as shown in Appendix \ref{sec:DNFFs}. In the following, we use the convention that we do not include the factor of $\mathcal{B}_{\Lambda^*}=\mathcal{B}(\Lambda^* \to p K^-)$ in the angular coefficients $L_i$, which means that the integral of Eq.~(\ref{eq:d4G}) over $\cos{\theta_\ell}$, $\cos{\theta_{\Lambda^*}}$, and $\phi$ is equal to $d\Gamma/dq^2$ for the primary decay $\Lambda_b \to \Lambda^*(1520)\ell^+\ell^-$.
We consider the CP-averaged, normalized angular observables \cite{Descotes-Genon:2019dbw}
\begin{equation}
 S_i = \frac{L_i + \overline{L}_i}{\mathrm{d}(\Gamma+\overline{\Gamma})/\mathrm{d}q^2}.
\end{equation}
Our predictions for  $S_{1c}$, $S_{1cc}$, $S_{1ss}$, $S_{2c}$, $S_{2cc}$, $S_{2ss}$, $S_{3ss}$, $S_{5s}$, and $S_{5sc}$ are shown in Figs.~\ref{fig:angular1} and \ref{fig:angular2}. Two further combinations of interest are the fraction of longitudinally polarized dileptons
\begin{equation}
F_L = 1-\frac{2 (L_{1cc}+2\, L_{2cc})}{3\, \mathrm{d}\Gamma/\mathrm{d}q^2}
\end{equation}
and the lepton-side forward-backward asymmetry
\begin{equation}
A_{FB}^\ell = \frac{L_{1c}+2\,L_{2c}}{2\, \mathrm{d}\Gamma/\mathrm{d}q^2};
\end{equation}
these are shown in Fig.~\ref{fig:angular3}. In the kinematic region considered here, our results for all angular observables are qualitatively similar to those predicted using quark-model form factors \cite{Mott:2011cx}, shown in Refs.~\cite{Descotes-Genon:2019dbw} and \cite{Das:2020cpv}, but there are substantial numerical differences. For example, the zero crossing in the forward-backward asymmetry is more than twice as far away from $q^2_{\rm max}$ as predicted by the quark model.

\FloatBarrier
\section{Conclusions}
\label{sec:conclusions}
\FloatBarrier

We have presented the first lattice-QCD calculation of the form factors describing the $\Lambda_b \to \Lambda^*(1520)$ matrix elements of the vector, axial vector, and tensor $b\to s$ currents. Similarly to the lattice calculation of $B\to K^*(892)$ form factors in Ref.~\cite{Horgan:2013hoa}, this exploratory study treats the $\Lambda^*(1520)$ as a stable particle. Even in this approximation, our work required overcoming several challenges. The simplest choices of three-quark interpolating fields with $I=0$ and $J^P=\frac32^-$ dominantly couple to higher-lying states; a previous lattice-QCD study of $\Lambda$-baryon spectroscopy \cite{Engel:2012qp} in fact was unable to identify the $\Lambda^*(1520)$ for this reason. Here we solved this problem by including gauge-covariant spatial derivatives in the interpolating field, at the expense of having to compute additional propagators with derivative sources. We also used all-mode averaging \cite{Blum:2012uh,Shintani:2014vja} to overcome the poor signal-to-noise ratios in the correlation functions involving the $\Lambda^*(1520)$. Traditionally, lattice-QCD calculations of heavy-to-light form factors have been performed in the rest frame of the heavy hadron, giving the final-state light hadron nonzero momentum. However, at nonzero momentum an interpolating field that would have $J^P=\frac32^-$ in the continuum then also couples to $J^P=\frac32^+$, and in some cases even $J^P=\frac12^+$, which would make isolating the $\Lambda^*(1520)$ extremely difficult. For this reason, we performed the lattice calculation in the $\Lambda^*(1520)$ rest frame, giving nonzero momentum to the $\Lambda_b$ instead. While this choice eliminates the problem of mixing with unwanted lighter states, it also limits the accessible $q^2$ range to be very close to $q^2_{\rm max}$. We performed the calculation for two different $\Lambda_b$ momenta, $|\mathbf{p}|\approx 0.935$ GeV and $|\mathbf{p}|\approx 1.402$ GeV,  corresponding to $q^2/q^2_{\rm max}\approx 0.986$ and $q^2/q^2_{\rm max}\approx 0.969$, respectively. This only allowed linear fits of the $q^2$-dependence (or, equivalently, $w$-dependence), which yield the values of the form factors at $q^2_{\rm max}$ and their slopes. Using three different ensembles of gauge fields on lattices that all have approximately the same spatial volume, we performed extrapolations linear in $a^2$ and $m_\pi^2$, with independent coefficients for the slopes and intersects of the form factors, to the physical limit.

Looking ahead, lower values of $q^2$ could be reached using the moving-NRQCD action \cite{Horgan:2009ti} for the $b$ quark, which enables much higher $\Lambda_b$ momenta while keeping discretization errors under control, but requires a more complicated matching of the currents to continuum QCD. Furthermore, a more rigorous analysis of $\Lambda_b \to \Lambda^*(1520)$ form factors that treats the $\Lambda^*(1520)$ as a resonance in coupled-channel $p$-$K$, $\Sigma$-$\pi$ scattering may be possible using the finite-volume formalism of Refs.~\cite{Briceno:2014uqa,Briceno:2015csa}, but this would still not include $\Lambda$-$\pi$-$\pi$ three-particle contributions. 

Using our form factor results, we have obtained Standard-Model predictions for the $\Lambda_b \to \Lambda^*(1520)\ell^+\ell^-$ differential branching fraction and several  $\Lambda_b \to \Lambda^*(1520)(\to p K^-) \ell^+\ell^-$ angular observables at high $q^2$. The uncertainty in the differential branching fraction in the region considered is approximately 20 percent, while some angular observables are more precise due to their reduced dependence on the form factors and benefits from correlations. We find $d\mathcal{B}/dq^2$ to be lower by a factor of 2 than predicted using the quark-model form factors of Ref.~\cite{Mott:2011cx}. Our results for the angular observables are qualitatively similar to those computed using the quark-model form factors \cite{Descotes-Genon:2019dbw}, but show significant quantitative differences. We look forward to future experimental results for $\Lambda_b \to \Lambda^*(1520)\ell^+\ell^-$.

\section*{Acknowledgments}

We thank Marzia Bordone, Danny van Dyk, and Sébastien Descotes-Genon for discussions, and the RBC and UKQCD Collaborations for making their gauge field ensembles available. SM is supported by the U.S. Department of Energy, Office of Science, Office of High Energy Physics under Award Number D{E-S}{C0}009913. GR is supported by the U.S. Department of Energy, Office of Science, Office of Nuclear Physics, under Contract No.~D{E-S}C0012704 (BNL). The computations for this work were carried out on facilities at the National Energy Research Scientific Computing Center, a DOE Office of Science User Facility supported by the Office of Science of the U.S. Department of Energy under Contract No. DE-AC02-05CH1123, and on facilities of the Extreme Science and Engineering Discovery Environment (XSEDE) \cite{XSEDE}, which is supported by National Science Foundation grant number ACI-1548562. We acknowledge the use of Chroma \cite{Edwards:2004sx,Chroma}, QPhiX \cite{JOO2015139,QPhiX}, QLUA \cite{QLUA}, MDWF \cite{MDWF}, and related USQCD software \cite{USQCD}.

\appendix

\FloatBarrier
\section{Relations between different form factor definitions}
\label{sec:FFrelations}
\FloatBarrier

In this appendix we provide the relations between two other definitions of $\Lambda_b \to \Lambda^*(1520)$ form factors used in the literature and our definition.

\subsection{Non-helicity-based definition}

This definition is used in Refs.~\cite{Pervin:2005ve,Mott:2011cx}. For the vector and axial vector currents, it has the same structure as the definition of $\Lambda_b\to\Lambda_c^*(2625)$ form factors in Ref.~\cite{Leibovich:1997az}. In the notation of our Eq.~(\ref{eq:Glambda}), it is given by
\begin{eqnarray}
\mathscr{G}^{\lambda}[\gamma^\mu] &=& v^\lambda\left( F_1\,\gamma^\mu+F_2\, v^\mu+ F_3\, v^{\prime\mu} \right) + F_4\, g^{\lambda\mu}, \\
\mathscr{G}^{\lambda}[\gamma^\mu\gamma_5] &=& v^\lambda \left(  G_1\,\gamma^\mu+G_2\, v^\mu+ G_3\, v^{\prime\mu} \right)\gamma_5 + G_4\, g^{\lambda\mu}\gamma_5, \\
\mathscr{G}^{\lambda}[i\sigma^{\mu\nu}q_\nu] &=& v^\lambda\left(  F_1^{T}\,\gamma^\mu+F_2^{T}\, v^\mu+ F_3^{T}\, v^{\prime\mu} \right) + F_4^{T}\, g^{\lambda\mu}, \\
\mathscr{G}^{\lambda}[i\sigma^{\mu\nu} q_\nu \gamma_5] &=& v^\lambda \left(  G_1^{T}\,\gamma^\mu+G_2^{T}\, v^\mu+ G_3^{T}\, v^{\prime\mu} \right)\gamma_5 + G_4^{T}\, g^{\lambda\mu}\gamma_5.
\end{eqnarray}
Note that only six of the eight tensor form factors in this definition are independent. The relation to our definition is
\begin{eqnarray}
 F_1&=&  \frac{m_{\Lambda_b}m_{\Lambda^*}}{s_-} (f_\perp+f_{\perp^\prime}), \\
 F_2&=& \frac{m_{\Lambda_b}^2 m_{\Lambda^*}}{q^2 s_+ s_-}  \big[ (m_{\Lambda_b}-m_{\Lambda^*}) s_- f_0  - 2 m_{\Lambda^*} q^2 (f_\perp- f_{\perp^\prime})  -  (m_{\Lambda_b}+m_{\Lambda^*}) \left(m_{\Lambda_b}^2-m_{\Lambda^*}^2-q^2\right)f_+\big], \\
 \nonumber F_3&=&  \frac{m_{\Lambda_b}m_{\Lambda^*}}{q^2
   s_+ s_-}   \big[ -  m_{\Lambda^*} (m_{\Lambda_b}-m_{\Lambda^*}) s_- f_0-2  m_{\Lambda_b} m_{\Lambda^*}
   q^2 f_\perp + 2 q^2 (m_{\Lambda_b} m_{\Lambda^*}-s_+) f_{\perp^\prime} \\
   && \hspace{10.8ex} +\,  m_{\Lambda^*} (m_{\Lambda_b}+m_{\Lambda^*}) \left(m_{\Lambda_b}^2-m_{\Lambda^*}^2+q^2\right)f_+\big], \\
 F_4&=& f_{\perp^\prime},
\end{eqnarray}
\begin{eqnarray}
 G_1&=&  \frac{ m_{\Lambda_b}m_{\Lambda^*}}{s_+} (g_\perp+g_{\perp^\prime}), \\
 G_2&=& \frac{m_{\Lambda_b}^2 m_{\Lambda^*}}{q^2 s_+ s_-} \big[-(m_{\Lambda_b}+m_{\Lambda^*})s_+g_0 - 2 m_{\Lambda^*} q^2 (g_\perp-g_{\perp^\prime})+
   (m_{\Lambda_b}-m_{\Lambda^*}) \left(m_{\Lambda_b}^2-m_{\Lambda^*}^2-q^2\right)g_+\big], \\
\nonumber G_3&=& \frac{m_{\Lambda_b}m_{\Lambda^*}}{q^2 s_+ s_-} \big[ m_{\Lambda^*}  (m_{\Lambda_b}+m_{\Lambda^*})s_+g_0  +  2  m_{\Lambda_b} m_{\Lambda^*}
   q^2g_\perp - 2 q^2 (m_{\Lambda_b} m_{\Lambda^*}+s_-)g_{\perp^\prime} \\
   && \hspace{10.8ex} -\,  m_{\Lambda^*} (m_{\Lambda_b}-m_{\Lambda^*})\left(m_{\Lambda_b}^2-m_{\Lambda^*}^2+q^2\right)g_+ \big], \\
 G_4&=& g_{\perp^\prime}, 
\end{eqnarray}
\begin{eqnarray}
 F_1^{T}&=& - \frac{m_{\Lambda_b}m_{\Lambda^*}  (m_{\Lambda_b}+m_{\Lambda^*}) }{s_-}(h_\perp+h_{\perp^\prime}), \\
 F_2^{T}&=&  \frac{m_{\Lambda_b}^2 m_{\Lambda^*}}{s_+ s_-} \big[ 2  m_{\Lambda^*}(m_{\Lambda_b}+m_{\Lambda^*})(h_\perp- h_{\perp^\prime})  +
   (m_{\Lambda_b}^2- m_{\Lambda^*}^2-q^2)h_+ \big], \\
 F_3^{T}&=& \frac{ m_{\Lambda_b} m_{\Lambda^*}}{s_+ s_-} \big[2 (m_{\Lambda_b}+m_{\Lambda^*}) (m_{\Lambda_b} m_{\Lambda^*} h_\perp-(m_{\Lambda_b} m_{\Lambda^*}-s_+) h_{\perp^\prime}) -  m_{\Lambda^*} \left(m_{\Lambda_b}^2-m_{\Lambda^*}^2+q^2\right)h_+\big], \\
 F_4^{T}&=& - (m_{\Lambda_b}+m_{\Lambda^*}) h_{\perp^\prime},
\end{eqnarray}

\begin{eqnarray}
 G_1^{T}&=&  \frac{m_{\Lambda_b} m_{\Lambda^*}  (m_{\Lambda_b}-m_{\Lambda^*})}{s_+} (\tilde{h}_{\perp}+\tilde{h}_{\perp^\prime}), \\
 G_2^{T}&=&  \frac{m_{\Lambda_b}^2 m_{\Lambda^*}}{s_+ s_-} \big[ -2  m_{\Lambda^*}(m_{\Lambda_b}-m_{\Lambda^*})(\tilde{h}_{\perp}-\tilde{h}_{\perp^\prime})+
   (m_{\Lambda_b}^2-m_{\Lambda^*}^2-q^2)\tilde{h}_{+}\big], \\
 G_3^{T}&=& \frac{ m_{\Lambda_b} m_{\Lambda^*}}{s_+ s_-} \big[2 (m_{\Lambda_b}-m_{\Lambda^*}) ( m_{\Lambda_b} m_{\Lambda^*}\tilde{h}_{\perp}- (m_{\Lambda^*}
   m_{\Lambda_b}+s_-)\tilde{h}_{\perp^\prime})- m_{\Lambda^*} \left(m_{\Lambda_b}^2-m_{\Lambda^*}^2+q^2\right)\tilde{h}_{+}\big], \\
 G_4^{T}&=&  (m_{\Lambda_b}-m_{\Lambda^*}) \tilde{h}_{\perp^\prime}.
\end{eqnarray}

\subsection{Helicity-based definition used by Descotes-Genon and Novoa Brunet}
\label{sec:DNFFs}

Reference \cite{Descotes-Genon:2019dbw} uses a helicity-based definition that differs from ours only by simple kinematic factors:
\begin{eqnarray}
 f_t^V &=& \frac{m_{\Lambda^*}}{s_+} f_0 , \\
 f_0^V &=& \frac{m_{\Lambda^*}}{ s_-} f_+, \\
 f_\perp^V &=& \frac{m_{\Lambda^*}}{ s_-} f_\perp, \\
 f_g^V &=& f_{\perp^\prime}, \\
 f_t^A &=& \frac{m_{\Lambda^*}}{s_-} g_0, \\
 f_0^A &=& \frac{m_{\Lambda^*}}{ s_+} g_+, \\
 f_\perp^A &=& \frac{m_{\Lambda^*}}{ s_+} g_\perp, \\
 f_g^A &=& -g_{\perp^\prime}, \\ 
 f_0^T &=& \frac{m_{\Lambda^*}}{ s_-} h_+ , \\
 f_\perp^T &=& \frac{m_{\Lambda^*}}{ s_-} h_\perp, \\
 f_g^T &=& (m_{\Lambda_b}+m_{\Lambda^*}) h_{\perp^\prime}, \\
 f_0^{T5} &=& \frac{m_{\Lambda^*}}{ s_+} \tilde{h}_+ , \\
 f_\perp^{T5} &=& \frac{m_{\Lambda^*}}{ s_+} \tilde{h}_\perp, \\
 f_g^{T5} &=& -(m_{\Lambda_b}-m_{\Lambda^*}) \tilde{h}_{\perp^\prime}.
\end{eqnarray}
Similarly, Ref.~\cite{Boer:2018vpx}, which considers $\Lambda_b\to\Lambda_c^*$, contains another helicity-based definition (for the vector and axial-vector form factors only) that also differs from ours only by simple kinematic factors.

\FloatBarrier

\providecommand{\href}[2]{#2}\begingroup\raggedright\endgroup


\begin{thebibliography}{10}

\bibitem{Blake:2016olu}
T.~Blake, G.~Lanfranchi, and D.~M. Straub, ``{Rare $B$ Decays as Tests of the
  Standard Model},'' \href{http://dx.doi.org/10.1016/j.ppnp.2016.10.001}{Prog.
  Part. Nucl. Phys. {\bfseries 92} (2017) 50--91},
\href{http://arxiv.org/abs/1606.00916}{{\ttfamily arXiv:1606.00916 [hep-ph]}}.
%%CITATION = ARXIV:1606.00916;%%.

\bibitem{Alguero:2019ptt}
M.~Algueró, B.~Capdevila, A.~Crivellin, S.~Descotes-Genon, P.~Masjuan,
  J.~Matias, and J.~Virto, ``{Emerging patterns of New Physics with and without
  Lepton Flavour Universal contributions},''
  \href{http://dx.doi.org/10.1140/epjc/s10052-019-7216-3}{Eur. Phys. J.
  {\bfseries C79} no.~8, (2019) 714},
\href{http://arxiv.org/abs/1903.09578}{{\ttfamily arXiv:1903.09578 [hep-ph]}}.
%%CITATION = ARXIV:1903.09578;%%.

\bibitem{Aebischer:2019mlg}
J.~Aebischer, W.~Altmannshofer, D.~Guadagnoli, M.~Reboud, P.~Stangl, and D.~M.
  Straub, ``{$B$-decay discrepancies after Moriond 2019},''
  \href{http://dx.doi.org/10.1140/epjc/s10052-020-7817-x}{Eur. Phys. J. C
  {\bfseries 80} no.~3, (2020) 252},
  \href{http://arxiv.org/abs/1903.10434}{{\ttfamily arXiv:1903.10434
  [hep-ph]}}.

\bibitem{Buttazzo:2017ixm}
D.~Buttazzo, A.~Greljo, G.~Isidori, and D.~Marzocca, ``{$B$-physics anomalies:
  a guide to combined explanations},''
  \href{http://dx.doi.org/10.1007/JHEP11(2017)044}{JHEP {\bfseries 11} (2017)
  044},
\href{http://arxiv.org/abs/1706.07808}{{\ttfamily arXiv:1706.07808 [hep-ph]}}.
%%CITATION = ARXIV:1706.07808;%%.

\bibitem{Gremm:1995nx}
M.~Gremm, F.~Kruger, and L.~M. Sehgal, ``{Angular distribution and polarization
  of photons in the inclusive decay $\Lambda_b \to X_s \gamma$},''
  \href{http://dx.doi.org/10.1016/0370-2693(95)00722-W}{Phys. Lett. {\bfseries
  B355} (1995) 579--583},
\href{http://arxiv.org/abs/hep-ph/9505354}{{\ttfamily arXiv:hep-ph/9505354
  [hep-ph]}}.
%%CITATION = HEP-PH/9505354;%%.

\bibitem{Mannel:1997xy}
T.~Mannel and S.~Recksiegel, ``{Flavor changing neutral current decays of heavy
  baryons: The Case $\Lambda_b \to \Lambda \gamma$},''
  \href{http://dx.doi.org/10.1088/0954-3899/24/5/006}{J. Phys. {\bfseries G24}
  (1998) 979--990},
\href{http://arxiv.org/abs/hep-ph/9701399}{{\ttfamily arXiv:hep-ph/9701399
  [hep-ph]}}.
%%CITATION = HEP-PH/9701399;%%.

\bibitem{Huang:1998ek}
C.-S. Huang and H.-G. Yan, ``{Exclusive rare decays of heavy baryons to light
  baryons: $\Lambda_b \to \Lambda \gamma$ and $\Lambda_b \to \Lambda \ell^+
  \ell^-$},'' \href{http://dx.doi.org/10.1103/PhysRevD.59.114022,
  10.1103/PhysRevD.61.039901}{Phys. Rev. {\bfseries D59} (1999) 114022},
  \href{http://arxiv.org/abs/hep-ph/9811303}{{\ttfamily arXiv:hep-ph/9811303
  [hep-ph]}}.
[Erratum: Phys. Rev.D61,039901(2000)].
%%CITATION = HEP-PH/9811303;%%.

\bibitem{Hiller:2001zj}
G.~Hiller and A.~Kagan, ``{Probing for new physics in polarized $\Lambda_b$
  decays at the $Z$},''
  \href{http://dx.doi.org/10.1103/PhysRevD.65.074038}{Phys. Rev. {\bfseries
  D65} (2002) 074038},
\href{http://arxiv.org/abs/hep-ph/0108074}{{\ttfamily arXiv:hep-ph/0108074
  [hep-ph]}}.
%%CITATION = HEP-PH/0108074;%%.

\bibitem{Chen:2002rg}
C.-H. Chen, C.~Q. Geng, and J.~N. Ng, ``{T violation in $\Lambda_b \to \Lambda
  \ell^+ \ell^-$ decays with polarized $\Lambda$},''
  \href{http://dx.doi.org/10.1103/PhysRevD.65.091502}{Phys. Rev. {\bfseries
  D65} (2002) 091502},
\href{http://arxiv.org/abs/hep-ph/0202103}{{\ttfamily arXiv:hep-ph/0202103
  [hep-ph]}}.
%%CITATION = HEP-PH/0202103;%%.

\bibitem{Legger:2006cq}
F.~Legger and T.~Schietinger, ``{Photon helicity in $\Lambda_b \to p K \gamma$
  decays},'' \href{http://dx.doi.org/10.1016/j.physletb.2006.12.011,
  10.1016/j.physletb.2007.02.044}{Phys. Lett. {\bfseries B645} (2007)
  204--212}, \href{http://arxiv.org/abs/hep-ph/0605245}{{\ttfamily
  arXiv:hep-ph/0605245 [hep-ph]}}.
[Erratum: Phys. Lett.B647,527(2007)].
%%CITATION = HEP-PH/0605245;%%.

\bibitem{Hiller:2007ur}
G.~Hiller, M.~Knecht, F.~Legger, and T.~Schietinger, ``{Photon polarization
  from helicity suppression in radiative decays of polarized $\Lambda_b$ to
  spin-3/2 baryons},''
  \href{http://dx.doi.org/10.1016/j.physletb.2007.03.056}{Phys. Lett.
  {\bfseries B649} (2007) 152--158},
\href{http://arxiv.org/abs/hep-ph/0702191}{{\ttfamily arXiv:hep-ph/0702191
  [hep-ph]}}.
%%CITATION = HEP-PH/0702191;%%.

\bibitem{Boer:2014kda}
P.~Böer, T.~Feldmann, and D.~van Dyk, ``{Angular Analysis of the Decay
  $\Lambda_b \to \Lambda (\to N \pi) \ell^+\ell^-$},''
  \href{http://dx.doi.org/10.1007/JHEP01(2015)155}{JHEP {\bfseries 01} (2015)
  155},
\href{http://arxiv.org/abs/1410.2115}{{\ttfamily arXiv:1410.2115 [hep-ph]}}.
%%CITATION = ARXIV:1410.2115;%%.

\bibitem{Meinel:2016grj}
S.~Meinel and D.~van Dyk, ``{Using $\Lambda_b\to \Lambda\mu^+\mu^-$ data within
  a Bayesian analysis of $|\Delta B| = |\Delta S| = 1$ decays},''
  \href{http://dx.doi.org/10.1103/PhysRevD.94.013007}{Phys. Rev. {\bfseries
  D94} no.~1, (2016) 013007},
\href{http://arxiv.org/abs/1603.02974}{{\ttfamily arXiv:1603.02974 [hep-ph]}}.
%%CITATION = ARXIV:1603.02974;%%.

\bibitem{Blake:2017une}
T.~Blake and M.~Kreps, ``{Angular distribution of polarised $\Lambda_b$ baryons
  decaying to $\Lambda \ell^+\ell^-$},''
  \href{http://dx.doi.org/10.1007/JHEP11(2017)138}{JHEP {\bfseries 11} (2017)
  138},
\href{http://arxiv.org/abs/1710.00746}{{\ttfamily arXiv:1710.00746 [hep-ph]}}.
%%CITATION = ARXIV:1710.00746;%%.

\bibitem{Das:2018sms}
D.~Das, ``{Model independent New Physics analysis in
  $\Lambda_b\to\Lambda\mu^+\mu^-$ decay},''
  \href{http://dx.doi.org/10.1140/epjc/s10052-018-5731-2}{Eur. Phys. J.
  {\bfseries C78} no.~3, (2018) 230},
\href{http://arxiv.org/abs/1802.09404}{{\ttfamily arXiv:1802.09404 [hep-ph]}}.
%%CITATION = ARXIV:1802.09404;%%.

\bibitem{Yan:2019tgn}
H.~Yan, ``{Angular distribution of the rare decay $\Lambda_b \to \Lambda(\to N
  \pi) \ell^+\ell^-$},''
\href{http://arxiv.org/abs/1911.11568}{{\ttfamily arXiv:1911.11568 [hep-ph]}}.
%%CITATION = ARXIV:1911.11568;%%.

\bibitem{Descotes-Genon:2019dbw}
S.~Descotes-Genon and M.~Novoa~Brunet, ``{Angular analysis of the rare decay
  $\Lambda_b\to \Lambda(1520)(\to NK)\ell^+\ell^-$},''
  \href{http://dx.doi.org/10.1007/JHEP06(2019)136}{JHEP {\bfseries 06} (2019)
  136},
\href{http://arxiv.org/abs/1903.00448}{{\ttfamily arXiv:1903.00448 [hep-ph]}}.
%%CITATION = ARXIV:1903.00448;%%.

\bibitem{Blake:2019guk}
T.~Blake, S.~Meinel, and D.~van Dyk, ``{Bayesian Analysis of $b\to s\mu^+\mu^-$
  Wilson Coefficients using the Full Angular Distribution of $\Lambda_b\to
  \Lambda(\to p\, \pi^-)\mu^+\mu^-$ Decays},''
  \href{http://dx.doi.org/10.1103/PhysRevD.101.035023}{Phys. Rev. D {\bfseries
  101} no.~3, (2020) 035023}, \href{http://arxiv.org/abs/1912.05811}{{\ttfamily
  arXiv:1912.05811 [hep-ph]}}.

\bibitem{Das:2020cpv}
D.~Das and J.~Das, ``{The $\Lambda_b\to\Lambda^\ast(1520)(\to
  N\!\bar{K})\ell^+\ell^-$ decay at low-recoil in HQET},''
  \href{http://dx.doi.org/10.1007/JHEP07(2020)002}{JHEP {\bfseries 07} (2020)
  002}, \href{http://arxiv.org/abs/2003.08366}{{\ttfamily arXiv:2003.08366
  [hep-ph]}}.

\bibitem{Aaltonen:2011qs}
{\bfseries CDF} Collaboration, T.~Aaltonen {\em et~al.}, ``{Observation of the
  Baryonic Flavor-Changing Neutral Current Decay $\Lambda_{b} \to \Lambda
  \mu^{+} \mu^{-}$},''
  \href{http://dx.doi.org/10.1103/PhysRevLett.107.201802}{Phys. Rev. Lett.
  {\bfseries 107} (2011) 201802},
\href{http://arxiv.org/abs/1107.3753}{{\ttfamily arXiv:1107.3753 [hep-ex]}}.
%%CITATION = ARXIV:1107.3753;%%.

\bibitem{Aaij:2013mna}
{\bfseries LHCb} Collaboration, R.~Aaij {\em et~al.}, ``{Measurement of the
  differential branching fraction of the decay
  $\Lambda_b^0\rightarrow\Lambda\mu^+\mu^-$},''
  \href{http://dx.doi.org/10.1016/j.physletb.2013.06.060}{Phys. Lett.
  {\bfseries B725} (2013) 25--35},
\href{http://arxiv.org/abs/1306.2577}{{\ttfamily arXiv:1306.2577 [hep-ex]}}.
%%CITATION = ARXIV:1306.2577;%%.

\bibitem{Aaij:2015xza}
{\bfseries LHCb} Collaboration, R.~Aaij {\em et~al.}, ``{Differential branching
  fraction and angular analysis of $\Lambda^{0}_{b} \rightarrow \Lambda
  \mu^+\mu^-$ decays},'' \href{http://dx.doi.org/10.1007/JHEP09(2018)145,
  10.1007/JHEP06(2015)115}{JHEP {\bfseries 06} (2015) 115},
  \href{http://arxiv.org/abs/1503.07138}{{\ttfamily arXiv:1503.07138
  [hep-ex]}}.
[Erratum: JHEP09,145(2018)].
%%CITATION = ARXIV:1503.07138;%%.

\bibitem{Aaij:2018gwm}
{\bfseries LHCb} Collaboration, R.~Aaij {\em et~al.}, ``{Angular moments of the
  decay $\Lambda_b^0 \rightarrow \Lambda \mu^{+} \mu^{-}$ at low hadronic
  recoil},'' \href{http://dx.doi.org/10.1007/JHEP09(2018)146}{JHEP {\bfseries
  09} (2018) 146},
\href{http://arxiv.org/abs/1808.00264}{{\ttfamily arXiv:1808.00264 [hep-ex]}}.
%%CITATION = ARXIV:1808.00264;%%.

\bibitem{Aaij:2019hhx}
{\bfseries LHCb} Collaboration, R.~Aaij {\em et~al.}, ``{First Observation of
  the Radiative Decay $\Lambda_{b}^{0} \to \Lambda \gamma$},''
  \href{http://dx.doi.org/10.1103/PhysRevLett.123.031801}{Phys. Rev. Lett.
  {\bfseries 123} no.~3, (2019) 031801},
\href{http://arxiv.org/abs/1904.06697}{{\ttfamily arXiv:1904.06697 [hep-ex]}}.
%%CITATION = ARXIV:1904.06697;%%.

\bibitem{Detmold:2016pkz}
W.~Detmold and S.~Meinel, ``{$\Lambda_b \to \Lambda \ell^+ \ell^-$ form
  factors, differential branching fraction, and angular observables from
  lattice QCD with relativistic $b$ quarks},''
  \href{http://dx.doi.org/10.1103/PhysRevD.93.074501}{Phys. Rev. {\bfseries
  D93} no.~7, (2016) 074501},
\href{http://arxiv.org/abs/1602.01399}{{\ttfamily arXiv:1602.01399 [hep-lat]}}.
%%CITATION = ARXIV:1602.01399;%%.

\bibitem{Aaij:2017mib}
{\bfseries LHCb} Collaboration, R.~Aaij {\em et~al.}, ``{Observation of the
  decay $\Lambda^0_b \to p K^- \mu^+ \mu^-$ and a search for $C\!P$
  violation},'' \href{http://dx.doi.org/10.1007/JHEP06(2017)108}{JHEP
  {\bfseries 06} (2017) 108},
\href{http://arxiv.org/abs/1703.00256}{{\ttfamily arXiv:1703.00256 [hep-ex]}}.
%%CITATION = ARXIV:1703.00256;%%.

\bibitem{Aaij:2019bzx}
{\bfseries LHCb} Collaboration, R.~Aaij {\em et~al.}, ``{Test of lepton
  universality with $ {\Lambda}_b^0\to
  {pK}^{-}{\mathrm{\ell}}^{+}{\mathrm{\ell}}^{-} $ decays},''
  \href{http://dx.doi.org/10.1007/JHEP05(2020)040}{JHEP {\bfseries 05} (2020)
  040}, \href{http://arxiv.org/abs/1912.08139}{{\ttfamily arXiv:1912.08139
  [hep-ex]}}.

\bibitem{Aaij:2015tga}
{\bfseries LHCb} Collaboration, R.~Aaij {\em et~al.}, ``{Observation of $J/\psi
  p$ Resonances Consistent with Pentaquark States in $\Lambda_b^0 \to J/\psi
  K^- p$ Decays},''
  \href{http://dx.doi.org/10.1103/PhysRevLett.115.072001}{Phys. Rev. Lett.
  {\bfseries 115} (2015) 072001},
\href{http://arxiv.org/abs/1507.03414}{{\ttfamily arXiv:1507.03414 [hep-ex]}}.
%%CITATION = ARXIV:1507.03414;%%.

\bibitem{Tanabashi:2018oca}
{\bfseries Particle Data Group} Collaboration, M.~Tanabashi {\em et~al.},
  ``{Review of Particle Physics},''
\href{http://dx.doi.org/10.1103/PhysRevD.98.030001}{Phys. Rev. {\bfseries D98}
  no.~3, (2018) 030001}.
%%CITATION = PHRVA,D98,030001;%%.

\bibitem{Mott:2011cx}
L.~Mott and W.~Roberts, ``{Rare dileptonic decays of $\Lambda_b$ in a quark
  model},'' \href{http://dx.doi.org/10.1142/S0217751X12500169}{Int. J. Mod.
  Phys. {\bfseries A27} (2012) 1250016},
\href{http://arxiv.org/abs/1108.6129}{{\ttfamily arXiv:1108.6129 [nucl-th]}}.
%%CITATION = ARXIV:1108.6129;%%.

\bibitem{Amhis:2020phx}
Y.~Amhis, S.~Descotes-Genon, C.~Marin~Benito, M.~Novoa-Brunet, and M.-H.
  Schune, ``{Prospects for New Physics searches with $\Lambda_b \to
  \Lambda(1520)\ell^+\ell^-$ decays},''
  \href{http://arxiv.org/abs/2005.09602}{{\ttfamily arXiv:2005.09602
  [hep-ph]}}.

\bibitem{Albrecht:2020azd}
J.~Albrecht, Y.~Amhis, A.~Beck, and C.~Marin~Benito, ``{Towards an amplitude
  analysis of the decay $\Lambda_b^0\to pK^-\gamma$},''
  \href{http://dx.doi.org/10.1007/JHEP06(2020)116}{JHEP {\bfseries 06} (2020)
  116}, \href{http://arxiv.org/abs/2002.02692}{{\ttfamily arXiv:2002.02692
  [hep-ph]}}.

\bibitem{Pervin:2005ve}
M.~Pervin, W.~Roberts, and S.~Capstick, ``{Semileptonic decays of heavy
  $\Lambda$ baryons in a quark model},''
  \href{http://dx.doi.org/10.1103/PhysRevC.72.035201}{Phys. Rev. {\bfseries
  C72} (2005) 035201},
\href{http://arxiv.org/abs/nucl-th/0503030}{{\ttfamily arXiv:nucl-th/0503030
  [nucl-th]}}.
%%CITATION = NUCL-TH/0503030;%%.

\bibitem{Meinel:2016cxo}
S.~Meinel and G.~Rendon, ``{Lattice QCD calculation of form factors for
  $\Lambda_b \to \Lambda(1520) \ell^+ \ell^-$ decays},''
  \href{http://dx.doi.org/10.22323/1.256.0299}{PoS {\bfseries LATTICE2016}
  (2016) 299},
\href{http://arxiv.org/abs/1608.08110}{{\ttfamily arXiv:1608.08110 [hep-lat]}}.
%%CITATION = ARXIV:1608.08110;%%.

\bibitem{Leibovich:1997az}
A.~K. Leibovich and I.~W. Stewart, ``{Semileptonic $\Lambda_b$ decay to excited
  $\Lambda_c$ baryons at order $\Lambda_{\rm QCD} / m_Q$},''
  \href{http://dx.doi.org/10.1103/PhysRevD.57.5620}{Phys. Rev. {\bfseries D57}
  (1998) 5620--5631},
\href{http://arxiv.org/abs/hep-ph/9711257}{{\ttfamily arXiv:hep-ph/9711257
  [hep-ph]}}.
%%CITATION = HEP-PH/9711257;%%.

\bibitem{Boer:2018vpx}
P.~Böer, M.~Bordone, E.~Graverini, P.~Owen, M.~Rotondo, and D.~Van~Dyk,
  ``{Testing lepton flavour universality in semileptonic $\Lambda_b \to
  \Lambda_c^*$ decays},'' \href{http://dx.doi.org/10.1007/JHEP06(2018)155}{JHEP
  {\bfseries 06} (2018) 155},
\href{http://arxiv.org/abs/1801.08367}{{\ttfamily arXiv:1801.08367 [hep-ph]}}.
%%CITATION = ARXIV:1801.08367;%%.

\bibitem{Aoki:2010dy}
{\bfseries RBC, UKQCD} Collaboration, Y.~Aoki {\em et~al.}, ``{Continuum Limit
  Physics from 2+1 Flavor Domain Wall QCD},''
  \href{http://dx.doi.org/10.1103/PhysRevD.83.074508}{Phys. Rev. {\bfseries
  D83} (2011) 074508},
\href{http://arxiv.org/abs/1011.0892}{{\ttfamily arXiv:1011.0892 [hep-lat]}}.
%%CITATION = ARXIV:1011.0892;%%.

\bibitem{Blum:2014tka}
{\bfseries RBC, UKQCD} Collaboration, T.~Blum {\em et~al.}, ``{Domain wall QCD
  with physical quark masses},''
  \href{http://dx.doi.org/10.1103/PhysRevD.93.074505}{Phys. Rev. {\bfseries
  D93} no.~7, (2016) 074505},
\href{http://arxiv.org/abs/1411.7017}{{\ttfamily arXiv:1411.7017 [hep-lat]}}.
%%CITATION = ARXIV:1411.7017;%%.

\bibitem{Kaplan:1992bt}
D.~B. Kaplan, ``{A Method for simulating chiral fermions on the lattice},''
  \href{http://dx.doi.org/10.1016/0370-2693(92)91112-M}{Phys. Lett. {\bfseries
  B288} (1992) 342--347},
\href{http://arxiv.org/abs/hep-lat/9206013}{{\ttfamily arXiv:hep-lat/9206013
  [hep-lat]}}.
%%CITATION = HEP-LAT/9206013;%%.

\bibitem{Furman:1994ky}
V.~Furman and Y.~Shamir, ``{Axial symmetries in lattice QCD with Kaplan
  fermions},'' \href{http://dx.doi.org/10.1016/0550-3213(95)00031-M}{Nucl.
  Phys. {\bfseries B439} (1995) 54--78},
\href{http://arxiv.org/abs/hep-lat/9405004}{{\ttfamily arXiv:hep-lat/9405004
  [hep-lat]}}.
%%CITATION = HEP-LAT/9405004;%%.

\bibitem{Shamir:1993zy}
Y.~Shamir, ``{Chiral fermions from lattice boundaries},''
  \href{http://dx.doi.org/10.1016/0550-3213(93)90162-I}{Nucl. Phys. {\bfseries
  B406} (1993) 90--106},
\href{http://arxiv.org/abs/hep-lat/9303005}{{\ttfamily arXiv:hep-lat/9303005
  [hep-lat]}}.
%%CITATION = HEP-LAT/9303005;%%.

\bibitem{Iwasaki:1984cj}
Y.~Iwasaki and T.~Yoshie, ``{Renormalization Group Improved Action for SU(3)
  Lattice Gauge Theory and the String Tension},''
\href{http://dx.doi.org/10.1016/0370-2693(84)91500-4}{Phys. Lett. {\bfseries
  143B} (1984) 449--452}.
%%CITATION = PHLTA,143B,449;%%.

\bibitem{Aoki:2012xaa}
{\bfseries RBC, UKQCD} Collaboration, Y.~Aoki, N.~H. Christ, J.~M. Flynn,
  T.~Izubuchi, C.~Lehner, M.~Li, H.~Peng, A.~Soni, R.~S. Van~de Water, and
  O.~Witzel, ``{Nonperturbative tuning of an improved relativistic heavy-quark
  action with application to bottom spectroscopy},''
  \href{http://dx.doi.org/10.1103/PhysRevD.86.116003}{Phys. Rev. {\bfseries
  D86} (2012) 116003},
\href{http://arxiv.org/abs/1206.2554}{{\ttfamily arXiv:1206.2554 [hep-lat]}}.
%%CITATION = ARXIV:1206.2554;%%.

\bibitem{Blum:2012uh}
T.~Blum, T.~Izubuchi, and E.~Shintani, ``{New class of variance-reduction
  techniques using lattice symmetries},''
  \href{http://dx.doi.org/10.1103/PhysRevD.88.094503}{Phys. Rev. {\bfseries
  D88} no.~9, (2013) 094503},
\href{http://arxiv.org/abs/1208.4349}{{\ttfamily arXiv:1208.4349 [hep-lat]}}.
%%CITATION = ARXIV:1208.4349;%%.

\bibitem{Shintani:2014vja}
E.~Shintani, R.~Arthur, T.~Blum, T.~Izubuchi, C.~Jung, and C.~Lehner,
  ``{Covariant approximation averaging},''
  \href{http://dx.doi.org/10.1103/PhysRevD.91.114511}{Phys. Rev. {\bfseries
  D91} no.~11, (2015) 114511},
\href{http://arxiv.org/abs/1402.0244}{{\ttfamily arXiv:1402.0244 [hep-lat]}}.
%%CITATION = ARXIV:1402.0244;%%.

\bibitem{Johnson:1982yq}
R.~C. Johnson, ``{Angular momentum on a lattice},''
\href{http://dx.doi.org/10.1016/0370-2693(82)90134-4}{Phys. Lett. {\bfseries
  114B} (1982) 147--151}.
%%CITATION = PHLTA,114B,147;%%.

\bibitem{Gockeler:2012yj}
M.~Gockeler, R.~Horsley, M.~Lage, U.~G. Meissner, P.~E.~L. Rakow, A.~Rusetsky,
  G.~Schierholz, and J.~M. Zanotti, ``{Scattering phases for meson and baryon
  resonances on general moving-frame lattices},''
  \href{http://dx.doi.org/10.1103/PhysRevD.86.094513}{Phys. Rev. {\bfseries
  D86} (2012) 094513},
\href{http://arxiv.org/abs/1206.4141}{{\ttfamily arXiv:1206.4141 [hep-lat]}}.
%%CITATION = ARXIV:1206.4141;%%.

\bibitem{Morningstar:2013bda}
C.~Morningstar, J.~Bulava, B.~Fahy, J.~Foley, Y.~C. Jhang, K.~J. Juge,
  D.~Lenkner, and C.~H. Wong, ``{Extended hadron and two-hadron operators of
  definite momentum for spectrum calculations in lattice QCD},''
  \href{http://dx.doi.org/10.1103/PhysRevD.88.014511}{Phys. Rev. {\bfseries
  D88} no.~1, (2013) 014511},
\href{http://arxiv.org/abs/1303.6816}{{\ttfamily arXiv:1303.6816 [hep-lat]}}.
%%CITATION = ARXIV:1303.6816;%%.

\bibitem{Paul:2018yev}
S.~Paul {\em et~al.}, ``{Towards the P-wave nucleon-pion scattering amplitude
  in the $\Delta (1232)$ channel},''
  \href{http://dx.doi.org/10.22323/1.334.0089}{PoS {\bfseries LATTICE2018}
  (2018) 089},
\href{http://arxiv.org/abs/1812.01059}{{\ttfamily arXiv:1812.01059 [hep-lat]}}.
%%CITATION = ARXIV:1812.01059;%%.

\bibitem{Albanese:1987ds}
{\bfseries APE} Collaboration, M.~Albanese {\em et~al.}, ``{Glueball Masses and
  String Tension in Lattice QCD},''
\href{http://dx.doi.org/10.1016/0370-2693(87)91160-9}{Phys. Lett. {\bfseries
  B192} (1987) 163--169}.
%%CITATION = PHLTA,B192,163;%%.

\bibitem{Bonnet:2000dc}
F.~D.~R. Bonnet, P.~Fitzhenry, D.~B. Leinweber, M.~R. Stanford, and A.~G.
  Williams, ``{Calibration of smearing and cooling algorithms in SU(3): Color
  gauge theory},'' \href{http://dx.doi.org/10.1103/PhysRevD.62.094509}{Phys.
  Rev. {\bfseries D62} (2000) 094509},
\href{http://arxiv.org/abs/hep-lat/0001018}{{\ttfamily arXiv:hep-lat/0001018
  [hep-lat]}}.
%%CITATION = HEP-LAT/0001018;%%.

\bibitem{Morningstar:2003gk}
C.~Morningstar and M.~J. Peardon, ``{Analytic smearing of SU(3) link variables
  in lattice QCD},'' \href{http://dx.doi.org/10.1103/PhysRevD.69.054501}{Phys.
  Rev. {\bfseries D69} (2004) 054501},
\href{http://arxiv.org/abs/hep-lat/0311018}{{\ttfamily arXiv:hep-lat/0311018
  [hep-lat]}}.
%%CITATION = HEP-LAT/0311018;%%.

\bibitem{Engel:2012qp}
{\bfseries BGR (Bern-Graz-Regensburg)} Collaboration, G.~P. Engel, C.~B. Lang,
  and A.~Schäfer, ``{Low-lying $\Lambda$ baryons from the lattice},''
  \href{http://dx.doi.org/10.1103/PhysRevD.87.034502}{Phys. Rev. {\bfseries
  D87} no.~3, (2013) 034502},
\href{http://arxiv.org/abs/1212.2032}{{\ttfamily arXiv:1212.2032 [hep-lat]}}.
%%CITATION = ARXIV:1212.2032;%%.

\bibitem{Gromes:1982ze}
D.~Gromes, ``{The Mysterious Spin Orbit Interactions in Baryons, Nonlocal
  Forces and the $P$ Wave Resonances},''
\href{http://dx.doi.org/10.1007/BF01571366}{Z. Phys. {\bfseries C18} (1983)
  249}.
%%CITATION = ZEPYA,C18,249;%%.

\bibitem{Edwards:2012fx}
{\bfseries Hadron Spectrum} Collaboration, R.~G. Edwards, N.~Mathur, D.~G.
  Richards, and S.~J. Wallace, ``{Flavor structure of the excited baryon
  spectra from lattice QCD},''
  \href{http://dx.doi.org/10.1103/PhysRevD.87.054506}{Phys. Rev. {\bfseries
  D87} no.~5, (2013) 054506},
\href{http://arxiv.org/abs/1212.5236}{{\ttfamily arXiv:1212.5236 [hep-ph]}}.
%%CITATION = ARXIV:1212.5236;%%.

\bibitem{Briceno:2017max}
R.~A. Briceno, J.~J. Dudek, and R.~D. Young, ``{Scattering processes and
  resonances from lattice QCD},''
  \href{http://dx.doi.org/10.1103/RevModPhys.90.025001}{Rev. Mod. Phys.
  {\bfseries 90} no.~2, (2018) 025001},
\href{http://arxiv.org/abs/1706.06223}{{\ttfamily arXiv:1706.06223 [hep-lat]}}.
%%CITATION = ARXIV:1706.06223;%%.

\bibitem{Hashimoto:1999yp}
S.~Hashimoto, A.~X. El-Khadra, A.~S. Kronfeld, P.~B. Mackenzie, S.~M. Ryan, and
  J.~N. Simone, ``{Lattice QCD calculation of $\bar{B} \to D \ell \bar{\nu}$
  decay form-factors at zero recoil},''
  \href{http://dx.doi.org/10.1103/PhysRevD.61.014502}{Phys. Rev. {\bfseries
  D61} (1999) 014502},
\href{http://arxiv.org/abs/hep-ph/9906376}{{\ttfamily arXiv:hep-ph/9906376
  [hep-ph]}}.
%%CITATION = HEP-PH/9906376;%%.

\bibitem{ElKhadra:2001rv}
A.~X. El-Khadra, A.~S. Kronfeld, P.~B. Mackenzie, S.~M. Ryan, and J.~N. Simone,
  ``{The Semileptonic decays $B \to \pi \ell \nu$ and $D \to \pi \ell \nu$ from
  lattice QCD},'' \href{http://dx.doi.org/10.1103/PhysRevD.64.014502}{Phys.
  Rev. {\bfseries D64} (2001) 014502},
\href{http://arxiv.org/abs/hep-ph/0101023}{{\ttfamily arXiv:hep-ph/0101023
  [hep-ph]}}.
%%CITATION = HEP-PH/0101023;%%.

\bibitem{Lehner:2012bt}
C.~Lehner, ``{Automated lattice perturbation theory and relativistic heavy
  quarks in the Columbia formulation},''
  \href{http://dx.doi.org/10.22323/1.164.0126}{PoS {\bfseries LATTICE2012}
  (2012) 126},
\href{http://arxiv.org/abs/1211.4013}{{\ttfamily arXiv:1211.4013 [hep-lat]}}.
%%CITATION = ARXIV:1211.4013;%%.

\bibitem{Detmold:2015aaa}
W.~Detmold, C.~Lehner, and S.~Meinel, ``{$\Lambda_b \to p \ell^-
  \bar{\nu}_\ell$ and $\Lambda_b \to \Lambda_c \ell^- \bar{\nu}_\ell$ form
  factors from lattice QCD with relativistic heavy quarks},''
  \href{http://dx.doi.org/10.1103/PhysRevD.92.034503}{Phys. Rev. {\bfseries
  D92} no.~3, (2015) 034503},
\href{http://arxiv.org/abs/1503.01421}{{\ttfamily arXiv:1503.01421 [hep-lat]}}.
%%CITATION = ARXIV:1503.01421;%%.

\bibitem{Briceno:2014uqa}
R.~A. Brice\~no, M.~T. Hansen, and A.~Walker-Loud, ``{Multichannel 1
  $\rightarrow$ 2 transition amplitudes in a finite volume},''
  \href{http://dx.doi.org/10.1103/PhysRevD.91.034501}{Phys. Rev. D {\bfseries
  91} no.~3, (2015) 034501}, \href{http://arxiv.org/abs/1406.5965}{{\ttfamily
  arXiv:1406.5965 [hep-lat]}}.

\bibitem{Briceno:2015csa}
R.~A. Brice\~no and M.~T. Hansen, ``{Multichannel 0 $\to$ 2 and 1 $\to$ 2
  transition amplitudes for arbitrary spin particles in a finite volume},''
  \href{http://dx.doi.org/10.1103/PhysRevD.92.074509}{Phys. Rev. D {\bfseries
  92} no.~7, (2015) 074509}, \href{http://arxiv.org/abs/1502.04314}{{\ttfamily
  arXiv:1502.04314 [hep-lat]}}.

\bibitem{Beylich:2011aq}
M.~Beylich, G.~Buchalla, and T.~Feldmann, ``{Theory of $B \to K^{(*)}\ell^+
  \ell^-$ decays at high $q^2$: OPE and quark-hadron duality},''
  \href{http://dx.doi.org/10.1140/epjc/s10052-011-1635-0}{Eur. Phys. J. C
  {\bfseries 71} (2011) 1635}, \href{http://arxiv.org/abs/1101.5118}{{\ttfamily
  arXiv:1101.5118 [hep-ph]}}.

\bibitem{UTfit}
{\bfseries UTfit} Collaboration.
\newblock \url{http://www.utfit.org/UTfit/ResultsSummer2018SM}.

\bibitem{Horgan:2013hoa}
R.~R. Horgan, Z.~Liu, S.~Meinel, and M.~Wingate, ``{Lattice QCD calculation of
  form factors describing the rare decays $B \to K^* \ell^+ \ell^-$ and $B_s
  \to \phi \ell^+ \ell^-$},''
  \href{http://dx.doi.org/10.1103/PhysRevD.89.094501}{Phys. Rev. D {\bfseries
  89} no.~9, (2014) 094501}, \href{http://arxiv.org/abs/1310.3722}{{\ttfamily
  arXiv:1310.3722 [hep-lat]}}.

\bibitem{Horgan:2009ti}
R.~Horgan {\em et~al.}, ``{Moving NRQCD for heavy-to-light form factors on the
  lattice},'' \href{http://dx.doi.org/10.1103/PhysRevD.80.074505}{Phys. Rev. D
  {\bfseries 80} (2009) 074505},
  \href{http://arxiv.org/abs/0906.0945}{{\ttfamily arXiv:0906.0945 [hep-lat]}}.

\bibitem{XSEDE}
J.~{Towns}, T.~{Cockerill}, M.~{Dahan}, I.~{Foster}, K.~{Gaither},
  A.~{Grimshaw}, V.~{Hazlewood}, S.~{Lathrop}, D.~{Lifka}, G.~D. {Peterson},
  R.~{Roskies}, J.~R. {Scott}, and N.~{Wilkins-Diehr}, ``{XSEDE: Accelerating
  Scientific Discovery},''
  \href{http://dx.doi.org/10.1109/MCSE.2014.80}{Computing in Science
  Engineering {\bfseries 16} no.~5, (2014) 62--74}.

\bibitem{Edwards:2004sx}
{\bfseries SciDAC, LHPC, UKQCD} Collaboration, R.~G. Edwards and B.~Joo, ``{The
  Chroma software system for lattice QCD},''
  \href{http://dx.doi.org/10.1016/j.nuclphysbps.2004.11.254}{Nucl. Phys. B
  Proc. Suppl. {\bfseries 140} (2005) 832},
  \href{http://arxiv.org/abs/hep-lat/0409003}{{\ttfamily
  arXiv:hep-lat/0409003}}.

\bibitem{Chroma}
R.~G. Edwards, B.~Joó, {\em et~al.}, ``{Chroma}.''
\newblock \url{https://github.com/JeffersonLab/chroma}.

\bibitem{JOO2015139}
B.~Joó, M.~Smelyanskiy, D.~D. Kalamkar, and K.~Vaidyanathan,
  \href{http://dx.doi.org/10.1016/B978-0-12-803819-2.00023-9}{``{Chapter 9 -
  Wilson Dslash Kernel From Lattice QCD Optimization},''} in {\em {High
  Performance Parallelism Pearls}}, pp.~139 -- 170.
\newblock Morgan Kaufmann, Boston, 2015.

\bibitem{QPhiX}
B.~Joó {\em et~al.}, ``{QPhiX Dslash and Solver Library}.''
\newblock \url{https://github.com/jeffersonlab/qphix}.

\bibitem{QLUA}
A.~Pochinsky, S.~Syritsyn, {\em et~al.}, ``{QLUA}.''
\newblock \url{https://usqcd.lns.mit.edu/w/index.php/QLUA}.

\bibitem{MDWF}
A.~Pochinsky, S.~Syritsyn, {\em et~al.}, ``{Möbius Domain Wall inverter}.''
\newblock \url{https://github.com/usqcd-software/mdwf}.

\bibitem{USQCD}
{\bfseries USQCD} Collaboration, ``{USQCD Software}.''
\newblock \url{http://usqcd-software.github.io}.

\end{thebibliography}
\end{document}